%% file: Main.tex
\documentclass[12pt,a4paper]{article} 
\pdfoutput=1
\usepackage[usenames,dvipsnames]{xcolor}
\usepackage{enumerate}
\usepackage{multirow}
\usepackage{rotating}
\usepackage{float}
\usepackage[compress]{cite}
\usepackage[colorlinks=true,pagebackref=true,urlcolor=Emerald,citecolor=blue,linkcolor=red]{hyperref}
\usepackage{amsmath,amssymb}
\usepackage[margin=10pt]{caption}
\usepackage{graphicx,subfigure}

\def\dlev{0}
\numberwithin{equation}{section}
\floatstyle{boxed}
\newfloat{floatbox}{t}{aux2}
\floatname{floatbox}{Box}
\newcommand{\eq}[1]{\begin{equation}#1\end{equation}}
\newcommand{\eqa}[1]{\begin{eqnarray}#1\end{eqnarray}}

\newcommand{\sssec}[1]{\subsubsection{#1}}

\newcommand{\ssecs}[1]{\subsection{#1\label{ssec-#1}}}
\newcommand{\sssecs}[1]{\subsubsection{#1\label{sssec-#1}}}

\newcommand{\ssecsx}[2]{\subsection{#1\label{ssec-#2-#1}}}
\newcommand{\sssecsx}[2]{\subsubsection{#1\label{sssec-#2-#1}}}

\ifnum\dlev=0
\usepackage[a4paper, includefoot, top=2.5cm, bottom=2.2cm, left=2.8cm, right=1.5cm]{geometry}

\newcommand{\ea}[1]{#1}
\newcommand{\ed}[1]{}
\fi\ifnum\dlev>0
\usepackage[top=2cm, bottom=2cm, left=1cm, right=1cm]{geometry}
\usepackage[normalem]{ulem}
\usepackage{soul}

\newcommand{\ea}[1]{\textcolor{ForestGreen}{#1}}
\newcommand{\ed}[1]{\textcolor{red}{\sout{#1}}}
\fi
\newcommand{\fig}[4][0.9]{\begin{figure}[tbp]\centering\captionsetup{width=#1\textwidth}\includegraphics[width=#1\textwidth]{\figname{#2}}\caption[#3]{#4}\label{fig-#2}\end{figure}}

\newcommand{\figi}[3][]{\subfigure[#3]{\includegraphics[width=#1\textwidth]{\figname{#2}}\label{fig-#2}}}
\newcommand{\figa}[5][0.9]{\begin{figure}[tbp]\centering #5\captionsetup{width=#1\textwidth}\caption[#3]{\label{fig-#2}#4}\end{figure}}
\newcommand{\figla}[4]{\begin{sidewaysfigure}[tbp]\centering #4\captionsetup{width=\textwidth}\caption[#2]{\label{fig-#1}#3}\end{sidewaysfigure}}
\ifnum\dlev>1

\renewcommand{\fig}[4][0.9]{\begin{figure}[tbp]\centering\captionsetup{width=#1\textwidth}\caption[#3]{#4}\label{fig-#2}\end{figure}}

\renewcommand{\figi}[3][]{\subfigure[#3]{\includegraphics[width=#1\textwidth,height=1px]{Graph/Graph-Zero.pdf}\label{fig-#2}}}
\fi
\newcommand{\tab}[4]{\begin{table}[tbp]\centering\caption[#2]{#3}#4\label{tab-#1}\end{table}}

\newcommand{\refeqraw}[1]{Eq.\ (#1)}
\newcommand{\Refeqraw}[1]{Eq.\ (#1)}
\newcommand{\refeq}[1]{\refeqraw{\ref{eq-#1}}}
\newcommand{\refsec}[1]{Chapter \ref{sec-#1}}
\newcommand{\refssec}[1]{Section \ref{ssec-#1}}
\newcommand{\refsssec}[1]{Section \ref{sssec-#1}}
\newcommand{\refig}[1]{Figure \ref{fig-#1}}
\newcommand{\reftab}[1]{Table\ \ref{tab-#1}}
\newcommand{\Refeq}[1]{\Refeqraw{\ref{eq-#1}}}

\newcommand{\Reftab}[1]{Table\ \ref{tab-#1}}
\newcommand{\subs}[1]{_\mathrm{#1}}
\newcommand{\sups}[1]{^\mathrm{#1}}
\newcommand{\dd}[1]{\mathrm{d}#1}
\newcommand{\mpl}{M\subs{p}}

\newcommand{\hf}{\frac{1}{2}}
\newcommand{\cm}[1]{}
\newcommand{\Gaunt}[6]{G_{#1}^{#4}{}_{#2}^{#5}{}_{#3}^{#6}}
\newcommand{\Ga}[3]{\Gaunt{l_{#1}}{l_{#2}}{l_{#3}}{m_{#1}}{m_{#2}}{m_{#3}}}
\newcommand{\Wigner}[6]{\left(\begin{array}{ccc}#1&#2&#3\\#4&#5&#6\end{array}\right)}

\newcommand{\Wj}[3]{\Wigner{l_{#1}}{l_{#2}}{l_{#3}}{m_{#1}}{m_{#2}}{m_{#3}}}
\newcommand{\Wjz}[3]{\Wigner{l_{#1}}{l_{#2}}{l_{#3}}{0}{0}{0}}
\newcommand{\vs}[1]{{\mathbf{#1}}}
\newcommand{\uv}[1]{\vs{\hat{#1}}}

\def\fNL{f\subs{NL}}
\def\tNL{\tau\subs{NL}}
\def\gNL{g\subs{NL}}
\def\hf{\frac{1}{2}}
\def\cO{{\cal O}}
\def\cH{{\cal H}}
\def\P{\mathcal P}
\def\ho{\mathrm{higher\ order}}
\def\title{Spectator fields and their imprints on the Cosmic Microwave Background}
\def\me{Lingfei Wang}

\hypersetup{pdftitle={\title}, pdfsubject={Doctoral thesis of \me{} at Lancaster University.}, pdfauthor={\me}, pdfcreator={\me}}

\begin{document}
\ifnum\dlev=0
\linespread{2}
\renewcommand{\footnotesize}{\small}
\thispagestyle{empty}
\begin{center}
{\Huge {\bfseries {\title}} \par}\vspace*{10mm}
{\Large {\me\\B.Sc.} \par}
{\large
{{Physics\\Department of Physics}\par}
{{Lancaster University}\par}\vspace*{5mm}
{{June 2016}\par}\vspace*{6cm}
{{A thesis submitted to Lancaster University for the degree of\\Doctor of Philosophy in the Faculty of Science and Technology} \par}}
  \end{center}
\else
\title\\
\me, B.Sc.\\
Physics Department, Lancaster University, Lancaster LA1 4YB, UK
\fi
\abstract{}
\newpage
\section*{Abstract}
When a subdominant light scalar field ends slow roll during inflation, but well after the Hubble exit of the pivot scales, it may determine the cosmological perturbations. This thesis investigates how such a scalar field, the \emph{spectator}, may leave its impact on the Cosmic Microwave Background (CMB) radiation and be consequently constrained. We first introduce the observables of the CMB, namely the power spectrum $P_\zeta$, spectral index $n_s$ and its running $\dd n_s/\dd\ln k$, the non-Gaussianities $\fNL$, $\gNL$ and $\tNL$, and the lack of isocurvature and polarization modes. Based on these studies, we derive the cosmological predictions for the spectator scenario, revealing its consistency with the CMB for inflection point potentials, hyperbolic tangent potentials, and those with a sudden phase transition. In the end, we utilize the spectator scenario to explain the CMB power asymmetry, with a brief tachyonic fast-roll phase.
\newpage
\section*{Acknowledgements}
I would like to thank my supervisor Anupam Mazumdar for his support and encouragement during my whole PhD period. I also appreciate very much the guidance and support in career development from my former supervisor Yeuk-Kwan Edna Cheung. I am very grateful to every colleague I have interacted with for the insightful discussions either within the cosmology group at Lancaster, or at various conferences/seminars, or even merely by email.

Thank all the friends I met in UK for bringing so much fun to my life in the distant land. And thanks to all my friends and relatives back in China whose supports have been vital during my PhD. On top of everything, I would like to express my unreserved thanks to my parents, without whom anything in my life would be deemed impossible.

\newpage
\section*{Declaration}
This thesis is my own work and no portion of the work referred to in this thesis has been submitted in support of an application for another degree or qualification at this or any other institute of learning.\vspace{1em}

\noindent\ea{This thesis is based on the author's contribution to the following publications:
\begin{itemize}
\item $\delta N$ formalism (\refsssec{Multi-field evolution} to \refsssec{Separable potentials}) are based partially on the author's contribution to \cite{Mazumdar:2012jj} and partially on the author's unpublished work.
\item Spectator scenario (\refssec{spe-Spectator scenario}) and its perturbative calculations (\refssec{spe-Perturbations}) are based on \cite{Mazumdar:2012rs,Wang:2013oea}.
\item Single inflaton with a perfect fluid (\refssec{Single inflaton with a perfect fluid}), step function spectator (\refsssec{Step function spectator}), and inflection point spectator (\refsssec{Inflection point spectator}) are based on \cite{Wang:2013oea}.
\item Inflection point inflation (\refsssec{Inflection point inflation}) and hyperbolic tangent spectator (\refsssec{Hyperbolic tangent spectator}) are based on \cite{Wang:2013hva}.
\item CMB power asymmetry (\refssec{CMB power asymmetry}) and its origin from spectator fields (\refsec{Asymmetry}) are based on \cite{Wang:2013lda,Mazumdar:2013yta}.
\end{itemize}
The author of the thesis contributes at least half to each and every one of the above publications.}
\newpage
\tableofcontents
\ifnum\dlev=0
\newpage\listoffigures
\newpage\listoftables
\newpage
\fi
\newpage\section{Introduction\label{sec-Introduction}}\input{Chapters/Introduction.tex}

\newpage\section{Observations of the Cosmic Microwave Background (CMB)\label{sec-CMB}}\input{Chapters/CMB.tex}
\newpage\section{Single-field slow-roll inflation\label{sec-Single}}\input{Chapters/Single.tex}
\newpage\section{Multi-component inflation\label{sec-Multi}}\input{Chapters/Multi.tex}
\newpage\section{Spectator scenario\label{sec-Spectator}}\input{Chapters/Spectator.tex}

\newpage\section{CMB power asymmetry from spectator fields\label{sec-Asymmetry}}\input{Chapters/Asymmetry.tex}
\newpage\section{Conclusions\label{sec-Conclusions}}\input{Chapters/Conclusions.tex}

\appendix
\newpage\section{Special functions\label{sec-Funcs}}\input{Chapters/Funcs.tex}

\newpage
\bibliographystyle{jcap}
\bibliography{Main}

\end{document}

%% file: Chapters/Introduction.tex
There are countless mysteries in the universe -- dark matter, dark energy, blackholes, and the early universe, to list a few. We have been constantly pursuing the mysteries, and discovering new ones as well -- two hundred years ago these terminologies did not even exist. As old mysteries are solved, and new ones are discovered, we understand the universe progressively.

Thousands of years ago, we hardly knew anything about the universe, let alone its origin. In the Chinese ancient myths, the earth is a flat square and the sky an inverted bowl over the earth \cite{IChing:1000}, and was born from a giant egg. Numerous alternative beliefs co-existed, such as the earth should be a floating disc between heaven and underworld \cite{Aune:2003}, or a globe instead \cite{Dicks:1985}, perhaps with a habitable interior \cite{Verne:2008}. At that time, nobody was able to verify any of the proposals, simply because no one could travel afar from the tribe and witness the boundary.

Of course the earth is a globe, as we have now reached the consensus. In history, however, it took us thousands of year to confirm that. When the sailors were conquering the seas and oceans,  map-making and astronomy were essential for navigation. As our ancestors sailed afar, such advances enabled them to realize that the earth could not fit into a flat map. That was the first indication that the earth is a globe, which eventually led to the confirmation by Magellan. Our extended scope has hence drawn the conclusion that the earth is round.

On the other hand, the accuracy and precision of our measurements determine with the equal importance, if not more, the advances of our philosophical views of the universe. Through the precise observations, we have become aware that the earth is not the centre of the universe, and neither is the sun. On the contrary, it is now believed that the universe is statistically isotropic \cite{Ade:2015kab}, favoring no special location or direction, and therefore does not possess any astronomical centre.

We have also realized the universe is not static: stars form and and die; galaxies can also form or merge. Even the whole observable universe was discovered to be expanding. Consequently, this raised a series of philosophical questions. In theory, how and why does the universe expand? In reality, what is the evolutionary history of the universe? The first question was answered by the Friedmann-Robertson-Walker (FRW) metric \cite{Friedman:1922kd,Robertson:1933zz,Walker:1944}, adding an additional degree of freedom into the spacetime metric for a homogeneous universe, known as the scale factor. According to the Hilbert-Einstein action, the relative expansion rate of the universe -- the \emph{Hubble rate} -- is then found to be proportional to the square root of energy density of the universe which, on the other hand, also evolves as the universe expands -- energy densities of non-relativistic and ultra-relativistic particles are both diluted by the universe expansion, though at different rates (see \cite{Mukhanov:1990me}). The FRW metric also allows us to solve the latter question by tracing the universe backwards in time. The universe should hence be made of denser and hotter plasma in the past, suggesting a time-finite evolutionary history in the minimal scenario, which has been entitled the \emph{Hot Big Bang Theory}.

The Hot Big Bang Theory also predicts another observational consequence -- the \emph{Cosmic Microwave Background} (CMB) radiation \cite{Alpher:1948}. If the universe started from high-temperature plasma, at the \emph{recombination} it should cool down sufficiently, so that electrons and protons are able to form hydrogen atoms, which are electrically neutral. Before recombination, the plasma is opaque to photons, with extensive interactions which prevent photons from travelling freely. After recombination, hydrogen atoms hardly interact with photons, allowing them to free-stream. Such cosmic photons are then gradually redshifted to the microwave level today as the universe expands, filling the cosmos in all directions and forming a background. Since the speed of light is constant, the CMB photons we see today should have originated from a sphere, which centres on us. This sphere is known as the \emph{Last Scattering Surface} (LSS).

\fig{cmb-PlanckCMB}{The CMB temperature fluctuation map by Planck}{The observed CMB temperature fluctuation map by the Planck satellite. \cite{Planck:CMB}}
The CMB was observed for the first time in 1964, by Penzias and Wilson accidentally \cite{Penzias:1965wn}. The observed CMB has a blackbody spectrum at the temperature $\approx2.7$ Kelvin. It also demonstrated a suprisingly high isotropy from all directions of the universe. This very first observation has been confirmed by more recent ones, such as the Cosmic Background Explorer (COBE) \cite{Bennett:1996ce}, the Wilkinson Microwave Anisotropy Probe (WMAP) \cite{Bennett:2012zja,Hinshaw:2012fq}, the Atacama Cosmology Telescope (ACT) \cite{Sievers:2013ica}, the South Pole Telescope (SPT) \cite{Story:2012wx,Hou:2012xq}, Planck \cite{Planck:2013kta,Ade:2015lrj,Ade:2015ava,Ade:2015kab,Ade:2015xua,Adam:2015rua}, and the Background Imaging of Cosmic Extragalactic Polarization (BICEP) \cite{Ade:2014xna,Ade:2014gua,Ade:2015tva}. Meanwhile, the more recent observations have also detected tiny anisotropies in the CMB ($\sim10^{-5}$), or the \emph{CMB temperature fluctuations}, as shown in \refig{cmb-PlanckCMB}, which correspond to the primordial perturbation that is nearly scale invariant and Gaussian. The properties of the CMB anisotropies will be discussed in \refsec{CMB}.

The CMB observations are difficult to explain with the Hot Big Bang Theory. First, it fails to explain naturally why the universe is mostly isotropic, because in Hot Big Bang, the opposite sides of the LSS can never form causal contact or reach thermal equilibrium. On the other hand, it also lacks a mechanism to produce the almost scale invariant primordial perturbation.

The failure of the Hot Big Bang Theory saw the birth of the Cold Big Bang Theory, which is the currently most accepted theory of the early universe. It prefixes the Hot Big Bang with an exponential expansion phase, known as \emph{inflation} \cite{Kazanas:1980tx,Starobinsky:1980te,Guth:1980zm,Linde:1981mu,Linde:1983p1108}. In order to pre-establish the causal contacts and the thermal equilibrium, during inflation we need the universe to be dominated in energy density by one or more components whose equations of state are smaller than $-1/3$\ea{, or equilvalently for scalar fields, whose kinetic energies remain weaker than potential energies}. Besides, these components should be able to drive inflation for at least $\sim50$ e-folds of the universe expansion \cite{Alabidi:2005qi,Ade:2015lrj}\footnote{
One e-fold of the universe expansion is the period where the universe size becomes $e$ times of its original value (measured by the scale factor). And $e=2.718\cdots$ is the mathematical constant corresponding to the base of natural logarithm.}
. However, the negative equation of state cannot be achieved with ordinary non-relativistic matter or relativistic particles, whose equations of state are $0$ and $1/3$ respectively. For this reason, inflation should have emerged from some component(s) other than matter or radiation. After inflation, such component(s) should decay into the hot plasma which signals the start of the Hot Big Bang, during a stage known as \emph{reheating} \cite{Albrecht:1982mp,Kofman:1994rk,Shtanov:1994ce,Kofman:1997yn,Greene:1997fu,Felder:2000hj,Bassett:2005xm}.

The initial success of inflation utilizes just a scalar field, which moves very slowly, unlike the oscillating scalar fields in a static universe \cite{Kazanas:1980tx,Starobinsky:1980te,Guth:1980zm,Sato:1980yn,Linde:1981mu,Linde:1983p1108}. The slow motion can take place naturally for scalar fields in an expanding universe, where the Hubble rate of universe expansion enters the equation of motion as a friction term, resembling a very viscous fluid holding a harmonic oscillator. The field will perform an over-damped slow motion instead of oscillations if the Hubble rate is much larger than the mass of the scalar field, which can be achieved typically when the field exceeds the Planck scale\footnote{
The gravitational constant is defined as $G=6.674\times10^{-11}\mathrm{N\cdot m^2/kg^2}$. We use natural units in the thesis where the speed of light $c$, the Planck constant $\hbar$, and the Boltzmann constant $k_B$ are set to unity.}
$\mpl\equiv1/\sqrt G\approx1.2209\times10^{19}\,\mathrm{GeV}$, or when it is \emph{super-Planckian}. During the slow motion, the equation of state of the scalar field is close to $-1$, behaving as a \emph{cosmological constant}, whose equation of state is exactly $-1$ and whose energy density does not depend on the universe expansion. Due to the nature of the over-damped slow motion, this scenario has hence been named \emph{slow-roll inflation}. Slow-roll inflation can be terminated safely when the scalar field reaches sub-Planckian values as opposed to super-Planckian, which is known as the \emph{graceful exit} of slow-roll inflation. It is also shown to produce the right amount of perturbations through quantum fluctuations. (For a review on cosmological perturbations, see \cite{Mukhanov:1990me,Riotto:2002yw,Malik:2008im}.) Due to its simplicity and its very good agreement with observations, \emph{single-field slow-roll inflation} has become a major success in modern physics, and the inflation scenario has been crowned as the ``inflation paradigm''. We will discuss single-field slow-roll inflation in \refsec{Single}.

Providing the exponential expansion alone does not guarantee a model's success. It should also predict correctly all the other observables, such as the nearly scale invariant and Gaussian primordial perturbation. Due to its simplicity, single-field slow-roll inflation has its limitations in producing all the possible features in the cosmological observables, known as the \emph{consistency relations} (see \refssec{Testing single-field slow-roll inflation with the CMB}). This motivates cosmologists to look for alternative scenarios or extensions of the single-field slow-roll inflation, which can provide a broader range of predictions hoping to cover more possible features in the future CMB observations. Various scenarios of inflation have been proposed in recent years. Non-canonical scalar fields can enforce slow roll with the speed limit from non-canonical kinetic terms without having to reach super-Planckian values, such as in DBI inflation \cite{Alishahiha:2004eh}. When multiple fields coexist, they can induce numerous scenarios of inflation, such as hybrid inflation \cite{Linde:1993cn,Copeland:1994vg,GarciaBellido:1996ke,GarciaBellido:1996qt,Fonseca:2010nk}, assisted inflation\cite{Liddle:1998jc,Malik:1998gy,Copeland:1999cs,Mazumdar:2001mm,Jokinen:2004bp,Dimopoulos:2005ac,Easther:2005zr}, and many more \cite{Stewart:1994ts,Gherghetta:1995dv,ArmendarizPicon:1999rj,Dvali:1998pa,Dimopoulos:2000md,Linde:2001ae,Dimopoulos:2001ix,Kachru:2003sx,Dvali:2003vv,Easther:2004qs,Dimopoulos:2004uf,Dimopoulos:2002kt,Burgess:2005sb,Dimopoulos:2004yb,Dimopoulos:2003ce,BuenoSanchez:2006eq,Dimopoulos:2005bp,Alabidi:2005qi,Bassett:2005xm,Dimopoulos:2005bx,Allahverdi:2006iq,Dimopoulos:2007bp,Allahverdi:2006we,Allahverdi:2007wh,Allahverdi:2006cx,BuenoSanchez:2006xk,BuenoSanchez:2008nc,Sasaki:2008uc,Cai:2008qb,Chen:2009zp,Enqvist:2010vd,Chatterjee:2011qr,Mazumdar:2011ih,Mazumdar:2010sa,Frazer:2011br,Allahverdi:2011su,Downes:2011gi,Mazumdar:2012jj,Duplessis:2012nb,Wang:2013hva,Martin:2013tda,Dimopoulos:2014boa}. In these scenarios, fields responsible for driving inflation are called the \emph{inflatons}. Most of such scenarios are outside the scope of the thesis, so we will only briefly mention the relevant multi-component inflation with canonical scalar fields in \refsec{Multi}.

Among the multi-field inflation scenarios, the minimal scenario is that one field (the inflaton) leads inflation but produces negligible perturbations, while the other field (the \emph{spectator}) is only responsible for generating the primordial perturbation but has absolutely no role in inflation, as discussed in \refsec{Spectator}. The two fields do not need to interact with each other, except minimally by gravity. In this sense, if the spectator field is perturbed or even removed from the model, inflation can still proceed without any change. The only difference is that the primordial perturbation would be much weaker.

The first realization of these separate roles is the curvaton scenario \cite{Moroi:2001ct,Lyth:2001nq,Enqvist:2001zp,Lyth:2002my}, as will be discussed briefly in \refssec{Curvaton scenario}. In the curvaton scenario, the scalar field responsible for the curvature perturbation is called a \emph{curvaton}. The curvaton field can be as simple as a light field without any coupling. It behaves as a\ea{n effective} cosmological constant during inflation, and only decays after inflation ends. The curvaton field can take up a significant part of the total energy density in the post-inflationary evolution. This greatly limits its parameter space because observations fail to see any isocurvature perturbations \cite{Lyth:2003ip,Ade:2015lrj}.

A more recent development of the separate roles is the spectator scenario \cite{Mazumdar:2012rs,Wang:2013oea}, in which the spectator ends slow roll well before the end of inflation. Therefore, the energy density of the spectator field or its decay products is redshifted away in the rest of inflation, leaving negligible contributions to the contents of the current universe, as discussed in \refssec{spe-Spectator scenario}. The only signature left from the spectator field is the primordial perturbation, which originated from the spectator perturbation at the Hubble exit (see \refssec{spe-Perturbations}). The price to pay is that spectator field potentials, such as the typical ones in \refssec{Spectator models}, are more complicated than a bare non-interacting light field.

A recent CMB feature which has come into people's attention is the \emph{CMB power asymmetry}. The CMB power asymmetry is the amount of asymmetry in the CMB power spectrum. For example, we may observe that the amplitude of the CMB perturbation is stronger on one hemisphere than that on the other. The CMB power asymmetry was first noticed in the WMAP data \cite{Eriksen:2003db,Eriksen:2007pc,Hoftuft:2009rq}, and later confirmed with a higher precision by the Planck satellite \cite{Ade:2013nlj,Ade:2015kab}. After modelling the CMB power asymmetry phenomenologically in \refssec{CMB power asymmetry}, we attempt to address its primordial origins in \refssec{Primordial origins of the CMB power asymmetry}, which turns extremely difficult if the primordial perturbation is (almost) scale invariant and sufficiently Gaussian (\refssec{Lack of asymmetry from scale invariant perturbations}). The observed amount of CMB power asymmetry, on the other hand, can be obtained in the presence of a brief tachyonic fast roll phase, through enhancing the very large scale perturbations, as shown in \refssec{Enhancing very large scale perturbations from a tachyonic fast-roll phase}. Therefore, the CMB power asymmetry can be explained by the spectator scenario with a tachyonic fast roll phase (\refssec{CMB power asymmetry from spectator scenario}), whilst satisfying other observational constraints in \refssec{Other observational constraints}.

The conclusion of the thesis is drawn in \refsec{Conclusions}.

%% file: Chapters/CMB.tex
In this chapter, we derive the statistical properties of the CMB temperature fluctuation. They shall constrain the early universe models in the forthcoming chapters.
\ssecs{CMB angular power spectrum}
Let us define the CMB temperature map as $T(\uv n)$, where $\uv n$ is the unit spatial three-vector for the 3-dimensional incoming direction of the observed CMB photons. The CMB temperature anisotropy is then defined as
\eq{\Delta T(\uv n)\equiv T(\uv n)-\overline T,\label{eq-cmb-T0}}
where the mean temperature of the CMB  is given by
\eq{\overline T\equiv\frac{1}{4\pi^2}\int\dd^2\uv n\,T(\uv n).}
The statistical information of the CMB temperature fluctuation can be extracted by decomposing $\Delta T(\uv n)$ into spherical harmonics $Y_{lm}(\uv n)$, with $l=0,1,2,\dots$ and $m=-l,-l+1,\dots,l$, as
\eq{\Delta T(\uv n)=\sum_{lm}a_{lm}Y_{lm}(\uv n).\label{eq-cmb-a0}}
The coefficients $a_{lm}$ are then called the \emph{angular multipoles} of the CMB temperature fluctuation. From the orthogonality condition of the spherical harmonics (see \refssec{Spherical harmonic functions}), we can solve $a_{lm}$ inversely, as
\eq{a_{lm}\equiv\int\dd^2\uv n\,\Delta T(\uv n)Y_{lm}^*(\uv n).\label{eq-cmb-a}}
The $a_{00}$ component should be exactly zero by definition, \refeq{cmb-T0}. The temperature fluctuation $\Delta T(\uv n)$ is a real function. This enforces the relation
\eq{a_{lm}^*=(-1)^ma_{l-m}.}

In the simplest scenario, we assume that no point is statistically preferred or different over the others in the universe, so that every point is statistically equivalent by nature. It also means there is no preferred direction from any point in the universe, including the earth. This is called \emph{statistical isotropy} (see for example \cite{Dodelson:2003ft}). Therefore the expectation value of any correlation function should remain invariant under spatial rotations \cite{Luo:1993xx,Hu:2001fa}. The statistical isotropy has been well tested \emph{for our observable universe patch} through the CMB \cite{Ade:2015kab}.

For the two-point correlation function, statistical isotropy means that it should only depend on the angle between the two directions, i.e.
\eq{\langle\Delta T(\uv n)\Delta T(\uv n')\rangle=F(\uv n\cdot\uv n').\label{eq-cmb-co2iso}}
This relation enforces $a_{lm}$ to satisfy
\eq{\langle a_{lm}^*a_{l'm'}\rangle=C_l\delta_{l,l'}\delta_{m,m'},\label{eq-cmb-C}}
where $C_l$ is the \emph{angular power spectrum} of the CMB, and is the independent of $m$ and $m'$. Here $\langle\ \rangle$ takes the expectation value over all the possible configurations of the universe arising from the quantum fluctuations in the early universe.

From the observed $a_{lm}$, an unbiased estimator for $C_l$ can be constructed as
\eq{\widehat C_l\equiv\frac{1}{2l+1}\sum_ma_{lm}^*a_{lm},}
whose variance is given by \cite{Dodelson:2003ft}
\eq{\mathrm{Var}\,\widehat C_l\equiv\left\langle\left(\widehat C_l-\langle\widehat C_l\rangle\right)^2\right\rangle=\frac{C_l^2}{2l+1},}
which is known as the \emph{cosmic variance}, and which cannot be lessened via multiple measurements.

The quantum fluctuations during inflation can be parameterized by the gauge-invariant\footnote{
The gauge invariance will be discussed further in \refssec{ssi-First order perturbations}.}
scalar quantity, the (primordial) \emph{curvature perturbation} $\zeta(\vs x)$, or its Fourier transformation partner $\zeta(\vs k)$, which is defined as\footnote{
We use the bold form to indicate spatial three-vectors in this thesis. The time dependence of $\zeta$ is implicit here.}
\eq{\zeta(\vs k)\equiv\int\zeta(\vs x)e^{-i\vs k\cdot \vs x}\dd^3\vs x.}
Since $\zeta(\vs x)$ is a scalar, it is also called the \emph{scalar perturbations} or the CMB temperature fluctuations. (See \refsssec{ssi-Scalar perturbations}.)

\fig{cmb-PlanckDl}{The $D_l$ curve from Planck}{The $D_l$ observation from Planck \cite{Ade:2015xua}. The blue points are the observed $D_l$ with error bars at $1\sigma$ confidence level. The red curve shows the best fit curve of $D_l$. The residues w.r.t the best fit are shown in the lower figure.}
We can write the transfer function from the curvature perturbation to the CMB temperature perturbation as $g_l(k)$, which leads to
\eq{a_{lm}=(-i)^l\int\frac{\dd^3\vs k}{2\pi^2}\,Y_{lm}^*(\uv k)g_l(k)\zeta(\vs k).\label{eq-cmb-alm}}
The detailed derivation can be found in \cite{Dodelson:2003ft}. The two-point correlation function of $\zeta(\vs k)$ can be written as
\eq{\langle\zeta(\vs k)\zeta^\dag(\vs k')\rangle=(2\pi)^3\delta(\vs k-\vs k')P_\zeta(k),\label{eq-cmb-zeta2}}
where $P_\zeta(k)$ is the \emph{power spectrum} of the curvature perturbation $\zeta(\vs k)$. With the above relations, we can derive $C_l$ as
\eq{C_l=\frac{2}{\pi}\int\dd k\,k^2|g_l(k)|^2P_\zeta(k).\label{eq-cmb-ClPz}}
The Planck satellite has given its observation of the $C_l$ values as well as the best-fit curve in \refig{cmb-PlanckDl}, in terms of \cite{Planck:2013kta}
\eq{D_l\equiv\frac{l(l+1)}{2\pi}C_l.}

We can parameterize $P_\zeta(k)$ order by order around a reference scale $k_0$, as
\eq{k^3P_\zeta(k)=k_0^3P_\zeta(k_0)\left(\frac{k}{k_0}\right)^{n_s-1+\frac{1}{2}\frac{\dd n_s}{\dd\ln k}\ln\frac{k}{k_0}+\ho},\label{eq-cmb-Pz}}
where the Planck observation has chosen \cite{Ade:2013zuv,Ade:2015xua}
\eq{k_0\equiv0.05\mathrm{Mpc}^{-1}.}
In \refeq{cmb-Pz}, the parameter $n_s$ is called the \emph{spectral index} of scalar perturbations, where $s$ stands for ``scalar'', and $\frac{\dd n_s}{\dd\ln k}$ is the \emph{running of the spectral index}. They are both taken as constant values at the reference scale $k=k_0$. The running is compatible with zero by
\eq{\frac{\dd n_s}{\dd\ln k}=-0.0065\pm0.0076.}
After taking the zero running, Planck reports other parameters and their errors \cite{Ade:2015lrj,Ade:2015xua}
\eqa{\P_\zeta&=&\frac{k_0^3}{2\pi^2}P_\zeta(k_0)=(2.142\pm0.048)\times10^{-9},\label{eq-cmb-Pz0}\\
n_s&=&0.9667\pm0.0040.}
All errors are at $1\sigma$ confidence level unless otherwise noted. The term $P_\zeta$ characterizes the overall strength of the curvature perturbations, as defined in \refsssec{Separate universe approach}.

\ssecs{CMB angular bi-spectrum}
The curvature perturbation $\zeta(\vs k)$ may deviate from independent Gaussian distributions. This \emph{non-Gaussianity} can provide extra statistical information through correlation functions in the CMB. (See \cite{Bartolo:2004if} for a review.) The CMB \emph{angular bi-spectrum} is hence defined as
\eq{B_{l_1l_2l_3}^{m_1m_2m_3}\equiv\langle a_{l_1m_1}a_{l_2m_2}a_{l_3m_3}\rangle.\label{eq-cmb-B6}}
Statistical isotropy requires the expectation values of three-point correlation functions to be invariant under spatial rotations. Therefore, we can extract the rotational invariant part of the CMB angular bi-spectrum, $B_{l_1l_2l_3}$, as \cite{Luo:1993xx,Hu:2001fa}
\eq{B_{l_1l_2l_3}^{m_1m_2m_3}=B_{l_1l_2l_3}\Wj{1}{2}{3},\label{eq-cmb-B33}}
where \ea{the pair of parentheses correspond to }the $3j$ symbol\ea{,}\ed{ is} defined in \refssec{3j symbols}.

The three-point correlation function of $\zeta(\vs k)$ is defined as
\eq{\langle\zeta(\vs k_1)\zeta(\vs k_2)\zeta(\vs k_3)\rangle=(2\pi)^3\delta^3(\vs k_1+\vs k_2+\vs k_3)B(k_1,k_2,k_3),\label{eq-cmb-Bk}}
where, due to statistical isotropy, $B(k_1,k_2,k_3)$ does not depend on the directions. For independent Gaussian $\zeta(\vs k)$, we have $B(k_1,k_2,k_3)=0$.

Then we are able to calculate \refeq{cmb-B6} and \refeq{cmb-B33}. Noting the relations \cite{Wang:1999vf}
\eq{\delta^3(\vs k_1+\vs k_2+\vs k_3)=\frac{1}{(2\pi)^3}\int\dd^3\vs xe^{i(\vs k_1+\vs k_2+\vs k_3)\cdot\vs x},}
\eq{e^{i\vs k\cdot\vs x}=4\pi\sum_{lm}i^lj_l(kx)Y_{lm}(\uv k)Y_{lm}^*(\uv x),}
we can find
\eqa{B_{l_1l_2l_3}&=&\sqrt\frac{(2l_1+1)(2l_2+1)(2l_3+1)}{4\pi}\left(\frac{2}{\pi}\right)^3\Wjz{1}{2}{3}\nonumber\\
&&\times\int k_1^2k_2^2k_3^2x^2\dd k_1\dd k_2\dd k_3\dd xB(k_1,k_2,k_3)\nonumber\\
&&\times g_{l_1}(k_1)g_{l_2}(k_2)g_{l_3}(k_3)j_{l_1}(k_1x)j_{l_2}(k_2x)j_{l_3}(k_3x).}
Here $j_l(x)$ is the spherical Bessel function.

Inflation may generate non-Gaussianities \emph{locally}. The local non-Gaussianity in the curvature perturbation, if any, is known to be small \cite{Ade:2015ava}. Such near-Gaussian local effects can be written as a series expansion of the perfect Gaussian variable
\eq{\zeta(\vs x)=\zeta_G(\vs x)+\frac{3}{5}\fNL(\zeta_G^2(\vs x)-\langle\zeta_G^2(\vs x)\rangle)+\ho,\label{eq-cmb-zetang}}
where $\zeta_G(\vs x)$ is the perfect Gaussian variable, and $\fNL$ is the parameter to indicate the amount of deviation from perfect Gaussian distributions, or the amount of \emph{local non-Gaussianity}. This expansion is only valid when
\eq{\zeta_G(\vs x)\ll1,}
\eq{\fNL\zeta_G(\vs x)\ll1,}
and similarly for higher order terms. From \refeq{cmb-zetang}, we can find the leading order expectation values of $\zeta(\vs x)$
\eqa{\langle\zeta(\vs x)\rangle&=&0,\\
\langle\zeta^2(\vs x)\rangle&=&\langle\zeta_G^2(\vs x)\rangle,\label{eq-cmb-nGz2}\\
\langle\zeta^3(\vs x)\rangle&=&\frac{18}{5}\fNL\langle\zeta_G^2(\vs x)\rangle^2,\label{eq-cmb-fnlzeta3}\\
\dots.&&\nonumber}

The Fourier transformation of \refeq{cmb-zetang} yields,
\eq{\zeta(\vs k)=\zeta_G(\vs k)+\frac{3}{5}\fNL\int\frac{\dd^3\vs k'}{(2\pi)^3}\zeta_G(\vs k')\zeta_G(\vs k-\vs k').\label{eq-cmb-znG}}
This allows us to solve $B(k_1,k_2,k_3)$ for the local non-Gaussianity according to \refeq{cmb-Bk}, as\footnote{
Here we do not distinguish between $P_\zeta(k)$ and $P_{\zeta_G}(k)$ because they are equal at the leading order.} 
\eq{B(k_1,k_2,k_3)=\frac{6}{5}\fNL\Bigl(P_\zeta(k_1)P_\zeta(k_2)+P_\zeta(k_1)P_\zeta(k_3)+P_\zeta(k_2)P_\zeta(k_3)\Bigr).\label{eq-cmb-Bk3}}

The CMB angular bi-spectrum is then \cite{Bartolo:2004if,Wang:2014qda}
\eqa{B_{l_1l_2l_3}&=&\frac{6}{5}\fNL\sqrt\frac{(2l_1+1)(2l_2+1)(2l_3+1)}{4\pi}\Wjz{1}{2}{3}\nonumber\\
&&\times\int\dd x\,x^2\Bigl[\alpha_{l_1}(x)\beta_{l_2}(x)\beta_{l_3}(x)\nonumber\\
&&+\alpha_{l_2}(x)\beta_{l_1}(x)\beta_{l_3}(x)+\alpha_{l_3}(x)\beta_{l_1}(x)\beta_{l_2}(x)\Bigr],\label{eq-cmb-B3}}
where we have defined
\eqa{\alpha_l(x)&\equiv&\frac{2}{\pi}\int k^2\dd k\,g_l(k)j_l(kx),\\
\beta_l(x)&\equiv&\frac{2}{\pi}\int k^2\dd k\,P_\zeta(k)g_l(k)j_l(kx).}

Planck has given the latest constraint on the primordial \emph{local} bi-spectrum \cite{Ade:2015ava}
\eq{\fNL=0.8\pm5.0.}
We will only be interested in the local type bi-spectrum in this thesis. Other types of the primordial bi-spectra have also been constrained by Planck, such as the equilateral type ($\fNL=-4\pm43$) and the orthogonal type ($\fNL=-26\pm21$), which can be found in \cite{Ade:2015ava}.

\ssecs{CMB angular tri-spectrum}
Similarly, the CMB angular tri-spectrum can be studied with $\langle a_{l_1m_1}a_{l_2m_2}a_{l_3m_3}a_{l_4m_4}\rangle$, which can be decomposed into a Gaussian part and a non-Gaussian part
\eq{\langle a_{l_1m_1}a_{l_2m_2}a_{l_3m_3}a_{l_4m_4}\rangle=\langle a_{l_1m_1}a_{l_2m_2}a_{l_3m_3}a_{l_4m_4}\rangle\subs G+\langle a_{l_1m_1}a_{l_2m_2}a_{l_3m_3}a_{l_4m_4}\rangle\subs{nG}.\label{eq-cmb-a4dec}}

The non-Gaussian part $\langle a_{l_1m_1}a_{l_2m_2}a_{l_3m_3}a_{l_4m_4}\rangle\subs{nG}$ only gains contribution from interactions between the perturbation modes, so it vanishes for purely Gaussian CMB perturbations, and is also called the \emph{connected} part. The Gaussian part is also called the \emph{disconnected} part, contributing a constant amount to the angular tri-spectrum at the leading order \cite{Bartolo:2004if}\footnote{
Alternatively it can be expressed in the rotational invariant way as in \cite{Hu:2001fa}.}
\eqa{\langle a_{l_1m_1}a_{l_2m_2}a_{l_3m_3}a_{l_4m_4}\rangle\subs G&=&(-1)^{m_1+m_3}C_{l_1}C_{l_3}\delta_{l_1,l_2}\delta_{l_3,l_4}\delta_{m_1,-m_2}\delta_{m_3,-m_4}\nonumber\\
&&+(-1)^{m_1+m_2}C_{l_1}C_{l_2}\delta_{l_1,l_3}\delta_{l_2,l_4}\delta_{m_1,-m_3}\delta_{m_2,-m_4}\nonumber\\
&&+(-1)^{m_2+m_4}C_{l_2}C_{l_4}\delta_{l_1,l_4}\delta_{l_2,l_3}\delta_{m_1,-m_4}\delta_{m_2,-m_3}.\label{eq-cmb-a4nG}}

Given the CMB power spectrum $C_l$, we can then calculate the Gaussian part of the CMB angular tri-spectrum. Any significant excess observed would imply non-Gaussianities in the angular tri-spectrum. Only considering the inflationary effects on the CMB tri-spectrum, we can write the non-Gaussian part of the four-point correlation function of the curvature perturbations, as
\eq{\langle\zeta(\vs k_1)\zeta(\vs k_2)\zeta(\vs k_3)\zeta(\vs k_4)\rangle\subs{nG}=(2\pi)^3\delta^3(\vs k_1+\vs k_2+\vs k_3+\vs k_4)T\subs{nG}(\vs k_1,\vs k_2,\vs k_3,\vs k_4).\label{eq-cmb-Tk}}

The tri-spectrum also allows various shapes, two of which have received most attention are parameterized in terms of $\tNL$ and $\gNL$, as \cite{Bartolo:2004if}
\eqa{&&T\subs{nG}(\vs k_1,\vs k_2,\vs k_3,\vs k_4)\nonumber\\
&\displaystyle=&\frac{1}{2}\tNL\Bigl(P_\zeta(k_1)P_\zeta(k_{12})P_\zeta(k_3)+P_\zeta(k_1)P_\zeta(k_{12})P_\zeta(k_4)+P_\zeta(k_1)P_\zeta(k_{13})P_\zeta(k_2)\nonumber\\
&&+P_\zeta(k_1)P_\zeta(k_{13})P_\zeta(k_4)+P_\zeta(k_1)P_\zeta(k_{14})P_\zeta(k_2)+P_\zeta(k_1)P_\zeta(k_{14})P_\zeta(k_3)\nonumber\\
&&+P_\zeta(k_2)P_\zeta(k_{12})P_\zeta(k_3)+P_\zeta(k_2)P_\zeta(k_{12})P_\zeta(k_4)+P_\zeta(k_2)P_\zeta(k_{23})P_\zeta(k_1)\nonumber\\
&&+P_\zeta(k_2)P_\zeta(k_{23})P_\zeta(k_4)+P_\zeta(k_2)P_\zeta(k_{24})P_\zeta(k_1)+P_\zeta(k_2)P_\zeta(k_{24})P_\zeta(k_3)\nonumber\\
&&+P_\zeta(k_3)P_\zeta(k_{13})P_\zeta(k_2)+P_\zeta(k_3)P_\zeta(k_{13})P_\zeta(k_4)+P_\zeta(k_3)P_\zeta(k_{23})P_\zeta(k_1)\nonumber\\
&&+P_\zeta(k_3)P_\zeta(k_{23})P_\zeta(k_4)+P_\zeta(k_3)P_\zeta(k_{34})P_\zeta(k_1)+P_\zeta(k_3)P_\zeta(k_{34})P_\zeta(k_2)\nonumber\\
&&+P_\zeta(k_4)P_\zeta(k_{14})P_\zeta(k_2)+P_\zeta(k_4)P_\zeta(k_{14})P_\zeta(k_3)+P_\zeta(k_4)P_\zeta(k_{24})P_\zeta(k_1)\nonumber\\
&&+P_\zeta(k_4)P_\zeta(k_{24})P_\zeta(k_3)+P_\zeta(k_4)P_\zeta(k_{34})P_\zeta(k_1)+P_\zeta(k_4)P_\zeta(k_{34})P_\zeta(k_2)\Bigr)\nonumber\\
&&\displaystyle+\frac{54}{25}\gNL\Bigl(P_\zeta(k_1)P_\zeta(k_2)P_\zeta(k_3)+P_\zeta(k_1)P_\zeta(k_2)P_\zeta(k_4)+P_\zeta(k_1)P_\zeta(k_3)P_\zeta(k_4)\nonumber\\
&&+P_\zeta(k_2)P_\zeta(k_3)P_\zeta(k_4)\Bigr),\label{eq-cmb-Fc}}
where $\vs k_{12}\equiv\vs k_1+\vs k_2$.

The parameter $\gNL$ comes similarly with $\fNL$. When the curvature perturbations are not perfectly Gaussian, the higher order terms can be parameterized as
\eq{\zeta(\vs x)=\zeta_G(\vs x)+\frac{3}{5}\fNL(\zeta_G^2(\vs x)-\langle\zeta_G^2(\vs x)\rangle)+\frac{9}{25}\gNL\zeta_G^3(\vs x)+\ho.\label{eq-cmb-zet3}}
Therefore $\gNL$ characterizes the strength of the cubic correction term. The parameter $\tNL$ does not appear directly in \refeq{cmb-zet3}, but it corresponds to the second order contribution from the CMB local bi-spectrum.

Similar to the bi-spectrum calculations, we can start from \cite{Munshi:2009wy,Wang:2014qda}
\eqa{\langle a_{l_1m_1}a_{l_2m_2}a_{l_3m_3}a_{l_4m_4}\rangle\subs{nG}&=&\frac{1}{2\pi^5}\int\delta^3(\vs k_1+\vs k_2+\vs k_3+\vs k_4)T\subs{nG}(\vs k_1,\vs k_2,\vs k_3,\vs k_4)\nonumber\\
&&\times\prod_{n=1}^4(-i)^{l_n}Y^*_{l_nm_n}(\uv{k}_n)g_{l_n}(k_n)\dd^3\vs k_n,\label{eq-cmb-ac}}
and calculate the contributions from the non-vanishing $\gNL$ and $\tNL$ terms, as \cite{Wang:2014qda}
\eqa{&&\langle a_{l_1m_1}a_{l_2m_2}a_{l_3m_3}a_{l_4m_4}\rangle\subs{nG}\nonumber\\
&=&\tNL\sum_{LM}(-1)^{M+l_1+l_2+l_3+l_4}\biggl(A^{l_1l_2}_{l_3l_4}(L)G^{m_1m_2M}_{l_1\ \,l_2\ \,L}G^{m_3m_4-M}_{l_3\ \,l_4\ \ L}\nonumber\\
&&+A^{l_1l_3}_{l_2l_4}(L)G^{m_1m_3M}_{l_1\ \,l_3\ \,L}G^{m_2m_4-M}_{l_2\ \,l_4\ \ L}+A^{l_1l_4}_{l_2l_3}(L)G^{m_1m_4M}_{l_1\ \,l_4\ \,L}G^{m_2m_3-M}_{l_2\ \,l_3\ \ L}\biggr)\nonumber\\
&&+\frac{27}{25}\pi\gNL\sum_{LM}(-1)^MB_{l_1l_2l_3l_4}G^{m_1m_2M}_{l_1\ \,l_2\ \,L}G^{m_3m_4-M}_{l_3\ \,l_4\ \ L},\label{eq-cmb-a4}}
where
\eqa{A^{l_1l_2}_{l_3l_4}(L)&\equiv&\int k^2\dd k\,P_\zeta(k)\gamma_{l_1l_2,L}(k)\gamma_{l_3l_4,L}(k),\\
B_{l_1l_2l_3l_4}&\equiv&\int x^2\dd x\Bigl(\alpha_{l_1}(x)\beta_{l_2}(x)\beta_{l_3}(x)\beta_{l_4}(x)+\alpha_{l_2}(x)\beta_{l_1}(x)\beta_{l_3}(x)\beta_{l_4}(x)\nonumber\\
&&+\alpha_{l_3}(x)\beta_{l_1}(x)\beta_{l_2}(x)\beta_{l_4}(x)+\alpha_{l_4}(x)\beta_{l_1}(x)\beta_{l_2}(x)\beta_{l_3}(x)\Bigr),\\
\gamma_{l_1l_2,l}(k)&\equiv&\sqrt\frac{2}{\pi}\int x^2\dd x\,j_l(kx)\Bigl(\alpha_{l_1}(x)\beta_{l_2}(x)+\alpha_{l_2}(x)\beta_{l_1}(x)\Bigr).}
The Gaunt integral $\Ga{1}{2}{3}$ is defined in \refeq{asf-Gaunt}.

The Planck observations have constrained the angular tri-spectra of the shapes $\tNL<2800$ at $95\%$ confidence level \cite{Ade:2013ydc} and $\gNL=(-9.0\pm7.7)\times10^4$ \ea{at $1\sigma$ confidence level }\cite{Ade:2015ava}.

\ssecs{CMB polarization modes}
So far, we have discussed the CMB \emph{temperature} anisotropies. Besides the temperature, each CMB photon also contains an additional degree of freedom. This results in the possible polarization in the CMB, whose fluctuations can also be measured, and can \ea{in theory}\ed{surely} provide extra information for the early universe. \ea{However, the CMB polarization also gains contribution from other sources, such as lensing and foreground dust, which make the measurement of contribution from primordial effects particularly difficult.}

The CMB polarization can be decomposed into two separate modes, $E$ and $B$, as discussed in \cite{Zaldarriaga:1996xe}. The $B$ mode \ea{gains contribution from}\ed{corresponds to} the (primordial) tensor perturbations. Its strength is determined by the primordial tensor perturbation, which can be similarly parameterized as in \refeq{cmb-Pz} and \refeq{cmb-Pz0}. \ed{In stead of $P_\zeta$, f}\ea{F}or the primordial tensor perturbation\ea{,} we use the power spectrum symbol $\P_t$. The relative strength of the tensor perturbation compared with that of the curvature perturbation is defined as the tensor-to-scalar ratio:
\eq{r\equiv\frac{\P_t}{\P_\zeta}.}
Fluctuations in the CMB polarization has yet to be observed, suggesting a weak tensor perturbation. The current constraints are given by $r_{0.002}<0.07$ at $95\%$ CL \cite{Array:2015xqh} and $r<0.12$ at $95\%$ CL \cite{Ade:2015tva} respectively.\footnote{
Here $r_{0.002}$ indicates the value $r$ measured at the scale $0.002\mathrm{Mpc}^{-1}$, instead of the previously chosen reference scale $k_0\equiv 0.05\mathrm{Mpc}^{-1}$. We will omit the subscript $0.002$ for the remaining part of the thesis.}

\ssecs{Isocurvature perturbations}
Besides the perturbations in the CMB, there are also other types of perturbations in the universe, such as the energy density perturbations in visible matter, cold dark matter, and neutrinos, as well as the velocity perturbations in neutrinos. In principle, these perturbations can be either independent of each other, or have some correlations. However, the simplest scenario would be that all these perturbations originated from the same curvature perturbation $\zeta$ in early universe. In this simplest scenario, the universe would be called \emph{adiabatic}.

The adiabaticity of the universe can be violated when the equation of state of our universe is not a mere function of energy density, which may take place if there is more than one degree of freedom in the early universe (see \refsec{Multi}). Perturbations perpendicular to the adiabatic perturbation are referred to as \emph{isocurvature perturbations}, which can leave their signatures in the CMB \cite{Peebles:1970ag,Efstathiou:1986pba,Bucher:1999re}.

By observing the CMB, the Planck satellite has not seen any isocurvature perturbations. Therefore, multi-field models should not produce isocurvature perturbations more than what Planck could have observed. This regulates the parameter spaces of multi-field models. One specific example of multi-field models, the curvaton scenario \cite{Lyth:2001nq,Lyth:2002my} (see \refssec{Curvaton scenario}), produces correlated or anti-correlated isocurvature perturbations whose amplitude is hence required to be small (see \cite{Ade:2015lrj}).

In this thesis, we will not be involved in any calculations of the isocurvature perturbations. Instead, quantitative results are only referred from existing publications. For this reason, numerical details of the isocurvature perturbations are irrelevant and omitted in the thesis.

\ssecs{CMB power asymmetry}
In previous sections, we have assumed the universe is statistically isotropic, so the expectation values of correlation functions are invariant under spatial rotations, such as \refeq{cmb-co2iso} for CMB angular spectrum, and \refeq{cmb-B33} for CMB angular bi-spectrum. This assumption is too ideal to be fully tested, simply because we cannot jump out of our current universe patch. However, what we \emph{can} test is the statistical isotropy in our \emph{observable patch}.

\figa{cmb-PlanckDip0}{CMB power asymmetries along the maximal and the temperature dipole directions}{The two $C_l$ curves and their relative differences are calculated by the Planck group, from two opposite patches of the CMB map with an angular diameter $90^\circ$ \cite{Ade:2013nljv1}. The directions have been picked to match the maximal power asymmetry, and the temperature dipole.}{
\figi[0.8]{cmb-PlanckDip}{The power asymmetry along the maximal power asymmetry direction.}\\
\figi[0.8]{cmb-PlanckDip3}{The power asymmetry along the temperature dipole direction.}}
The Planck satellite has tested statistical isotropy by various means in \cite{Ade:2013nlj}, highlighting the power asymmetry of the CMB\footnote{
In \cite{Ade:2015kab}, the latest Planck release has confirmed their previous results in \cite{Ade:2013nlj}.}
. The \emph{power asymmetry} of the CMB suggests the amplitude of CMB temperature perturbations can be asymmetric, so perturbations on one side can be stronger than those on the opposite side. This signal was also found by the previous observations in \cite{Eriksen:2003db,Eriksen:2007pc}, and may as well show up in the CMB polarization perturbations \cite{Namjoo:2014pqa}. This might suggest a ``preferred direction'' in our observable universe patch.

In \refig{cmb-PlanckDip0}, the Planck group have processed two opposite patches of the CMB map, and have shown the two $C_l$ curves together with their relative differences. The preferred direction from the maximal CMB power asymmetry and that from the CMB temperature dipole were found not to match. The CMB power asymmetry thus may have a different origin with that of the temperature dipole.

\figa{cmb-PlanckDip02}{Subsequent observational studies on CMB power asymmetry}{The relative differences of the opposite $C_l$ curves are plotted in more recent studies.}{\figi[0.85]{cmb-PlanckDip2}{The measured relative differences are shown in the lines with different masks. The simulated relative differences from pure Gaussian universes are shown with the shades, at $1\sigma$ significance. \cite{Flender:2013jja}}\\
\figi[0.85]{cmb-PlanckDip4}{The measured relative differences are shown in the coloured lines. The simulated relative differences from pure Gaussian universes are shown in gray lines (each), and also shown with the shade at $1\sigma$ significance. \cite{Ade:2013nlj}}}
In the later analyses \cite{Ade:2013nlj,Flender:2013jja}, the power asymmetry becomes less significant for high $l$, as shown in \refig{cmb-PlanckDip02}. Still, the power asymmetry remains at low $l$, overcoming cosmic variance effect at $\sim3\sigma$.

To address the possible power asymmetry in the CMB, let us first consider the \emph{scale independent} case. We can start from the symmetric and unmodulated CMB temperature fluctuations $\overline{\Delta T}(\uv n)$, which by itself is statistically isotropic. The power asymmetry then can be modelled at the leading order as a dipole modulation multiplier along direction $\uv p$ with strength $A$. The CMB temperature fluctuations after modulation then become: (see for instance \cite{Eriksen:2003db,Eriksen:2007pc})
\eq{\Delta T(\uv{n})=(1+A\,\uv{p}\cdot\uv{n})\overline{\Delta T}(\uv{n}).}
The currently observed CMB power asymmetry is weak, so we expect $0<A\ll1$. The 2013 Planck observation \cite{Ade:2013nlj} sees $A=0.07\pm0.02$, which confirms the previous analyses on WMAP data \cite{Eriksen:2003db,Eriksen:2007pc}. The Planck 2015 results are ``essentially identical'' \cite{Ade:2015kab}.

If we pick a small local patch in the direction $\uv{n}$ on the CMB map, and calculate the CMB power spectrum only in this patch, it will also acquire a directional dependence (neglecting $\cO(A^2)$),
\eq{\P_{\Delta T}(\uv{n})=(1+2A\,\uv{p}\cdot\uv{n})\P_{\overline{\Delta T}}.}
This directional dependence becomes most significant when we compare two opposite directions, $\uv{n}=\uv{p}$ and $\uv{n}=-\uv{p}$. Their relative difference is given by \cite{Mazumdar:2013yta}
\eq{\frac{\P_{\Delta T}(\uv{p})-\P_{\Delta T}(-\uv{p})}{\frac{1}{2}(\P_{\Delta T}(\uv{p})+\P_{\Delta T}(-\uv{p}))}=4A.\label{eq-ca-PTd}}

Since the CMB temperature fluctuations are seeded by curvature perturbations, it is straightforward to think that the power asymmetry may share the same origin and also come from inflation. In fact, because of the large scale of CMB power asymmetry, it is difficult to be seeded by any post-inflationary mechanism. Further discussions about its inflationary origins and scale dependences are covered in \refsec{Asymmetry}.

Other than power asymmetry, the CMB tempearture fluctuation exhibits a simple pattern -- weak and almost scale invariant perturbations with negligible running, small and insignificant non-Gaussianities, and lack of $B$ polarization mode or isocurvature perturbations. Naively, a single scalar field in the early universe would suffice to produce all of them. This will be the topic of next chapter.

%% file: Chapters/Single.tex
Single-field slow-roll inflation is the earliest successful attempt in explaining CMB temperature fluctuations. In this chapter, we look into single-field slow-roll inflation at background and perturbation levels. We then examine power-law and inflection point potentials for single-field slow-roll inflation, which are compared against CMB observations.
\ssecsx{Background evolution}{ssi}
In the case of a single-field slow-roll inflation, we can start from the Einstein-Hilbert action with a real scalar inflaton $\phi$:
\eq{S\equiv\int\sqrt{-g}\,\dd^4x\left(\frac{\mpl^2}{16\pi}R-\frac{1}{2}\partial^\mu\phi\partial_\mu\phi-V(\phi)\right),\label{eq-ssi-S0}}
where \ea{the determinant is defined as}
\eq{g\equiv\det(g_{\mu\nu}),}
and the Planck mass is defined as $\mpl\equiv1/\sqrt G$.
The definition of the Ricci scalar $R$ can be found in many general relativity or cosmology textbooks, such as \cite{Dodelson:2003ft}. We have picked the sign convention of space-time metric $g_{\mu\nu}$ as $(-,+,+,+)$. Greek indices go through $0,1,2,3$ for the four space-time dimensions, and Latin ones correspond to $1,2,3$ only for the three spatial dimensions.

The metric $g_{\mu\nu}$ and the inflaton $\phi$ are functions of space and time in general. Assuming a mostly homogenous initial condition for single-field slow-roll inflation, where the perturbative approach is feasible, we can expand spatial perturbations order by order, as\footnote{
Inflation may also take place if the inhomogeneity is large, for example in the case of chaotic inflation and multi-verse \cite{Linde:1983p1108,Linde:1986fc,Burgess:2005sb}. We will not discuss these possibilities in the thesis.}
\eqa{g_{\mu\nu}(x^\mu)&=&\bar g_{\mu\nu}(t)+\delta g_{\mu\nu}(x^\mu),\\
\phi(x^\mu)&=&\bar\phi(t)+\delta\phi(x^\mu),\label{eq-ssi-dphidef}}
where the bars on top indicate background solution, and the $\delta$'s in front indicate perturbations which are space-time dependent. Throughout this thesis, we will simply drop the bars on top for background evolutions. Symbols without a $\delta$ in front would automatically indicate the background, unless explicit spatial dependences are specified. The background FRW metric can be written as \ea{(in flat space)}
\eq{g_{\mu\nu}\equiv\left(\begin{array}{cccc}-1&0&0&0\\0&a^2(t)&0&0\\0&0&a^2(t)&0\\0&0&0&a^2(t)\end{array}\right).\label{eq-ssi-gmunu0}}
The parameterization $a(t)$ is known as the \emph{scale factor}, parameterizing the ``size'' of the universe.

Now we solve the background dynamics of inflation, whose review can be found in \cite{Bassett:2005xm}. We define the relative expansion rate of the universe as
\eq{H\equiv\frac{\dd a}{a\dd t},}
which is known as the \emph{Hubble rate}. The Hubble rate \ea{is determined by}\ed{yields to} the Friedmann equation
\eq{H^2=\frac{8\pi\rho}{3\mpl^2},}
where $\rho$ is the total energy density of the universe. During single-field slow-roll inflation, we have
\eq{\rho=\frac{1}{2}\dot\phi^2+V(\phi).\label{eq-ssi-rho0}}
The equation of motion of the inflaton $\phi$ at background level can be derived from \refeq{ssi-S0} as
\eq{\ddot\phi+3H\dot\phi+V'(\phi)=0,\label{eq-ssi-eom0}}
where $\dot{}\equiv\dd/\dd t$, and primes on potentials indicate derivatives w.r.t the field.

In an expanding universe, the Hubble rate acts as a friction force for oscillating fields. When the friction is strong enough, $\phi$ becomes over-damped. This requires the conditions
\eqa{\epsilon_\phi&\equiv&\frac{4\pi V'{}^2}{9\mpl^2H^4}=\frac{\mpl^2V'{}^2}{16\pi\rho^2}<1,\label{eq-ssi-epsilon0}\\
\eta_\phi&\equiv&\frac{V''}{3H^2}=\frac{\mpl^2V''}{8\pi\rho},\hspace{0.5in}|\eta_\phi|<1.\label{eq-ssi-eta0}}

The parameters $\epsilon_\phi$ and $\eta_\phi$ are called the \emph{first} and \emph{second slow-roll parameters} for $\phi$ respectively. Similarly, \refeq{ssi-epsilon0} and \refeq{ssi-eta0} are the \emph{slow-roll conditions} for $\phi$. When the slow-roll conditions are well satisfied, i.e. $\epsilon_\phi\ll1$ and $|\eta_\phi|\ll1$, the slow-roll parameters $\epsilon_\phi$ and $\eta_\phi$ can be regarded as small variables, which then enable $\phi$ to roll very slowly. In such cases, we only need to care about the leading order contributions from $\epsilon_\phi$ and $\eta_\phi$, which is the so called \emph{slow-roll approximation}.

When the slow-roll approximation holds, we can find
\eqa{\ddot\phi&\ll&3H\dot\phi\approx-V'(\phi),\label{eq-ssi-eomapprox}\\
\dot\phi^2&\ll&V(\phi).}
Therefore, potential energy will dominate over kinetic energy during slow roll. Potential energy then drives inflation because it is not diluted by universe expansion. The kinetic term also changes slowly enough, so the second order derivative in the equation of motion becomes negligible at background level. The Hubble rate and the equation of motion then become
\eq{H^2=\frac{8\pi V}{3\mpl^2},\label{eq-ssi-Hsr}}
and
\eq{3H\dot\phi+V'(\phi)=0.\label{eq-ssi-eomsr}}

For single-field slow-roll inflation, we can confirm the following relations
\eqa{\epsilon_\phi&=&\frac{\dd}{\dd t}\frac{1}{H}=\frac{\dd\ln H}{\dd N},\label{eq-ssi-epsilon1}\\
\frac{\dd\ln\epsilon_\phi}{\dd N}&=&-4\epsilon_\phi+2\eta_\phi,\label{eq-ssi-eta1}}
where $N$, the \emph{number of remaining e-folds of the universe expansion} \ea{till a specific point, e.g.\ the end of inflation}, is defined as
\eq{\dd N\equiv-H\dd t,}
or alternatively
\eq{a(N)=a(N_0)e^{N_0-N},}
for any reference point at $N=N_0$, where we have chosen the e-folding $N$ as the time measure. Since the slow-roll conditions are only well satisfied for $\epsilon_\phi\ll1$ and $|\eta_\phi|\ll1$, single-field slow-roll inflation is terminated when either $\epsilon_\phi\ge1$ or $|\eta_\phi|\ge1$ is satisfied.

One important concept in the expanding universe is the total distance a photon can travel in the future, assuming the Hubble rate is kept constant. It is known as the \emph{\ea{event horizon}\ed{Hubble radius}} of the universe\ed{, and the area enclosed is the \emph{Hubble patch}}. The \ea{comoving event horizon}\ed{comoving Hubble radius} at any time $t_0$ is
\eq{\int_{t_0}^\infty\frac{\dd t}{a(t)}=\int_{t_0}^\infty\frac{\dd t}{a_0e^{H(t-t_0)}}=\frac{1}{a_0H_0}.}

\ea{For inflation, the value of comoving event horizon coincides with the \emph{comoving Hubble radius}, $1/a_0H_0$, which is the universe expansion rate in length scale. The volume enclosed by the \ea{comoving} Hubble radius is the \emph{Hubble patch}.} Scales much smaller than the \ea{comoving} Hubble radius are called \emph{sub-Hubble}, and those much larger are \emph{super-Hubble}. As the universe expands, equilibrium can only be established on sub-Hubble scales (if not pre-established).

When the universe is dominated by a perfect fluid with \ea{constant} equation of state \ea{parameter} $w$, its energy density has the power-law relation \cite{Mukhanov:1990me}
\eq{\rho\propto a^{-3(1+w)}.}
This indicates the \ea{comoving} Hubble radius would follow
\eq{\frac{1}{aH}\propto a^{\frac{1}{2}(1+3w)}.\label{eq-ssi-HubbleR}}

In \ea{the} Hot Big Bang, the universe starts with relativistic particles, and cools down gradually during expansion. As the temperature drops\ed{ below the typical mass of particles}, the universe moves from radiation dominated era to matter dominated era. The radiation dominated and matter dominated eras correspond to the equations of state $w=1/3$ and $w=0$ respectively, so the \ea{comoving} Hubble radius would always be increasing as the universe expands. Consequently, \ea{the} Hot Big Bang does not have a convincing mechanism to form thermal equilibrium on the LSS. This is one of the major difficulties of the Hot Big Bang theory.

Inflation solves the difficulty with the (near) exponential expansion phase. According to \refeq{ssi-epsilon1}, for $\epsilon_\phi\ll1$, the Hubble rate (and hence the energy density) decreases very slowly per e-fold of universe expansion. This means the universe is dominated by a near cosmological constant during inflation, from $\phi$ whose the equation of state is almost $-1$. The \ea{comoving} Hubble radius therefore decreases during inflation, according to \refeq{ssi-HubbleR}, which allows a pre-established thermal equilibrium on the LSS if inflation lasts sufficiently long.

The evolution of $\phi$ can be solved from the background equation of motion \refeq{ssi-eomapprox}, as
\eq{\frac{8\pi V(\phi)}{\mpl^2V'(\phi)}\dd\phi=\dd N.\label{eq-ssi-dN}}
When either of the slow-roll parameters reaches ${\mathcal O}(1)$, the single-field slow-roll inflation will come to an end. This allows us to define the end of single-field slow-roll inflation as
\eq{N=N_e\mathrm{\ at\ }\max(\epsilon_\phi,|\eta_\phi|)=1,\label{eq-ssi-eoi}}
where we use the subscript $e$ for the end of inflation.

\ssecsx{First order perturbations}{ssi}
During inflation, perturbations can exist \ea{i}\ed{o}n fields and/or the metric. We are only interested in scalar fields, so their perturbations will be discussed in \refsssec{ssi-Scalar perturbations}. The metric is a $4\times4$ real symmetric matrix, with 10 total degrees of freedom. They can be decomposed into 4 scalar degrees of freedom (see \refeq{ssi-dg}), 4 vector degrees of freedom, and 2 tensor degrees of freedom. The scalar perturbations couple with the energy density and pressure inhomogeneities; the vector perturbations are normally redshifted away as the universe expands; the tensor perturbations do not couple with other inhomogeneities, but only depend on the energy scale of inflation. Therefore, we are interested in the scalar and tensor perturbations, and treat them separately in this section. (See \cite{Kodama:1985bj,Mukhanov:1990me,Riotto:2002yw,Bassett:2005xm,Malik:2008im} for a review.)

\sssecsx{Scalar perturbations}{ssi}
First order scalar perturbations appear in both the metric and the field. The field perturbation is simply $\delta\phi(x^\mu)$, as defined in \refeq{ssi-dphidef}. The most generic form of the metric perturbations can be written as\footnote{
Here we perturb the metric with physical time, not the conformal time.}
\eq{\delta g_{\mu\nu}=\left(\begin{array}{c@{\hspace{0.3in}}c}-2A&\partial_iB\\\partial_iB&a^2(-2C\delta_{i,j}+D_{ij}E)\end{array}\right),\label{eq-ssi-dg}}
where $A,B,C,E$ correspond to different scalar perturbation modes which are in general space-time dependent. Here
\eq{D_{ij}\equiv\partial_i\partial_j-\frac{1}{3}\delta_{i,j}\nabla^2.}

\fig[0.8]{ssi-gauge}{The freedom of gauge choice during inflation}{We are free to choose any gauge in the universe, such as those shown in blue, red or green. The choice of gauge should not affect any physical process or quantity. The solid and dashed black curves illustrate the exponential universe expansion during inflation. The horizontal and vertical axes are for the physical spatial and time dimensions respectively.}
Then we are left with five scalar perturbations $\delta\phi,A,B,C,$ and $E$. There are however unphysical degrees of freedom in them. The same physical process can be portrayed with different sets of the perturbations by choosing a different reference frame, or gauge/slicing, as shown in \refig{ssi-gauge}. We will use tilde for the variables after transformations. So consider the infinitesimal local space-time transformation in comoving coordinates:
\eq{x^\mu\rightarrow\tilde x^\mu=x^\mu+\delta x^\mu,\label{eq-ssi-trans}}
in which the transformation $\delta x^\mu$ can depend on the space-time coordinates $x^\mu$. 

The field perturbation $\delta\phi$ follows the single transformation rule as
\eq{\delta\phi\rightarrow\widetilde{\delta\phi}=\delta\phi-\dot\phi\delta x^0.\label{eq-ssi-phigauge}}
Regarding the metric perturbations, any space-time line element should remain invariant, \ea{$\widetilde{\dd s^2}=\dd s^2$,} from 
\eq{\dd s^2=(g_{\mu\nu}+\delta g_{\mu\nu})\dd x^\mu\dd x^\nu\label{eq-ssi-dsold}}
to
\eq{\widetilde{\dd s^2}=(g_{\mu\nu}+\widetilde{\delta g_{\mu\nu}})\dd\tilde x^\mu\dd\tilde x^\nu.\label{eq-ssi-dsnew}}
The transformation rule of $A,B,C,E$ can thus be solved as follows \cite{Kodama:1985bj,Mukhanov:1990me,Riotto:2002yw,Bassett:2005xm,Malik:2008im}.

Before the coordinate transformation, the original line element is expressed in \refeq{ssi-dsold}. It can be expanded in the form
\eqa{\dd s^2&=&(g_{\mu\nu}+\delta g_{\mu\nu})\dd x^\mu\dd x^\nu\nonumber\\
&=&-(1+2A)\bigl(\dd x^0\bigr)^2+a^2\sum_{i=1}^3\Bigl(1-2C+2\partial_i\partial_iE\Bigr)\bigl(\dd x^i\bigr)^2\nonumber\\
&&+2\partial_iB\dd x^0\dd x^i+a^2\sum_{i\ne j}\partial_i\partial_j E\dd x^i\dd x^j.\label{eq-gauge-lineold}}
After transformation, the new line element (\refeq{ssi-dsnew}) becomes
\eqa{\widetilde{\dd s^2}&=&\Bigl(g_{\mu\nu}+\widetilde{\delta g_{\mu\nu}}\Bigr)\dd\tilde x^\mu\dd\tilde x^\nu\nonumber\\
&=&-\Bigl(1+2\widetilde A+2\partial_0\delta x^0\Bigr)\bigl(\dd x^0\bigr)^2\nonumber\\
&&+a^2\sum_{i=1}^3\Bigl(1-2\widetilde C+2\partial_i\partial_i\widetilde E+2\partial_i\delta x^i+2H\delta x^0\Bigr)\bigl(\dd x^i\bigr)^2\nonumber\\
&&+\Bigl(2\partial_i\widetilde B-\partial_i\delta x^0+a^2\partial_0\delta x^i\Bigr)\dd x^0\dd x^i\nonumber\\
&&+2a^2\sum_{i>j}\Bigl(\partial_i\partial_j\widetilde E+\partial_i\delta x^j+\partial_j\delta x^i\Bigr)\dd x^i\dd x^j.\label{eq-gauge-linenew}}

For any line element $\dd x^\mu$, the transformation \refeq{ssi-trans} should leave the length invariant. This hence requires each of the coefficients of $\dd x^\mu\dd x^\nu$ to be identical in \refeq{gauge-lineold} and \refeq{gauge-linenew}, i.e.
\eqa{A&=&\widetilde A+\partial_0\delta x^0,\\
\forall i,\hspace{0.2in}-C+\partial_i\partial_i E&=&-\widetilde C+\partial_i\partial_i\widetilde E+\partial_i\delta x^i+H\delta x^0,\label{eq-gauge-C1}\\
\forall i,\hspace{0.2in}\partial_i B&=&\partial_i\widetilde B-\partial_i\delta x^0+a^2\partial_0\delta x^i,\label{eq-gauge-B1}\\
\forall i>j,\hspace{0.2in}\partial_i\partial_j E&=&\partial_i\partial_j\widetilde E+\partial_i\delta x^j+\partial_j\delta x^i.\label{eq-gauge-E1}}
The transformation rule for $A$ can then be solved as
\eq{\widetilde A=A-\partial_0\delta x^0.\label{eq-gauge-A2}}

We can decompose the spatial part of the coordinate transformation by defining $\beta$ and $v^i$ as
\eq{\delta x^i=\partial^i(a^2\beta)+v^i,\hspace{0.4in}\partial_iv^i=0.}
$\beta$ contributes to the change in the scalar metric perturbations, while the ``transverse vector'' $v^i$ only contributes to the vector perturbations, and thus is not of \ed{our }interest \ea{here}. The decomposition reshapes \refeq{gauge-B1} into
\eq{\partial_i\Bigl[\widetilde B-B-\delta x^0+\partial_0(a^2\beta)\Bigr]=-a^2\partial_0v^i,}
from which we can find the transformation rule for $B$, as
\eq{\widetilde B=B+\delta x^0-\partial_0(a^2\beta).\label{eq-gauge-B2}}
Similarly, we can derive the transformation rules for $C$ and $E$:
\eqa{\widetilde C&=&C-\frac{1}{3}\nabla^2\beta+H\delta x^0,\label{eq-gauge-C2}\\
\widetilde E&=&E-2\beta.\label{eq-gauge-E2}}

In this sense, it is possible to transform the coordinates and the scalar perturbations together, so the physical process remains the same but the perturbations ($A,B,C,E,\phi$) become different. This means we have unphysical gauge degrees of freedom in the representation.

There are typically two possible treatments\footnote{
For a review and/or lecture notes on cosmological perturbations, see \cite{Kodama:1985bj,Mukhanov:1990me,Riotto:2002yw,Bassett:2005xm,Malik:2008im}.
}:
\begin{itemize}
\item Construct gauge-invariant variables and limit the calculations to these gauge-invariant variables as much as possible \cite{Bardeen:1980kt,Bardeen:1983qw,Kodama:1985bj,Mukhanov:1990me,Stewart:1993bc}.
\item Choose the specific gauge that is most convenient for the calculations. The gauge invariance can be recovered later by combining the gauge dependent quantities.
\end{itemize}

In this thesis, we will employ the second treatment, because the relevant discussions will mostly take place in one gauge -- the \emph{spatially flat gauge}. In the spatially flat gauge, the spatial part of the metric is unperturbed, with $\delta g_{ij}=0$. This can be achieved from an arbitrary gauge, with the gauge transformation that eliminates $E$ and $C$ (defined in \refeq{ssi-dg}), as
\eq{\beta=\frac{1}{2}E,\hspace{0.5in}H\delta x^0=-C-\frac{1}{6}\nabla^2E.}
The gauge degrees of freedom are then completely fixed by the transformation.

In the spatially flat gauge, the spatial part of the metric is a multiple of the identity matrix \ea{in Cartesian spatial coordinates}. This allows the remaining metric perturbations $A,B$ to decouple from the field perturbation $\delta\phi$ in the perturbed action at leading order, as can be seen from \refeq{ssi-S0}. We can extract the leading order perturbed action coming from the field perturbation $\delta\phi$, as\footnote{
The first order perturbations are always zero because their coefficients are required to vanish by the equations of motion at background level.}
\eq{\delta S(\delta\phi)=-\frac{1}{2}\int a^3\dd^4x\left(\partial^\mu\delta\phi\partial_\mu\delta\phi+V''(\phi)\delta\phi^2\right).\label{eq-ssi-S1}}

This yields the equation of motion for the field perturbation
\eq{\ddot{\delta\phi}+3H\dot{\delta\phi}+V''(\phi)\delta\phi-\partial_i\partial^i\delta\phi=0.}
After Fourier transformation, in the momentum space it becomes
\eq{\ddot{\delta\phi}_{\vs k}+3H\dot{\delta\phi}_{\vs k}+\left(V''(\phi)+\frac{k^2}{a^2}\right)\delta\phi_{\vs k}=0,}
where
\eq{\delta\phi_{\vs k}\equiv\int\dd^3\vs x\,\delta\phi(\vs x)e^{-i\vs k\cdot\vs x}.\label{eq-ssi-Fourier}}

This is a harmonic oscillator with a friction force and a varying mass term, which cannot be quantized directly. We only know how to quantize a canonical harmonic oscillator with a constant mass in a flat space-time. So we switch to the variables
\eqa{\dd\tau&\equiv&\frac{\dd t}{a},\label{eq-ssi-tau}\\
\psi&\equiv&a\delta\phi,\label{eq-ssi-psi}}
where $\tau$ is called the \emph{conformal time}. The relevant action then becomes
\eq{\delta S(\psi)=\int\dd^3\vs x\dd\tau\left[\frac{1}{2}\psi'{}^2-aH\psi'\psi+\frac{1}{2}\left(H^2-V''(\phi)\right)a^2\psi^2-\frac{1}{2}\sum_{i=1}^3(\partial_i\psi)^2\right],\label{eq-ssi-S2}}
where the definition $'\equiv\dd/\dd\tau$ only holds in this section of the thesis.

Given an arbitrary function of perturbation $\psi$ and (conformal) space-time, $f(\psi,\vs x,\tau)$, we can add its total derivative w.r.t space $\vs x$ or time $\tau$ into the Lagrangian density. This does not change the physics of the system, so we add $\frac{1}{2}\int\dd^3\vs x\dd\tau[\dd(aH\psi^2)/\dd\tau]$  into \refeq{ssi-S2}, which gives the new action
\eq{\delta S(\psi)=\int\dd^3\vs x\dd\tau\left[\frac{1}{2}\psi'{}^2+\left(H^2+\frac{1}{2}\dot H-\frac{1}{2}V''(\phi)\right)a^2\psi^2-\frac{1}{2}\sum_{i=1}^3(\partial_i\psi)^2\right],\label{eq-ssi-S3}}
Now $\psi$ has become a canonical harmonic oscillator with the equation of motion
\eq{\psi''-\left(2H^2+\dot H-V''(\phi)\right)a^2\psi-\sum_{i=1}^3\partial_i\partial_i\psi=0.}
After Fourier transformation, it becomes
\eq{\psi_\vs k''+\left[k^2-(2H^2+\dot H-V''(\phi))a^2\right]\psi_\vs k=0.}
Using the slow-roll parameters $\epsilon_\phi$ and $\eta_\phi$, and according to \refeq{ssi-eta0} and \refeq{ssi-epsilon1}, we further simplify the equation of motion to
\eq{\psi_\vs k''+\left[k^2-a^2H^2(2-\epsilon_\phi-3\eta_\phi)\right]\psi_\vs k=0.\label{eq-ssi-eom4}}

We can fix the conformal time $\tau$ to be zero at the end of inflation. This would give rise to the simple relation
\eq{\int_\tau^0\dd\tau=\int_t^{t_e}\frac{\dd t}{a}=\int_a^{a_e}\frac{\dd a}{a^2H}\simeq\frac{1}{H}\int_a^{a_e}\frac{\dd a}{a^2}\approx\frac{1}{aH}.\label{eq-ssi-tauint}}
So with this choice of the conformal time, the conformal time becomes the negative \ea{comoving} Hubble radius
\eq{\tau\approx-\frac{1}{aH}.}
The approximations hold in \refeq{ssi-tauint} because well before the end of inflation ($\sim50$ e-folds), the slow-roll parameter $\epsilon_\phi\ll1$, so the universe expands exponentially while the Hubble rate remains almost constant. This approximation has relative error $\sim\epsilon_\phi$. For this reason, when we substitute it back into \refeq{ssi-eom4}, we also drop the slow-roll parameters in \refeq{ssi-eom4}. This leads to
\eq{\psi_\vs k''+\left(k^2-\frac{2}{\tau^2}\right)\psi_\vs k=0.\label{eq-ssi-eom5}}

Now focus on the mass term of the harmonic oscillator. The first term $k^2$ remains constant. The second term $2/\tau^2\approx2a^2H^2$ increases exponentially during inflation. Therefore, any specific perturbation mode with the momentum $k$ may experience two distinct phases of evolution during inflation:
\begin{itemize}
\item In the beginning, the momentum dominates the mass term with $|k\tau|\gg1$. The perturbation $\psi_\vs k$ is a perfect quantum harmonic oscillator.
\item As the universe expands exponentially during inflation, the \ea{comoving} Hubble radius of the universe decreases and the evolution enters $|k\tau|\ll1$. At this point, the perturbation $\psi_\vs k$ obtains a tachyonic mass and is no longer a quantum harmonic oscillator. The evolution has become classical and is determined by the harmonic oscillator initial conditions during the preceding quantum phase.
\end{itemize}

Since $1/k$ is the wavelength of the perturbation, and $|\tau|=1/aH$ is the \ea{comoving} Hubble radius of the universe, another way to distinguish the two phases is whether the perturbation mode can fit into one Hubble patch. Perturbations with $|k\tau|\gg1$ are thus also called \emph{sub-Hubble} perturbations, and those with $|k\tau|\ll1$ are \emph{super-Hubble} perturbations. In this sense, perturbations may start sub-Hubble during inflation, and gradually become super-Hubble. The time of switching from sub-Hubble to super-Hubble is known as the \emph{Hubble exit}, or \emph{leaving the Hubble patch}.

We use hats for operators in this chapter. The real harmonic oscillator $\psi_\vs k$ can then be quantized as
\eq{\hat\psi_\vs k=v_\vs k\hat a_\vs k+v_{-\vs k}^*\hat a_{-\vs k}^\dag,\label{eq-ssi-sol3}}
where $\hat a_\vs k$ and $\hat a_{\vs k'}^\dag$ conform with the commutation relation
\eq{[\hat a_\vs k,\hat a_{\vs k'}^\dag]=(2\pi)^3\delta^3(\vs k-\vs k').}
Also, $v_\vs k$ is a c-number yielding to
\eq{v_\vs k''+\left(k^2-\frac{2}{\tau^2}\right)v_\vs k=0.\label{eq-ssi-eom6}}

\Refeq{ssi-eom6} has the solution \cite{Dodelson:2003ft}
\eq{v_\vs k=\frac{e^{-ik\tau}}{\sqrt{2k}}\left(1-\frac{i}{k\tau}\right),\label{eq-ssi-sol}}
so in the early times where $t$ \ea{is sufficiently small or, equivalently, $-\tau$ is sufficiently large}\ed{or $\tau$ are sufficiently small}, \refeq{ssi-sol3} would reduce to the harmonic oscillator solution with
\eq{\hat\psi_\vs k\propto \frac{e^{-ik\tau}}{\sqrt{2k}}.}

The power spectrum of $\psi_\vs k$ is $P_\psi(k)$, defined as
\eq{\langle0|\hat\psi_\vs k\hat\psi_{\vs k'}^\dag|0\rangle=(2\pi)^3\delta^3(\vs k-\vs k')P_{\psi}(k),\label{eq-ssi-psipower}}
where, according to the quantization,
\eq{P_\psi(k)=|v_\vs k|^2=\frac{1}{2k}\left(1+\frac{1}{k^2\tau^2}\right).\label{eq-ssi-Ppsi}}
Here we have assumed the vacuum state of the universe $|0\rangle$ is the Bunch-Davis vacuum \cite{Bunch:1978yq}.

With $\tau=-1/aH$ and $\delta\phi=\psi/a$, from \refeq{ssi-Ppsi} we get the power spectrum of $\delta\phi$, (similarly defined,) as
\eq{P_{\delta\phi}(k)=\frac{H^2}{2k^3}\left(1+\frac{k^2}{a^2H^2}\right).\label{eq-ssi-PdPhi}}
Several e-folds after Hubble exit, the second term in the parentheses will become negligible, so the power spectrum of $\delta\phi(\vs k)$ reaches constant:
\eq{P_{\delta\phi}(k)\Big|\subs{super-Hubble}=\frac{H^2}{2k^3}.\label{eq-ssi-dphi}}
Therefore, after the perturbation mode leaves the Hubble patch, it gradually stops evolving and becomes \emph{frozen}.

We can recover the time part of the gauge invariance by considering only the time translation
\eq{x^0\rightarrow\tilde x^0=x^0+\delta x^0.\label{eq-ssi-trans0}}
From \refeq{ssi-phigauge} and \refeq{gauge-C2}, we can construct the gauge invariant curvature perturbation by compensating the changes under the time translation \refeq{ssi-trans0}. The \ea{time} gauge invariant curvature perturbation $\zeta$ can thus be defined as\footnote{
\ea{To further take into account the spatial transformations (for $\mu=1,2,3$ in \refeq{ssi-trans}), the spacetime gauge invariant curvature perturbation can be defined as $\zeta\sups{(st)}\equiv\zeta+\frac{1}{6}\nabla^2E$. The extra term is the same for a spacetime gauge invariant curvature perturbation for \refeq{ssi-zeta2}.}}
\eq{\zeta\equiv C+\frac{H}{\dot\phi}\delta\phi.\label{eq-ssi-zeta}}

Since now $\zeta$ is a gauge-invariant quantity, its power spectrum should not depend on the choice of gauge. After the mode $\vs k$ freezes, the power spectrum of $\zeta$ then becomes
\eq{P_\zeta(k)=P_\zeta(k)\Big|_{C=0}=\frac{H^2}{\dot\phi^2}P_{\delta\phi}(k)=\frac{2\pi}{k^3}\frac{H^2}{\epsilon_\phi\mpl^2}.\label{eq-ssi-Pz1}}

We then expand the power spectrum of $\zeta$ around a reference scale $k_0$ following \refeq{cmb-Pz}, which provides the spectral index of the scalar perturbations
\eqa{n_s&\equiv&1+\frac{\dd}{\dd\ln k}\ln\left(k^3P_\zeta(k)\right)\nonumber\\
&=&1-\frac{\dd}{\dd N}\ln\frac{H^2}{\epsilon_\phi}\nonumber\\
&=&1-6\epsilon_\phi+2\eta_\phi.\label{eq-ssi-ns1}}

The running of the spectral index can be derived similarly,
\eq{\frac{\dd n_s}{\dd\ln k}=8\epsilon_\phi(-3\epsilon_\phi+2\eta_\phi)-2\xi_\phi,\label{eq-ssi-nsh}}
where the third slow-roll parameter is defined as
\eq{\xi_\phi\equiv\frac{\mpl^4V'V'''}{(8\pi\rho)^2},\label{eq-ssi-xi}}
with the relation
\eq{\frac{\dd\eta_\phi}{\dd N}=\xi_\phi-2\epsilon_\phi\eta_\phi.}

The definition of $\zeta$ in \refeq{ssi-zeta} is obviously only applicable to single-field slow-roll inflation. In the more general case, we can consider an adiabatic universe. The energy density perturbation of the universe, $\delta\rho$, transforms under the time translation \refeq{ssi-trans0} as
\eq{\delta\rho\rightarrow\widetilde{\delta\rho}=\delta\rho-\dot{\delta\rho}\,\delta x^0.}
This allows us to define the gauge invariant curvature perturbation more generally as
\eq{\zeta\equiv C+\frac{H}{\dot\rho}\delta\rho.\label{eq-ssi-zeta2}}
It can be confirmed easily that \refeq{ssi-zeta2} reduces to \refeq{ssi-zeta} for single-field slow-roll inflation.

\sssecs{Separate universe approach}
The above section has explained how perturbations evolve before the Hubble exit. Now the question is how perturbations would evolve after the Hubble exit but before the beginning of Hot Big Bang. It has been known in general, that super-Hubble curvature perturbations are conserved:
\begin{itemize}
\item on uniform energy density hypersurfaces/slicings,
\item if the universe remains sufficiently adiabatic in the future evolution (i.e.\ the pressure or the equation of state is a mere function of energy density).
\end{itemize}

This is known as the \emph{separate universe approach} \cite{Sasaki:1995aw,Sasaki:1998ug,Wands:2000dp,Lyth:2004gb,Sasaki:2007ay}, which essentially regards each of the local Hubble patches as a homogenous patch. Each Hubble patch is then perturbed as a whole by the super-Hubble perturbations, and smoothened within so the sub-Hubble perturbations are neglected. When the above two conditions are satisfied, all the Hubble patches on the uniform energy density hypersurface are identical except that each has a different time shift, originated from the super-Hubble curvature perturbations. The separate universe approximation should be valid because the sub-Hubble perturbations, or the gradients from the super-Hubble perturbations, have a vanishing net effect when averaged over all the Hubble patches in the universe.

For single-field slow-roll inflation, the homogenous Hubble patch has only one degree of freedom and is always adiabatic. Therefore, super-Hubble curvature perturbations are always conserved during single-field slow-roll inflation, regardless of the uniform energy density hypersurface chosen.

For the more generic multi-component inflation, (\refsec{Multi},) extra complications can be involved. The universe may contain non-adiabatic perturbations, which can transfer to curvature perturbation gradually well after the Hubble exit. For this reason, the \ea{desired}\ed{valid} uniform energy density hypersurface can be significantly delayed.

Therefore, as a more generic method using separate universe approach, in this section we study how the field perturbation in the Hubble patch, $\delta\phi$, transfers to curvature perturbation $\zeta$, at or after the Hubble exit. For this purpose, we count the extra number of e-folds of universe expansion for the chosen Hubble patch up to the uniform energy density hypersurface, \ea{due to}\ed{taking into account} the initial super-Hubble field perturbation $\delta\phi$ \ea{on spatially flat hypersurfaces}. This extra amount of expansion should then correspond to the scalar perturbation $C$ on uniform energy density hypersurface, or equally the curvature perturbation $\zeta$. This is portrayed in \refig{ssi-sepu}. The so-called \emph{$\delta N$ formalism} thus reduces the problem into simply solving background evolutions, and then taking the differentiations.

\fig{ssi-sepu}{A schematic figure on $\delta N$ formalism}{In the spatially flat gauge (green), the perturbed patches (red) can be evolved to the uniform energy density hypersurface (blue) that fulfills the adiabatic condition. The resulting perturbation in the number of e-folds of the universe expansion then corresponds to the amount of curvature perturbation $\zeta$ for every patch. In single-field slow-roll inflation, the uniform energy density hypersurface can be chosen at the Hubble exit as shown. The solid and dashed black curves illustrate the exponential universe expansion during inflation. The horizontal and vertical axes are for the physicsl spatial and time dimensions respectively.}
For single-field slow-roll inflation, we have \ea{(on spatially flat slicing)}
\eq{\zeta(\vs x)=\delta N(\vs x)=N_\phi\delta\phi(\vs x),\label{eq-ssi-dNdphi}}
where \ea{$\delta\phi(\vs x)$ is the field perturbation on spatially flat hypersurfaces}. According to \refeq{ssi-dN},
\eq{N_\phi\equiv\frac{\partial N}{\partial\phi}=\frac{8\pi V(\phi)}{\mpl^2V'(\phi)}.\label{eq-ssi-Nphi}}
Since single-field slow-roll inflation is always adiabatic, we can pick the uniform energy density hypersurface right at the Hubble exit as shown in \refig{ssi-sepu}. Therefore the field perturbation $\delta\phi(\vs x)$ in the location space does not evolve after Hubble exit. This immediately gives
\eq{\zeta_\vs k=N_\phi\delta\phi_\vs k,\label{eq-ssi-zetax}}
with the power spectrum
\eq{P_\zeta(k)=N_\phi^2P_{\delta\phi}(k)=\frac{2\pi}{k^3}\frac{H^2}{\epsilon_\phi\mpl^2}.}
As we have seen, it easily reduces to \refeq{ssi-Pz1} in the single-field slow-roll inflation scenario.

The spectral index and its running can then be obtained straightforwardly. For example,
\eqa{n_s&\equiv&1+\frac{\dd\ln k^3P_\zeta(k)}{\dd\ln k}\nonumber\\
&=&1-\frac{\dd\ln k^3P_\zeta(k)}{\dd N}\nonumber\\
&=&1-\frac{2\,\dd\ln N_\phi}{\dd N}-\frac{\dd\ln k^3P_{\delta\phi}(k)}{\dd N}\nonumber\\
&=&1-\frac{2N_{\phi\phi}}{N_\phi}\phi'-\frac{\dd\ln H^2}{\dd N}\nonumber\\
&=&1-6\epsilon_\phi+2\eta_\phi,}
where $\phi'\equiv\dd\phi/\dd N$, and $N_{\phi\phi}\equiv\partial N_\phi/\partial\phi$. So it also recovers \refeq{ssi-ns1}.

We can also look at the problem in location space completely. Starting from \refeq{ssi-dNdphi}, the power spectrum of field perturbation in the location space is defined as\footnote{
Note again that hats are used for quantum operators.}
\eq{\langle0|\widehat{\delta\phi}(\vs x)\widehat{\delta\phi}^\dag(\vs x)|0\rangle=P_{\delta\phi}(\vs x).}
Therefore,
\eq{P_{\delta\phi}(\vs x)=\int\frac{\dd^3\vs k}{(2\pi)^3}P_{\delta\phi}(k)=\frac{1}{4\pi^2}\int H^2\dd\ln k.\label{eq-ssi-Pdphi}}
The curvature perturbation then follows
\eq{P_\zeta(\vs x)=\frac{1}{\pi\mpl^2}\int\frac{H^2}{\epsilon_\phi}\dd\ln k.\label{eq-ssi-zetax2}}

From the above equations, we can obviously see that perturbations receive contributions from all the super-Hubble modes. We will discuss the infrared and ultraviolet divergences of the integral shortly, but let us first look at the strength of quantum fluctuations compared with classical motion.

\Refeq{ssi-Pdphi} tells us that for every e-fold of inflation, the field perturbation $\delta\phi(\vs x)$ typically gains the extra amount $\sim\pm H/2\pi$. At the same time, the classical slow-roll motion of the inflaton $\phi$ contributes by the amount $\sim\dd\phi/\dd N$. The relative strength of quantum fluctuations can be written as
\eq{\frac{H}{2\pi}\left(\frac{\dd\phi}{\dd N}\right)^{-1}=\frac{1}{\sqrt{\pi\epsilon_\phi}}\frac{H}{\mpl}.}
When this ratio is much smaller than unity, quantum fluctuations are much weaker than classical slow roll. In the opposite limit, the motion of field is dominated by quantum fluctuations, and we can also say the field is \emph{frozen}.

It is worth noting that the above calculations are only applicable to single-field slow-roll inflation. When multiple components coexist in the universe, such as in \refsec{Multi}, the choice of uniform energy density hypersurface must agree with the adiabatic condition. The earliest possible choice may be well after the Hubble exit. In such case, we can either evolve the field perturbation $\delta\phi$ from the Hubble exit to the hypersurface, or the corresponding $N_\phi$ from the hypersurface back to the Hubble exit. In either way, their product $\delta N=N_\phi\delta\phi$ should follow the correct evolution.

Let us now turn back to the observational divergences in the integrals in \refeq{ssi-Pdphi} and \refeq{ssi-zetax2}. The ultraviolet divergence has several suppressions and cut-offs. For example, our limited resolution of the observations cannot observe perturbations of too small scales, and this introduces a hard cut-off to the ultraviolet divergence. The physical divergences, on the other hand, can only be resolved by fundamental theories, which are beyond the scope of the thesis.

To deal with the infrared divergence, let us think what $P_\zeta(\vs x)$ represents in the observations. We can only see perturbations within our current Hubble patch, so we cannot compare them with patches too distant away. In this sense, perturbations with scales much larger than the Hubble size \ea{today} should have only suppressed contributions to our observed $P_\zeta(\vs x)$, indicated with $P_\zeta\sups{(ob)}(\vs x)$. They only produce gradient effects within our Hubble patch, which are exponentially suppressed towards the infrared limit.

In particular, suppose only the limited spatial region $\mathcal{R}$ is visible to us. For any observable spatial function $f(\vs x)$, the observed power spectrum of perturbations in the location space can be written as
\eq{P_f\sups{(ob)}\equiv\frac{1}{(2\pi)^3}\left\langle\overline{|f-\bar f|^2}\right\rangle=\frac{1}{(2\pi)^3}\left(\left\langle|f|^2\right\rangle-\left\langle\left|\bar f\right|^2\right\rangle\right),\label{eq-ssi-PfR1}}
where the top bar means averaging over the visible region $\mathcal{R}$, as
\eq{\bar f=\overline{f(\vs x)}\equiv\frac{1}{|\mathcal{R}|}\int_{\vs x\in\mathcal{R}}f(\vs x)\dd\vs x,}
and the volume of $\mathcal{R}$ is defined as
\eq{|\mathcal{R}|\equiv\int_{\vs x\in\mathcal{R}}\dd\vs x.}
At this point we do not specify the dimension or geometry of $\mathcal{R}$.

Let us define the power spectrum of $f$ in the Fourier space as
\eq{\langle f_\vs kf_{\vs k'}^*\rangle=(2\pi)^3\delta^3(\vs k-\vs k')P_f(k).}
The first term of \refeq{ssi-PfR1} can then be calculated at ease
\eq{\left\langle|f|^2\right\rangle=\int\dd^3\vs kP_f(k).\label{eq-ssi-fx2}}
The second terms requires some calculation
\eqa{\left\langle\left|\bar f\right|^2\right\rangle&=&\left\langle\left|\overline{f(\vs x)}\right|^2\right\rangle\nonumber\\
&=&\frac{1}{2|\mathcal{R}|^2}\int_{\vs x,\vs x'\in\mathcal{R}}\dd\vs x\dd\vs x'\left\langle f(\vs x)f^*(\vs x')+h.c\right\rangle\nonumber\\
&=&\frac{1}{2|\mathcal{R}|^2}\int\dd^3\vs kP_f(k)\int_{\vs x,\vs x'\in\mathcal{R}}\dd\vs x\dd\vs x'\left[e^{i\vs k\cdot(\vs x-\vs x')}+c.c\right]\nonumber\\
&=&\frac{1}{|\mathcal{R}|^2}\int\dd^3\vs kP_f(k)\int_{\vs x,\vs x'\in\mathcal{R}}\dd\vs x\dd\vs x'\cos\vs k\cdot(\vs x-\vs x')\nonumber\\
&=&\int\dd^3\vs kP_f(k)\left[\left(\overline{\cos\vs k\cdot\vs x}\right)^2+\left(\overline{\sin\vs k\cdot\vs x}\right)^2\right],\label{eq-ssi-PfR2}}
where $h.c$ and $c.c$ indicate the Hermitian conjugate and the complex conjugate respectively.

In our universe, the visible region $\mathcal{R}$ is spherically symmetric. The second term in the parentheses of \refeq{ssi-PfR2} thus vanishes. For the visible and dark matter, we can see perturbations with different distances from us, within the ball of radius $r\subs{max}$, i.e.\ $\mathcal{R}=\{\vs x|x\le r\subs{max}\}$. For such cases, the average becomes
\eqa{\overline{\cos\vs k\cdot\vs x}&\equiv&\frac{1}{|\mathcal{R}|}\int_{x\le r\subs{max}}\cos(\vs k\cdot\vs x)\dd\vs x\nonumber\\
&=&\frac{3}{r\subs{max}^3}\int_0^{r\subs{max}}r^2\dd r\;\frac{\sin kr}{kr}\nonumber\\
&=&\frac{3}{\theta^3}(\sin\theta-\theta\cos\theta),\label{eq-ssi-cosav1}}
where we have defined the dimensionless variable
\eq{\theta\equiv kr\subs{max}.}
Therefore using \refeq{ssi-cosav1}, we get the observed power spectrum in a ball
\eqa{P_f\sups{(ob)}(\vs x)&=&\frac{1}{(2\pi)^3}\left(\left\langle|f|^2\right\rangle-\left\langle\left|\bar f\right|^2\right\rangle\right)\nonumber\\
&=&\frac{1}{2\pi^2}\int\dd k k^2P_f(k)\left[1-\frac{9}{\theta^6}(\sin\theta-\theta\cos\theta)^2\right]\nonumber\\
&=&\frac{k_0^3P_f(k_0)}{2\pi^2}\int\frac{\dd\theta}{\theta}\left[1-\frac{9}{\theta^6}(\sin\theta-\theta\cos\theta)^2\right],\label{eq-ssi-PfR3}}
where in the last step, we have assumed a scale-invariant power spectrum
\eq{k^3P_f(k)=k_0^3P_f(k_0),\hspace{0.5in}\mathrm{for\ any\ }k.}
Now in the infrared limit where $k\rightarrow0$ or equivalently $\theta\rightarrow0$, the term being integrated in \refeq{ssi-PfR3} is equal to $\theta/5$. So the observed power spectrum does not have any infrared singularity within our observed Hubble patch.

Similarly, the CMB is only emitted from the LSS sphere. Therefore the averaging should take place for all directions, but only at a fixed radius $r\subs{ls}$, which is our comoving distance to the LSS. This corresponds to $\mathcal{R}=\{\vs x|x=r\subs{ls}\}$, giving
\eq{\overline{\cos\vs k\cdot\vs x}\equiv\frac{1}{|\mathcal{R}|}\int_{x=r\subs{ls}}\cos(\vs k\cdot\vs x)\dd\vs x=\frac{\sin\theta}{\theta},\label{eq-ssi-cosav2}}
where we have redefined the dimensionless variable
\eq{\theta\equiv kr\subs{ls}.}
The observed power spectrum on a sphere then becomes
\eqa{P_f\sups{(ob)}(\vs x)&=&\frac{1}{2\pi^2}\int\dd k k^2P_f(k)\left(1-\frac{\sin^2\theta}{\theta^2}\right)\nonumber\\
&=&\frac{k_0^3P_f(k_0)}{2\pi^2}\int\frac{\dd\theta}{\theta}\left(1-\frac{\sin^2\theta}{\theta^2}\right),\label{eq-ssi-PfR4}}
where we have similarly assumed a scale invariant power spectrum. The term being integrated is proportional to $\theta/3$ in the $k\rightarrow0$ limit, so the infrared divergence is similarly resolved for the CMB. This can also be seen in \refig{ssi-infrared}.

\fig[0.8]{ssi-infrared}{The infrared safe integral for the CMB spectrum}{The integral in \refeq{ssi-PfR4} is shown as a function of its integral lower bound. The infrared cutoff, $k\subs{min}r\subs{ls}$, is infrared safe because it converges as $k\subs{min}\rightarrow0$. The upper bound of the integral is taken $e^2$ for demonstration.}
Since the integral in \refeq{ssi-PfR4} is an $\cO(1)$ model independent constant, we can abandon it by redefining the power spectrum as
\eq{\P_f\equiv\frac{1}{2\pi^2}k_0^3P_f(k_0).\label{eq-ssi-PfR6}}
Then the field perturbation and curvature perturbation will have the power spectra \cite{Sasaki:1995aw,Sasaki:1998ug,Wands:2000dp,Lyth:2004gb,Sasaki:2007ay}
\eq{\P_{\delta\phi}=\frac{H^2}{4\pi^2},\label{eq-ssi-dphix2}}
and
\eq{\P_\zeta=\frac{H^2}{\pi\epsilon_\phi\mpl^2}.\label{eq-ssi-zetax3}}

\sssecsx{Tensor perturbations}{ssi}
The primordial scalar perturbations are responsible for the CMB temperature fluctuations we see today. The tensor perturbations, on the other hand, produce polarization fluctuations in the CMB. The tensor perturbations have only two degrees of freedom. For a photon traveling in the $z$ direction, they act on the \ea{flat FRW} metric in the form of\footnote{
For a review or textbook, see \cite{Kodama:1985bj,Mukhanov:1990me,Riotto:2002yw,Dodelson:2003ft,Bassett:2005xm,Malik:2008im}.}
\eq{\delta g_{\mu\nu}=\frac{a^2}{2}\left(\begin{array}{c@{\hspace{0.2in}}c@{\hspace{0.2in}}c@{\hspace{0.2in}}c}
0&0&0&0\\0&h_+&h_\times&0\\0&h_\times&-h_+&0\\0&0&0&0\end{array}\right).}
The variables $h_+$ and $h_\times$ are the two independent modes of the tensor perturbations. Such perturbations will induce the perturbation in the action
\eq{\delta S(h_+,h_\times)=\frac{\mpl^2}{64\pi}\int a^3\dd^4x\sum_{i=+,\times}\left[\frac{1}{2}\dot h_i^2-\frac{1}{2}(\partial_zh_i)^2\right].\label{eq-ssi-St1}}

From \refeq{ssi-St1}, we find that each of the tensor perturbations corresponds to a plane wave solution traveling along the $z$ direction. We simply need to canonicalize the fields by switching to the conformal time $\tau$, and defining (while neglecting the subscripts because both tensor modes act identically and independently)
\eq{\tilde h\equiv\frac{a\mpl}{8\sqrt\pi}h.}
So the new perturbed action becomes (for each of the tensor perturbations)
\eq{\delta S(\tilde h)=\int\dd^3\vs x\dd\tau\left[\frac{1}{2}\tilde h'{}^2+\left(H^2+\frac{1}{2}\dot H\right)a^2\tilde h^2-\frac{1}{2}(\partial_z\tilde h)^2\right].\label{eq-ssi-St2}}

Now we are able to quantize $\tilde h$ following the routine in \refsssec{ssi-Scalar perturbations}. This gives the power spectrum of $\tilde h$
\eq{P_{\tilde h}(k)=\frac{1}{2k}\left(1+\frac{1}{k^2\tau^2}\right),\label{eq-ssi-Pt0}}
and hence the power spectrum of the tensor perturbations $h$ well after the Hubble exit
\eq{P_t(k)\equiv P_h(k)=\frac{32\pi H^2}{k^3\mpl^2},\label{eq-ssi-Pt1}}
where the subscript $t$ represents tensor perturbations. Since the power spectrum of the tensor perturbations only depends on the energy scale of inflation, the measurement of the tensor perturbations is very helpful in determining how early inflation took place \cite{Lyth:1996im}.

In practice, the power spectrum $P_t(k_0)$ is rarely used when referring to the strength of tensor perturbations. More often, the relative strength of tensor perturbations w.r.t scalar perturbation is used. This parameter is called the \emph{tensor-to-scalar ratio}, and is defined as
\eq{r\equiv\frac{\P_t}{\P_\zeta}=16\epsilon_\phi\label{eq-ssi-r},}
where the last equal sign holds only for single-field slow-roll inflation. Note the power spectrum we have used is similarly defined for tensor perturbations according to \refeq{ssi-PfR6}, as
\eq{\P_t=\frac{16H^2}{\pi\mpl^2}.\label{eq-ssi-Pt3}}

The Planck observations have not found any tensor perturbation with $r<0.11$ at $95\%$ CL, and neither has BICEP with $r<0.07$ \cite{Mortonson:2014bja,Ade:2015lrj,Array:2015xqh}. This immediately gives the upper bound for inflation energy scale
\eq{\rho^\frac{1}{4}=\left(\frac{3r\P_\zeta}{128}\right)^\frac{1}{4}\mpl<1.6\times10^{-3}\mpl.}

\ssecsx{Higher order perturbations}{ssi}
As explained in \refssec{CMB angular bi-spectrum} and \refssec{CMB angular tri-spectrum}, observations in the CMB temperature fluctuations may find possible deviations from pure Gaussian perturbations. These non-Gaussianities may come from primordial cosmology, such as inflation. In \refeq{cmb-B3}, we have shown how the non-Gaussian curvature perturbation $\zeta$ can induce a non-vanishing three-point correlation function on the CMB temperature map. In this section, we will discuss how single-field slow-roll inflation produces primordial non-Gaussianity, in terms of local $\fNL$, $\tNL$ and $\gNL$, and therefore how they will be constrained by recent observations. (See \refssec{CMB angular bi-spectrum} and \refssec{CMB angular tri-spectrum} for definitions, and \cite{Bartolo:2004if,Chen:2010xka} for a review.)

The $\delta N$ formalism proves to be an effective approach for higher order perturbations. For single-field slow-roll inflation, every smoothened Hubble patch has only one degree of freedom -- the inflaton $\phi(\vs x)$. Therefore we can always write the number of remaining e-folds of inflation as a function of inflaton in the background evolution $N(\phi)$. Previously in \refsssec{Separate universe approach}, we have expanded $N(\phi)$ up to linear order in \refeq{ssi-dNdphi}. More generically, it can be expanded to higher orders as \cite{Lyth:2005fi,Seery:2005gb,Byrnes:2006vq}
\eqa{\zeta(\vs x)&=&\delta N(\vs x)\nonumber\\
&=&N_\phi\delta\phi(\vs x)+\frac{1}{2}N_{\phi\phi}\left(\delta\phi^2(\vs x)-\langle\delta\phi^2(\vs x)\rangle\right)\nonumber\\
&&+\frac{1}{6}N_{\phi\phi\phi}\delta\phi^3(\vs x)+\ho.\label{eq-ssi-dNdphih}}

In \refeq{ssi-dNdphih}, including the $\langle\delta\phi^2(\vs x)\rangle$ term is to ensure the expectation value of curvature perturbation vanishes strictly $\langle\zeta(\vs x)\rangle=0$. Knowing\ed{ly} that $\delta\phi(\vs x)$ is a Gaussian variable, by comparing \refeq{cmb-znG} with \refeq{ssi-dNdphih}, we can find the Gaussian part of curvature perturbation easily
\eq{\zeta_G(\vs x)=N_\phi\delta\phi(\vs x).}
Replacing $\delta\phi(\vs x)$ with $\zeta_G(\vs x)/N_\phi$, \refeq{ssi-dNdphih} becomes
\eq{\zeta(\vs x)=\zeta_G(\vs x)+\frac{1}{2}\frac{N_{\phi\phi}}{N_\phi^2}\left(\zeta_G^2(\vs x)-\langle\zeta_G^2(\vs x)\rangle\right)+\frac{1}{6}\frac{N_{\phi\phi\phi}}{N_\phi^3}\zeta_G^3(\vs x)+\ho,\label{eq-ssi-zzG}}
which corresponds to \cite{Lyth:2005fi,Seery:2005gb,Byrnes:2006vq}
\eq{\fNL=\frac{5}{6}\frac{N_{\phi\phi}}{N_\phi^2},}
and
\eq{\gNL=\frac{25}{54}\frac{N_{\phi\phi\phi}}{N_\phi^3}.}

Therefore $\fNL$ and $\gNL$ can be calculated easily based on $\delta N$ formalism. Note however that the above expressions are in general scale dependent, because $N_\phi$, $N_{\phi\phi}$ and $N_{\phi\phi\phi}$ can change during inflation. This feature should be interpreted as the scale dependences of local non-Gaussianities, an extension of \refeq{cmb-zetang} or \refeq{cmb-zet3}.

The parameter $\tNL$ comes from the second order effect of local $\fNL$ on the CMB tri-spectrum. In single-field slow-roll inflation, there is only one degree of freedom $\phi$ that may produce the curvature perturbation. Therefore, the simple relation holds
\eq{\tNL=\frac{36}{25}\fNL^2.\label{eq-ssi-tNL}}

Higher order derivatives of $N$ can be derived from its first order derivative in \refeq{ssi-Nphi}. The local non-Gaussianities can then be expressed in terms of the slow-roll parameters \cite{Maldacena:2002vr,Seery:2005gb,Byrnes:2006vq}
\eqa{\fNL&=&\frac{5}{6}(2\epsilon_\phi-\eta_\phi),\\
\gNL&=&\frac{25}{54}\left[2\eta_\phi(\eta_\phi-\epsilon_\phi)-\xi_\phi\right],\\
\tNL&=&(2\epsilon_\phi-\eta_\phi)^2.}

\ssecs{Testing single-field slow-roll inflation with the CMB}
Before diving into the models of single-field slow-roll inflation, we first briefly summarize the cosmological observables that can be utilized to test inflationary models. Based on the previous contents in \refsec{CMB} and \refsec{Single}, we can construct \reftab{ssi-obs}.
\def\esp{0.08in}
\def\esps{0.16in}
\def\nline{\vspace{\esps}\\}
\tab{ssi-obs}{Single-field slow-roll inflation predictions and CMB observations}{Single-field slow-roll inflation predictions and the Planck CMB observations \cite{Ade:2015lrj,Ade:2015ava}. Errors are at $1\sigma$ unless otherwise noted.}{
\begin{tabular}{c@{\hspace{0.3in}}c@{\hspace{0.3in}}c}
\hline
Parameters&Predictions&Observations\\
\hline\vspace{-0.2in}\nline
$\P_\zeta$&$\displaystyle\frac{H^2}{\pi\epsilon_\phi\mpl^2}$&$(2.142\pm0.048)\times10^{-9}$\nline
$n_s$&$1-6\epsilon_\phi+2\eta_\phi$&$0.9667\pm0.0040$\nline
$\displaystyle\frac{\dd n_s}{\dd\ln k}$&$8\epsilon_\phi(-3\epsilon_\phi+2\eta_\phi)-2\xi_\phi$&$-0.0065\pm0.0076$\nline
$\fNL$&$\displaystyle\frac{5}{6}(2\epsilon_\phi-\eta_\phi)$&$0.8\pm5.0$\nline
$\gNL$&$\displaystyle\frac{25}{54}\left(2\eta_\phi(\eta_\phi-\epsilon_\phi)-\xi_\phi\right)$&$(-9.0\pm7.7)\times10^4$\nline
$\tNL$&$(2\epsilon_\phi-\eta_\phi)^2$&$<2800$ at $95\%$ CL\nline
$r$&$16\epsilon_\phi$&$<0.07$ at $95\%$ CL\nline
\hline
\end{tabular}}
\let\nline\undefined
\let\esps\undefined
\let\esp\undefined

Since all the energy scale free\footnote{
By energy scale free, we mean all the observables in \reftab{ssi-obs} except the power spectrum of the curvature perturbation. This is because none of them depend on the overall energy scale of inflation.}
cosmological observables of single-field slow-roll inflation can be expressed in terms of slow-roll parameters, single-field slow-roll inflation has to satisfy a series of consistency relations including \cite{Maldacena:2002vr, Byrnes:2006vq}
\eq{\left(n_s-1+\frac{1}{8}r\right)^2=4\tNL=\frac{144}{25}\fNL^2,\label{eq-ssi-cons2}}
\eq{\left(n_s-1+\frac{1}{8}r\right)(n_s-1)+\frac{\dd n_s}{\dd\ln k}=\frac{108}{25}\gNL.\label{eq-ssi-cons3}}
As a result, if the observations disagree with any of the above consistency relations, we can rule out single-field slow-roll inflation. In such cases, one would then have to introduce extra complexities in the model\footnote{
\Reftab{ssi-obs} can change when additional complexities are introduced, such as when the single-field slow-roll inflation has a non-trivial initial condition with a large momentum in $\phi$, far away from the slow-roll attractor solution near the Hubble exit of the CMB scales. Such scenarios are off the topic of this thesis, and will not be discussed.}
.

The second consistency relation (\refeq{ssi-cons3}) is far from practical use. The errors in $r$, $\gNL$ and $\frac{\dd n_s}{\dd\ln k}$ are currently large. However, our current observations are accurate enough to distinguish the first consistency relation (\refeq{ssi-cons2}). The curvature perturbation is almost scale invariant and the tensor-to-scalar ratio is small ($r<0.07$). Single-field slow-roll inflation is thus required to produce small non-Gaussianities in $\fNL\ll1$ and $\tNL\ll1$, regardless of the specific model \cite{Maldacena:2002vr, Byrnes:2006vq}. This can already be confirmed partly by the Planck observations.

\ssecs{Models of single-field slow-roll inflation}
There are hundreds of models of inflation, even just for single-field slow-roll inflation\cite{Martin:2013tda,Martin:2013nzq}. In this thesis, we will only discuss the inflation with a power-law potential, and the inflection point inflationary models. (For a review, see \cite{Linde:2005ht,Lidsey:1995np,Riotto:2002yw,Bassett:2005xm,Mazumdar:2010sa}.)

\sssecs{Inflation with a power-law potential}
Inflation can be realized with a very simple power-law potential of a real scalar $\phi$, in the form
\eq{V(\phi)=\frac{1}{p}\lambda\mpl^4\left(\frac{\phi}{\mpl}\right)^p.\label{eq-ssi-Vpl}}

This potential contains two parameters -- the exponent $p$ that determines the power of the potential, and the coupling constant $\lambda$. Typically, people are interested in the $p=2$ and $p=4$ models, corresponding to the quadratic and quartic potentials respectively. For $p=2$, we can define
\eq{m\equiv\sqrt{\lambda}\mpl}
as the bare mass of $\phi$. \Refeq{ssi-Vpl} then reduces to \emph{quadratic inflation}, corresponding to a single scalar field with mass $m$ which drives inflation with the potential
\eq{V(\phi)=\frac{1}{2}m^2\phi^2.}
The $p=4$ case is called \emph{chaotic inflation} \cite{Linde:1983p1108}, which is dominated by the self-coupling of a massless scalar. The potential is
\eq{V(\phi)=\frac{1}{4}\lambda\phi^4.}
Due to the simple potential form, the power-law potential is sometimes regarded as the simplest model of inflation.

We will still use the generic potential \refeq{ssi-Vpl}. The slow-roll parameters can then be calculated as
\eqa{\epsilon_\phi&=&\frac{p^2\mpl^2}{16\pi\phi^2},\\
\eta_\phi&=&\frac{p(p-1)\mpl^2}{8\pi\phi^2},\\
\xi_\phi&=&\frac{p^2(p-1)(p-2)\mpl^4}{64\pi^2\phi^4}.}
Taking the inflation condition as $\epsilon_\phi<1$, we find that the power-law potential can provide inflation easily as long as\footnote{
Although for $p>2$, $|\eta_\phi|=1$ is reached first during inflation, we will still use $\epsilon_\phi<1$ as the only measure for the end of inflation for simplicity.}
\eq{|\phi|>\frac{p}{4\sqrt\pi}\mpl.\label{eq-ssi-pireq}}

Since during inflation $\phi$ never changes sign, without loss of generality we take $\phi>0$. From \refeq{ssi-dN}, we can solve the background evolution of $\phi$ as a function of the number of remaining e-folds of inflation $N$, as
\eq{\phi(N)=\frac{\mpl}{4\sqrt\pi}\sqrt{p(p+4N)}.\label{eq-ssi-phiN1}}
Putting it back into the slow-roll parameters then gives
\eqa{\epsilon_\phi(N)&=&\frac{p}{p+4N},\\
\eta_\phi(N)&=&\frac{2(p-1)}{p+4N},\\
\xi_\phi(N)&=&\frac{4(p-1)(p-2)}{(p+4N)^2},}
and all the cosmological observables
\eqa{\P_\zeta&=&\frac{128\pi}{3}\lambda\left[\frac{p(p+4N)}{16\pi}\right]^{\frac{p}{2}+1},\label{eq-ssi-pPz1}\\
n_s&=&1-\frac{2(p+2)}{p+4N},\label{eq-ssi-pns}\\
\frac{\dd n_s}{\dd \ln k}&=&-\frac{8(p+2)}{(p+4N)^2},\label{eq-ssi-pdns}\\
r&=&\frac{16p}{p+4N}.\label{eq-ssi-pr}}
We have omitted the non-Gaussianity observables which are automatically determined by single-field slow-roll consistency relations.

The parameter $\lambda$ can be fixed by the power spectrum of scalar perturbations according to \refeq{ssi-pPz1}. If we also restrict the power parameter $p$ among certain discrete values, such as 2 and 4 in this thesis, the inflation model with a power-law potential is then left with no free parameter.

\def\esp{0.08in}
\def\esps{0.16in}
\def\nline{\vspace{\esps}\\}
\tab{ssi-pobs}{Inflationary predictions from a power-law potential and the observations}{Inflationary predictions from a power-law potential and the observations. Quadratic ($p=2$) and chaotic ($p=4$) inflations are demonstrated. The Planck 2015 data \cite{Ade:2015lrj,Ade:2015ava,Ade:2015xua} are used.}{
\begin{tabular}{cccccc}
\hline
\multirow{2}{*}{Parameters}&\multirow{2}{*}{Observations}&\multicolumn{2}{c}{Quadratic inflation}&\multicolumn{2}{c}{Chaotic inflation}\\
&&$N=50$&$N=60$&$N=50$&$N=60$\\
\hline\vspace{-0.2in}\nline
$m/\mpl$&\multirow{2}{*}[-\esp]{$\P_\zeta=2.142\times10^{-9}$}&$5.0\times 10^{-7}$&$4.2\times 10^{-7}$&N/A&N/A\nline
$\lambda$&&N/A&N/A&$3.7\times10^{-15}$&$2.2\times10^{-15}$\nline
$n_s$&$0.9667\pm0.0040$&0.960&0.967&0.941&0.951\nline
$\displaystyle\frac{\dd n_s}{\dd\ln k}$&$-0.0065\pm0.0076$&$-8\times10^{-4}$&$-5\times10^{-4}$&$-1\times10^{-3}$&$-8\times10^{-4}$\nline
$r$&$<0.07$ at $95\%$ CL&0.16&0.13&0.31&0.26\nline
\hline
\end{tabular}}
\let\nline\undefined
\let\esps\undefined
\let\esp\undefined
Then, the power-law potential for inflation is completely predictive. The only uncertainty comes from $N$. Since we are only able to know the physics at the relatively low energy scales, we cannot be certain about the post-inflationary dynamics before Hot Big Bang, such as how the inflaton $\phi$ decays into visible matter and dark sector. But in general, the CMB scales should correspond to $N\sim50$ to $60$. Therefore we can simply calculate the observables for $N=50$ and $N=60$ separately. In \reftab{ssi-pobs}, we show the results for $p=2$ and $p=4$.

\fig[1]{ssi-Planck}{Planck constraints for single-field slow-roll inflation.}{The Planck observations constrain inflationary models as above \cite{Ade:2015lrj}. The $x$ and $y$ coordinates correspond to the spectral index and the tensor-to-scalar ratio. The shaded grey, red, and blue regions show the Planck observational constraints when combined with different data sources. The colour depths indicate the $1\sigma$ and $2\sigma$ confidence levels. Other shaded regions and line segments correspond to the predictions of various inflationary models, for 50 and 60 e-folds of inflation before the Hubble exit on smaller and bigger circle sides respectively.}
From \reftab{ssi-pobs}, we see that quadratic inflation agrees very well with the observed scalar perturbations. It however produces some tensor perturbations which is in tension with Planck. The chaotic inflation model is in even stronger tension with Planck data in scalar spectral index and tensor-to-scalar ratio, as can be seen from \refig{ssi-Planck}.

In fact, the potential \refeq{ssi-Vpl} makes it difficult to produce large tensor-to-scalar ratio while keeping $n_s\approx1$. From \refeq{ssi-pns} and \refeq{ssi-pr}, the power-law potential needs to satisfy an additional consistency relation besides those in \refssec{Testing single-field slow-roll inflation with the CMB}:
\eq{\left(1+\frac{2}{p}\right)^2r^2=64(1-n_s)^2.}
Given $r<0.07$ and $n_s\approx0.96$, the consistency relation requires $p\lesssim1$, which is beyond the scope of the thesis. The power-law potential cannot reproduce our observations for $p=2$ or $p=4$, due to the consistency relation.

On the other hand, another fundamental problem of the power-law potential is that the inflaton $\phi$ typically needs to reach Planck scale $\mpl$ to produce inflation, as required in \refeq{ssi-pireq} \ea{\cite{Lyth:1998xn}}. \ed{This is sometimes called the \emph{trans-Planckian problem}.}

\sssecs{Inflection point inflation}
In this section, we will consider the single-field slow-roll inflation models that can be effectively regarded as an inflection or saddle point potential in the neighbourhood\footnote{
We only study inflection point here.}
. Such potentials may arise from Minimal Supersymmetric Standard Model (MSSM)~\cite{Chung:2003fi,BuenoSanchez:2006xk}, as an example. The inflection point brings about a plateau in the potential, which is flat enough locally to accommodate slow roll while the inflaton stays sub-Planckian ($\ll\mpl$).

The constructed scalar potential is \cite{Allahverdi:2006iq,Allahverdi:2006we}\footnote{
Inflection point potentials are possible in various forms \cite{Allahverdi:2006iq,Allahverdi:2007vy,Allahverdi:2006we,Enqvist:2010vd,Mazumdar:2010sa,Chatterjee:2011qr,Mazumdar:2011ih,Wang:2013hva}. Here we only discuss a specific one.}
\eq{V(\phi) = \hf m^2\phi^2-\lambda A\frac{\phi^6}{6\mpl^3}+\lambda^2\frac{\phi^{10}}{\mpl^6},\label{eq-ssi-f-V}}
where $m$ and $A>0$ are called the soft breaking mass and the $A$-term respectively. Let us define
\eq{4\alpha^2\equiv1-\frac{A^2}{40m^2}.\label{eq-ssi-f-alpha}}
There exists an inflection point $\phi_0$ in the potential $V(\phi)$ with $V''(\phi_0)=0$, which lies at
\eq{\phi_0=\left(\frac{m\mpl^3}{\lambda\sqrt{10}}\right)^\frac{1}{4}+ \cO(\alpha^2).\label{eq-ssi-f-phi0}}
At the inflection point $\phi_0$,
\eqa{V(\phi_0)&=&\frac{4}{15}m^2\phi_0^2+\cO(\alpha^2),\label{eq-ssi-f-V0}\\
V'(\phi_0)&=&4\alpha^2m^2\phi_0+\cO(\alpha^4),\label{eq-ssi-f-Vp0}\\
V'''(\phi_0)&=&32\frac{m^2}{\phi_0}+\cO(\alpha^2).}

\fig[0.8]{ssi-f-pot}{A schematic potential for inflection point inflation}{The inflection point inflation can be achieved typically around the inflection point $\phi_0$ because of the cancellation in $V'(\phi)$.}
Neglecting higher order expansion terms around the inflection point $\phi_0$, the effective potential around inflection point becomes
\eq{V(\phi)=V(\phi_0)+V'(\phi_0)(\phi-\phi_0)+\frac{V'''(\phi_0)}{6}(\phi-\phi_0)^3.}
The first and second slow-roll parameters then become
\eqa{\epsilon_\phi&=&\frac{225\mpl^2}{16\pi\phi_0^6}\left[\alpha^2\phi_0^2+4(\phi-\phi_0)^2\right]^2(1+\cO(\alpha^2)),\label{eq-ssi-f-epsilon}\\
\eta_\phi&=&\frac{15\mpl^2}{\pi\phi_0^3}(\phi-\phi_0)+\cO(\alpha^2).\label{eq-ssi-f-eta}}
From \refeq{ssi-f-epsilon} and \refeq{ssi-f-eta}, we can find inflation near the inflection point if the potential is flat enough with $\epsilon_\phi<1$, i.e.
\eq{\alpha^4\ll\frac{16\pi\phi_0^2}{225\mpl^2}\ll1.}
The second inequality comes from our wish to keep the inflaton sub-Planckian. The potential then has the shape as \refig{ssi-f-pot}.

From now on we keep only the leading order terms. The slow-roll parameters are then simplified to
\eqa{\epsilon_\phi&=&\frac{225\mpl^2}{16\pi\phi_0^6}\left[\alpha^2\phi_0^2+4(\phi-\phi_0)^2\right]^2,\label{eq-ssi-f-epsilon2}\\
\eta_\phi&=&\frac{15\mpl^2}{\pi\phi_0^3}(\phi-\phi_0).\label{eq-ssi-f-eta2}}
Inflation hence only occurs close enough to the inflection point with $|\eta_\phi|<1$, as\footnote{
The first slow-roll condition should also be satisfied to allow slow-roll inflation. However, during inflection point inflation, the violation of the second slow-roll condition is usually much earlier. For this reason, we don't consider the first slow-roll condition.}
\eq{|\phi-\phi_0|<\frac{\pi\phi^3_0}{15\mpl^2}.\label{eq-ssi-f-cond}}
The Hubble expansion rate is given by
\eq{H^2\approx\frac{32\pi m^2\phi_0^2}{45\mpl^2}.}

According to \refeq{ssi-f-cond}, we set the end of inflation at
\eq{\phi_e\equiv\phi_0-\frac{\pi\phi^3_0}{15\mpl^2},}
where the subscript $e$ indicates end of inflation. The dynamics of background solution can then be solved, which yields the power spectrum of curvature perturbation \cite{BuenoSanchez:2006xk, Allahverdi:2007wh, Enqvist:2010vd,Wang:2013hva}
\eq{\P_\zeta(N)=\frac{256\pi m^2\phi_0^4}{3\times15^3\alpha^4\mpl^6}\sin^4\left(\frac{15\alpha\mpl^2}{4\pi\phi_0^2}N\right),}
and the spectral index for scalar perturbations \cite{BuenoSanchez:2006xk, Allahverdi:2007wh, Enqvist:2010vd,Wang:2013hva}
\eq{n_s=1-\frac{15\alpha\mpl^2}{\pi\phi_0^2}\cot\left(\frac{15\alpha\mpl^2}{4\pi\phi_0^2}N\right).}

Since the potential in this setup is usually very flat around the inflection point, the second slow-roll parameter $|\eta_\phi|$ is usually much larger than the first slow-roll parameter $\epsilon_\phi$. The spectral tilt $n_s-1$ should then mostly come from $\eta_\phi$, while leaving $\epsilon_\phi$ very small. In such cases, one would expect a very small tensor-to-scalar ratio. Overall, the inflection point potential \refeq{ssi-f-V} has been shown to agree well with the Planck observations in \cite{Wang:2013hva}.

\ea{On the other hand, inflection and/or saddle point potentials of inflation require the model parameters to be substantially tuned\cite{BuenoSanchez:2006xk}. Here the parameters $m$ and $A$ must be tuned to bring about a very small $\alpha$. This raises the question why in nature the parameters would cancel so finely, and can be regarded disadvantageous for inflection/saddle point inflation. This is however beyond the scope of the thesis.}

To summarize, in this chapter we have derived generic predictions of single-field slow-roll inflation. We have studied power-law and inflection point potentials. The consistency relation in single-field slow-roll inflation forbids many features in the CMB, such as large non-Gaussianities, which will be investigated in the framework of multi-field inflation in the upcoming chapters.

%% file: Chapters/Multi.tex
In this chapter, we derive the cosmological predictions for multi-field inflation. Spectator fields can be regarded as the minimal multi-field inflation scenario. Single-field slow-roll inflation with an extra perfect fluid is also discussed. Conclusions of this chapter will provide assistance for the spectator calculations in the next chapter.

\ssecs{Generic multi-field slow-roll inflation}
\sssec{$\delta N$ formalism}
Consider $n$ slowly rolling real canonical scalar fields, indicated with $\phi^\mu$ where $\mu=0,1,\dots,n-1$. Assuming $\delta N$ formalism (see \refsssec{Separate universe approach} and \cite{Sasaki:1995aw,Wands:2000dp,Lyth:2005fi,Seery:2005gb,Byrnes:2006vq}) and perturbative calculations are applicable, we can always write the remaining e-folds of universe expansion as a function of the fields \ea{on spatially flat hypersurfaces}
\eq{N=N(\phi^0,\phi^1,\dots,\phi^{n-1})=N(\phi^\mu).}
According to separate universe approach \cite{Sasaki:1995aw,Wands:2000dp}, the perturbation in $N$ then corresponds to \ea{a} curvature perturbation, which can be expanded in terms of the field perturbations $\delta\phi^\mu$ \ea{on spatially flat hypersurfaces} as \cite{Lyth:2005fi,Seery:2005gb,Byrnes:2006vq}
\eq{\zeta(\vs x)=\delta N(\vs x)=N_\mu\delta\phi^\mu+\hf N_{\mu\nu}(\delta\phi^\mu\delta\phi^\nu-\langle\delta\phi^\mu\delta\phi^\nu\rangle)+\frac{1}{6}N_{\mu\nu\lambda}\delta\phi^\mu\delta\phi^\nu\delta\phi^\lambda+\ho,}
where
\eqa{N_\mu&\equiv&\frac{\dd N}{\dd\phi^\mu},\\
N_{\mu\nu}&\equiv&\frac{\dd^2 N}{\dd\phi^\mu\dd\phi^\nu},\\
N_{\mu\nu\lambda}&\equiv&\frac{\dd^3 N}{\dd\phi^\mu\dd\phi^\nu\dd\phi^\lambda},\\
\dots.\nonumber}
The space-time dependences are omitted for simplicity.

We know for slowly rolling scalars, their quantum fluctuations should be Gaussian, like \refeq{ssi-psipower}, which will satisfy
\eq{\langle\delta\phi^\mu(\vs k)\delta\phi^\nu(\vs k')\rangle=(2\pi)^3\delta^{\mu\nu}\delta^3(\vs k-\vs k')P_{\delta\phi}(k).\label{eq-mul-g-dphi2}}
For super-Hubble perturbations, \refeq{ssi-dphi} and \refeq{ssi-dphix2} still hold, giving rise to
\eq{P_{\delta\phi}(k)=\frac{H^2}{2k^3},\hspace{0.8in}\P_{\delta\phi}=\frac{H^2}{4\pi^2}.}
The leading order curvature perturbation is the Gaussian part
\eq{\zeta_G(\vs x)\equiv N_\mu\delta\phi^\mu(\vs x).}
The power spectrum of curvature perturbation can hence be calculated easily at leading order \cite{Sasaki:1995aw,Wands:2000dp}
\eq{\P_\zeta=\P_{\delta\phi}\sum_\mu N_\mu^2.}

The local bi-spectrum is measured from (\refssec{CMB angular bi-spectrum})
\eqa{\langle\zeta^3(\vs x)\rangle&=&3\left\langle\zeta_G^2\times\hf N_{\mu\nu}\Bigl(\delta\phi^\mu\delta\phi^\nu-\langle\delta\phi^\mu\delta\phi^\nu\rangle\Bigr)\right\rangle\nonumber\\
&=&\frac{3}{2}N_{\mu\nu} N_\lambda N_\eta\left\langle\delta\phi^\lambda\delta\phi^\eta\Bigl(\delta\phi^\mu\delta\phi^\nu-\langle\delta\phi^\mu\delta\phi^\nu\rangle\Bigr)\right\rangle\nonumber\\
&=&3N_{\mu\nu}N_\mu N_\nu \langle(\delta\phi)^2\rangle^2\nonumber\\
&=&\frac{3N_{\mu\nu}N_\mu N_\nu}{\Bigl(\sum_\mu N_\mu^2\Bigr)^2}\langle\zeta_G^2(\vs x)\rangle^2.\label{eq-mul-g-z31}}
Recall the definition of $\fNL$ from \refeq{cmb-zetang} and \refeq{cmb-fnlzeta3}. By comparing \refeq{cmb-fnlzeta3} and \refeq{mul-g-z31}, we find the expression of $\fNL$ for multi-field inflation, as \cite{Lyth:2005fi,Seery:2005gb}
\eq{\fNL=\frac{5}{6}\frac{N_{\mu\nu}N_\mu N_\nu}{\Bigl(\sum_\mu N_\mu^2\Bigr)^2}.\label{eq-mul-g-fnl}}

Similarly, the local tri-spectra can be derived as \cite{Byrnes:2006vq}
\eqa{\gNL&=&\frac{25}{54}\frac{N_{\mu\nu\lambda}N_\mu N_\nu N_\lambda}{{\Bigl(\sum_\mu N_\mu^2\Bigr)^3}}\;,\label{eq-mul-g-gnl}\\
\tNL&=&\frac{N_\mu N_{\mu\lambda}N_{\lambda\nu}N_\nu}{\Bigl(\sum_\mu N_\mu^2\Bigr)^3}\;.\label{eq-mul-g-tnl}}

When only one field contributes to \ea{the} curvature perturbation, such as when other fields are heavy and provide negligible perturbations, we can redefine the fields so that the only field contributing to curvature perturbation is named as $\phi^0$. Then the only non-vanishing term among $N_\mu$ for $\mu=0,1,\dots,n-1$ is $N_0$. The non-Gaussianity parameters reduce to 
\eqa{\fNL&=&\frac{5}{6}\frac{N_{00}}{N_0^2}\;,\label{eq-mul-g-fnls}\\
\gNL&=&\frac{25}{54}\frac{N_{000}}{N_0^3}\;,\label{eq-mul-g-gnls}\\
\tNL&=&\frac{N_{00}^2}{N_0^4}\;.\label{eq-mul-g-tnls}}
Therefore, the consistency relation \refeq{ssi-tNL} also holds in the multi-field scenarios where only one field produces curvature perturbation.

For the power spectrum of tensor perturbations, the same relation \refeq{ssi-Pt3} also holds for multi-field inflation. However, the tensor-to-scalar ratio changes because the scalar perturbations are different
\eq{r\equiv\frac{\P_t}{\P_\zeta}=\frac{64\pi}{\mpl^2\sum_\mu N_\mu^2}\;.}

\sssecs{Multi-field evolution}
In the above section, we have expressed cosmological observables in terms of $N_\mu$, $N_{\mu\nu}$, and $N_{\mu\nu\lambda}$. We will move on to computing $N_\mu$, $N_{\mu\nu}$, and $N_{\mu\nu\lambda}$ in this section.

\fig{mul-g-dec}{Adiabatic-entropy and curvature-isocurvature decompositions of multi-field perturbations}{The field perturbation $\delta\phi$ (black) is schematically decomposed into adiabatic + entropy perturbation modes (green), or into field perturbation modes (purple), or into curvature + isocurvature perturbation modes (red) on a two dimensional hypersurface of the full phase space. The adiabatic and curvature modes are along the trajectory (blue). The entropy mode is perpendicular to the trajectory while the isocurvature mode is on the uniform $N$ hypersurface (dashed red).}
Consider the generic multi-field inflation scenario with $n$ slowly rolling canonical scalar fields, $\phi^\mu$, where $\mu=0,1,\dots,n-1$. When the full action of the model and the initial conditions are given, the classical solution of the background evolution becomes a fixed trajectory in the $n$ dimensional phase space of the physical system, as in \refig{mul-g-dec}. Based on the perturbative calculations of separate universe approach \cite{Sasaki:1995aw,Sasaki:1998ug,Wands:2000dp,Lyth:2004gb,Sasaki:2007ay}, we can treat each Hubble patch as a homogenous separate universe which receives the field perturbations $\phi^\mu\rightarrow\phi^\mu+\delta\phi^\mu$. As shown in \refig{mul-g-dec}, the generic perturbations $\delta\phi^\mu$ can be decomposed in three typical ways, each of which then corresponds to a different approach in calculating the evolution of perturbations:
\begin{itemize}
\item The perturbation $\delta\phi^\mu$ is decomposed along the trajectory of the background evolution, and the $n-1$ dimensional hypersurface that is perpendicular to the trajectory. The \emph{adiabatic} mode is the component along the trajectory, and the $n-1$ \emph{entropy} modes are on the $n-1$ dimensional hypersurface. Together, they form a complete and orthogonal basis of the $n$ dimensional field space. The adiabatic mode can be regarded as a time shift, which is the same as the perturbation mode in single-field slow-roll inflation. The complexity arises from the entropy modes, which can still produce curvature perturbation after Hubble exit. From the Hubble exit, we need to keep track of all the perturbations and calculate how entropy modes transfer to adiabatic mode as the universe evolves, up to the point when entropy modes cease to transfer to adiabatic mode, known as the \emph{adiabatic limit}. After reaching the adiabatic limit, the adiabatic perturbation then corresponds to the amount of curvature perturbation that would be generated. Isocurvature perturbations can be calculated similarly \cite{GarciaBellido:1995qq,Gordon:2000hv}.
\item The perturbation $\delta\phi^\mu$ is decomposed along the $n$ field directions, into $n$ separate components of the field perturbations, $\delta\phi^\mu$. We can then evolve the field perturbations $\delta\phi^\mu$, or the distribution of perturbations $P(\delta\phi^\mu)$, from Hubble exit to the adiabatic limit. Then, field perturbations can be projected onto the curvature perturbation direction, and also isocurvature perturbation directions if needed, yielding the cosmological predictions straightforwardly \cite{Mulryne:2009kh,Mulryne:2010rp,Seery:2012vj}.
\item The perturbations $\delta\phi^\mu$ are decomposed along the trajectory of background evolution, and the $n-1$ dimensional hypersurface, on which any field perturbation does not lead to any curvature perturbation. The \emph{curvature} mode is the component along the trajectory, and the $n-1$ \emph{isocurvature} modes are on the $n-1$ dimensional hypersurface. Together, they form a complete but not necessarily orthogonal basis of the $n$ dimensional field space. Given any perturbation $\delta\phi^\mu$ on any point of the trajectory, we can instantly tell that $\delta\phi^\mu$ produces the curvature perturbation that is exactly equal to the curvature mode of the decomposition. This is because the isocurvature modes do not produce any curvature perturbation by definition (see \refig{mul-g-dec}). According to $\delta N$ formalism, such isocurvature modes do not produce any $\delta N$ in separate universes, so we can call the $n-1$ dimensionsal hypersurface as the \emph{uniform $N$ hypersurface}. Note that in our convention, adiabatic and curvature perturbations are not the same, and neither are isocurvature and entropy perturbations. To derive the $n-1$ dimensional isocurvature hypersurface, we need to evolve the isocurvature hypersurface from a known position, such as at the adiabatic limit or a known boundary, back to the Hubble exit of the perturbation mode of our concern. Isocurvature perturbations can be calculated similarly \cite{Vernizzi:2006ve,Yokoyama:2007uu,Yokoyama:2007dw,Mazumdar:2012jj}.
\end{itemize}
For the comparisons between these approaches, see \cite{Mulryne:2010rp,Mazumdar:2012jj,Seery:2012vj}.

For convenience, we will use the last two approach\ea{es} in the thesis. In order to parameterize the uniform $N$ hypersurface, we can use $N_\mu$ to determine the direction of the hypersurface, $N_{\mu\nu}$ for the geometrical curvature of the hypersurface, and so forth for higher order expansion if required. Therefore the question of evolving the uniform $N$ hypersurface now becomes the question of evolving $N_\mu$, $N_{\mu\nu}$, $\dots$, from a known boundary hypersurface at a later time to the Hubble exit of the perturbation mode of our interest. For example, we may pick the boundary \ed{at }when the universe reaches the adiabatic limit, where the uniform $N$ hypersurface overlaps with uniform energy density hypersurface.

According to \ea{the} separate universe approach, we can parameterize the phase space of \ea{a} homogenous Hubble patch with $\phi^\mu$, for the $n$ fields with $\mu=0,1,\dots,n-1$. Here we also define a different parameterization, using $p_\mu$ where $\mu=0,1,\dots,n-1$. The zeroth component corresponds to $p_0\equiv N$, the coordinate along curvature direction, and $p_i$ (for the $i=1,2,\dots,n-1$ components in $p_\mu$) are the isocurvature coordinates \ea{on uniform $N$ hypersurface}.

In the absence of any isocurvature perturbation concerns, $p_i$ can be chosen arbitrarily without exact definitions, as long as they form a complete but not necessarily orthogonal basis of uniform $N$ hypersurface. We can then write the local coordinate transformation $\phi^\mu\longleftrightarrow p_\mu$ as
\eq{\dd\phi^\mu=\sigma^{\mu\nu}\dd p_\nu,}
where the summation over $\nu$ is implicit, and the transformation matrix $\sigma^{\mu\nu}$ is defined as
\eq{\sigma^{\mu\nu}\equiv\frac{\dd\phi^\mu}{\dd p_\nu}.}
The inverse transformation is
\eq{\dd p_\mu=\sigma_{\mu\nu}\dd\phi^\nu,}
in which the inverse transformation matrix satisifies
\eq{\sigma_{\mu\nu}\equiv\frac{\dd p_\mu}{\dd\phi^\nu},}
with the consistency relation
\eq{\sigma_{\mu\alpha}\sigma^{\alpha\nu}=\sigma^{\nu\alpha}\sigma_{\alpha\mu}=\delta_\mu^\nu\;.\label{eq-mul-g-delta}}

Consider the action for multi-field slow-roll inflation
\eq{S\equiv\int\sqrt{-g}\,\dd^4x\left(\frac{\mpl^2}{16\pi}R-V+\sum_\mu\frac{1}{2}\partial^\nu\phi^\mu\partial_\nu\phi^\mu\right),\label{eq-mul-g-S0}}
where the \ed{total }potential is abbreviated from the form
\eq{V=V(\phi)=V(\phi^\mu)\equiv V(\phi^0,\phi^1,\dots,\phi^{n-1}).}
Assuming we have solved the background evolution for the fields $\phi^\mu$, we then have limited information about the transformation matrix, as
\eq{\sigma^{\mu0}=\frac{\dd\phi^\mu}{\dd N}=\frac{\mpl^2V_{,\mu}}{8\pi V}\;,\label{eq-mul-g-sm0}}
where we use superscripts and subscripts ``$,\mu$'' for total derivatives w.r.t $p_\mu$ and $\phi^\mu$ respectively. The slow-roll approximation has been applied at the second equality.

The differentiations satisfy
\eq{\frac{\dd}{\dd p_\mu}=\sigma^{\mu\nu}\frac{\dd}{\dd\phi^\nu},\hspace{0.5in}\frac{\dd}{\dd\phi^\mu}=\sigma_{\mu\nu}\frac{\dd}{\dd p_\nu}.\label{eq-mul-g-dd}}
It is also commutative on $\sigma$ by definition, as
\eq{\sigma^{\mu\nu,\lambda}=\sigma^{\mu\lambda,\nu}.\label{eq-mul-g-com}}
Applying \refeq{mul-g-dd} on \refeq{mul-g-com}, we find
\eq{\sigma^{\mu\nu,\lambda}=\sigma^{\mu\lambda,\nu}=\sigma^{\alpha\nu}\sigma^{\mu\lambda}{}_{,\alpha}.}

Since the inverse transformation matrix $\sigma_{\lambda\mu}$ is the inverse of $\sigma^{\lambda\mu}$, its derivative should follow the differential rule for inverse matrices, i.e.
\eq{\sigma_{\lambda\mu}{}^{,0}=-\sigma_{\lambda\alpha}\sigma_{\beta\mu}\sigma^{\alpha\beta,0}=-\sigma_{\lambda\alpha}\sigma_{\beta\mu}\sigma^{\gamma\beta}\sigma^{\alpha0}{}_{,\gamma}=-\sigma_{\lambda\alpha}\sigma^{\alpha0}{}_{,\mu}\;.\label{eq-mul-g-smn0}}
By substituting \refeq{mul-g-sm0} into \refeq{mul-g-smn0}, it reduces to
\eq{\sigma_{\lambda\mu}{}^{,0}=\frac{\mpl^2}{8\pi}\left(\frac{V_{,\alpha}V_{,\mu}}{V^2}-\frac{V_{,\alpha\mu}}{V}\right)\sigma_{\lambda\alpha}.\label{eq-mul-g-smn02}}
After the manipulation, \refeq{mul-g-smn02} has become a set of first order linear differential equations for $\sigma_{\lambda\mu}$, w.r.t the e-folding $p_0=N$. The equation set is separable for each $\lambda$. Since the background evolution $\phi^\mu$ is known, the coefficients of \refeq{mul-g-smn02} are also known.

By definition $N_\mu\equiv N_{,\mu}=\sigma_{0\mu}$, so the $\lambda=0$ part  of \refeq{mul-g-smn02} corresponds to
\eq{\frac{\dd N_\mu}{\dd N}=\frac{\mpl^2}{8\pi}\left(\frac{V_{,\alpha}V_{,\mu}}{V^2}-\frac{V_{,\alpha\mu}}{V}\right)N_\alpha.\label{eq-mul-g-dN1}}
Therefore we have formulated the set of differential equations that $N_\mu$ should satisfy, which can then be solved backwards on a per model basis.

For higher order perturbations, some calculations would reveal a similar relation
\eq{\sigma_{\lambda\mu,\nu}{}^{,0}=-\sigma_{\lambda\mu,\alpha}\sigma^{\alpha0}{}_{,\nu}-\sigma_{\lambda\nu,\alpha}\sigma^{\alpha0}{}_{,\mu}-\sigma_{\lambda\alpha}\sigma^{\alpha0}{}_{,\mu\nu}.\label{eq-mul-g-slmn}}
The $\lambda=0$ component corresponds to
\eqa{\frac{\dd N_{\mu\nu}}{\dd N}&=&\frac{\mpl^2}{8\pi}\left(\frac{V_{,\alpha}V_{,\nu}}{V^2}-\frac{V_{,\alpha\nu}}{V}\right)N_{\mu\alpha}+\frac{\mpl^2}{8\pi}\left(\frac{V_{,\alpha}V_{,\mu}}{V^2}-\frac{V_{,\alpha\mu}}{V}\right)N_{\nu\alpha}\nonumber\\
&&-\frac{\mpl^2}{8\pi}\left(\frac{2V_{,\alpha}V_{,\mu}V_{,\nu}}{V^3}-\frac{V_{,\alpha}V_{,\mu\nu}}{V^2}-\frac{V_{,\mu}V_{,\alpha\nu}}{V^2}-\frac{V_{,\nu}V_{,\alpha\mu}}{V^2}+\frac{V_{,\alpha\mu\nu}}{V}\right)N_\alpha.\label{eq-mul-g-dN2}}
After solving $N_\mu$ from \refeq{mul-g-dN1}, we are then able to solve $N_{\mu\nu}$ from \refeq{mul-g-dN2}.

From above, we can solve $N_\mu$ and $N_{\mu\nu}$, so the curvature perturbation $\zeta$ from any field perturbations $\delta\phi^\mu$ is predictable. For the isocurvature perturbations, if we know exactly how the inflationary model couples to the standard model degrees of freedom, we can then define isocurvature directions accordingly.

For example, when we are interested in the relative energy density perturbation of cold dark matter (CDM), we can specifically define $p_1$ along this direction. The rest of the components $p=2,\dots,n-1$ should be picked on the uniform $N(=p_0)$ and $p_1$ hypersurface with $n-2$ dimensions, so that these components are decoupled from CDM isocurvature perturbation in addition to curvature perturbation. Such isocurvature perturbations should follow the same evolution equations as \refeq{mul-g-dN1} and \refeq{mul-g-dN2}, corresponding to different $\lambda$ components of the same equations, \refeq{mul-g-smn02} and \refeq{mul-g-slmn}.

One substantial application of the local coordinate transformation $\phi^\mu\longleftrightarrow p_\mu$ is to obtain $N_0$ after knowing the rest, i.e. $N_i$, for $i=1,2,\dots,n-1$. This is useful, for example, when calculating the combined curvature perturbations from the inflaton and the curvaton/spectator field in \refsec{Spectator}.

For this purpose, consider the field perturbation $\delta\phi^0$, which can be decomposed into two components:
\begin{itemize}
\item the perturbation $\delta p_0$ along the trajectory direction, which projects onto the field perturbation space as $\{\delta\phi^\mu=\frac{\dd \phi^\mu}{\dd p_0}\delta p_0\}$, and
\item the field perturbations $\delta\phi^i$ which fill the remaining $n-1$ dimensions in the phase space.
\end{itemize}
The first component contributes to the curvature perturbation by the amount
\eq{\delta p_0=\left(\frac{\dd\phi^0}{\dd p_0}\right)^{-1}\delta\phi^0.}
The second component then cancels out the rest of the field perturbations $\delta\phi^i$ coming from the first component. Their net contribution to field space is exactly $\delta\phi^0$, whilst their total contribution to curvature perturbation can be obtained because we know $N_i$.

The mathematical approach is simple. From \refeq{mul-g-delta}, we get
\eq{\sigma_{0\mu}\sigma^{\mu0}=N_\mu\frac{\dd\phi^\mu}{\dd N}=1.}
This immediately gives
\eq{N_0=\left(\frac{\dd\phi^0}{\dd N}\right)^{-1}\left(1-N_i\frac{\dd\phi^i}{\dd N}\right),\label{eq-mul-g-N0}}
where $\dd\phi^\mu/\dd N$ is known from the background equations of motion.

\sssecs{Two-field inflation}
When only two fields are present, the conclusions can be further simplified. For $\mu=\nu=0$, \refeq{mul-g-delta} becomes
\eq{\frac{\mpl^2}{8\pi V}\sum_\alpha N_\alpha V_{,\alpha}=1.}
This allows us to eliminate $N_\alpha$ for $\alpha\ne\mu$ in \refeq{mul-g-dN1} because there are only two fields, which gives
\eq{\frac{\dd N_0}{\dd N}=uN_0+v,\label{eq-mul-g-2dN}}
where the temporary variables are defined as
\eqa{u&\equiv&\frac{\mpl^2}{8\pi}\left(\frac{V_{,0}V_{,01}}{VV_{,1}}-\frac{V_{,00}}{V}\right),\\
v&\equiv&\frac{V_{,0}}{V}-\frac{V_{,01}}{V_{,1}},}
and similarly for $N_1$.

Now the differential equations for $N_0$ and $N_1$ are separable, allowing us to provide a solution in the integral form \ea{with Mathematica}
\eq{N_0(N)=N_0(N\subs{ref})+e^{\int_{N\subs{ref}}^Nu\dd N}\int_{N\subs{ref}}^Nve^{-\int_{N\subs{ref}}^Nu\dd N}\dd N,\label{eq-mul-g-int}}
where the subscript ``ref'' corresponds to a reference point which can be picked at the time when the universe has become adiabatic, so $N_0(N\subs{ref})$ is known. It should stay on the same trajectory with that of background evolution, i.e. $p_i\sups{ref}=p_i$.

Similarly, since $N_{\mu\nu}\sigma^{\nu0}=\dd N_\mu/\dd N$ can be read out directly from \refeq{mul-g-2dN}, for second order perturbations we can also eliminate $N_{\mu\alpha}$ for $\alpha\ne\nu$ in \refeq{mul-g-dN2}. The second order equations will also \ea{fit within}\ed{surrender to} the form \refeq{mul-g-2dN}, but with different though more complicated $u$ and $v$. The second order solutions hence also have an integral form. The integral form is a further simplification that only applies on two-field models, because we are unable to separate the differential equations for $n>2$ in general.

\sssecs{Separable potentials}
There are other scenarios where the conclusions can be further simplified, even for more than two fields. One common case is when the potential adopts a specific form, such as the \emph{separable potentials} \cite{GarciaBellido:1995qq,Vernizzi:2006ve,Choi:2007su}. For example, very often we may encounter potentials that can be decomposed for each field, as
\eq{V(\phi^\mu)\equiv\sum_\mu V_\mu(\phi^\mu),\label{eq-mul-Vsep}}
so the fields do not couple with each other. Note here that $V_\mu$ indicates the separated potentials for each field $\phi^\mu$, not the derivative of $\phi^\mu$ which would contain a comma.

Since the potential is separable, we have the relation
\eq{V_{,\mu\nu}=0,\hspace{0.5in}\mathrm{for\ }\mu\ne\nu.}
This allows us to greatly simplify \refeq{mul-g-dN1}, giving
\eqa{\frac{\dd N_\mu}{\dd N}&=&\frac{\mpl^2}{8\pi}\frac{V_{,\alpha}V_{,\mu}}{V^2}N_\alpha-\frac{\mpl^2V_{,\mu\mu}}{8\pi V}N_\mu\\
&=&\frac{V_{,\mu}}{V}N_\alpha\frac{\dd\phi^\alpha}{\dd N}-\eta_\mu N_\mu\\
&=&-\eta_\mu N_\mu+\frac{V_{,\mu}}{V},\label{eq-mul-g-sN1}}
where in the last step, we have applied the relation $\sigma_{0\alpha}\sigma^{\alpha0}=1$. The slow-roll parameters for multi-field inflation with the separable potential \refeq{mul-Vsep} are defined as
\eqa{\epsilon_\mu&\equiv&\frac{\mpl^2V_{,\mu}^2}{16\pi\rho^2},\\
\eta_\mu&\equiv&\frac{\mpl^2V_{,\mu\mu}}{8\pi\rho},}
and the total energy density is just $\rho\approx V$.

For separable potentials, the slow-roll parameters satisfy the relation
\eq{\frac{\dd\ln\epsilon_\mu}{\dd N}=2\eta_\mu-4\epsilon,\label{eq-mul-g-sep1}}
where the total first slow-roll parameter is defined as
\eq{\epsilon\equiv-\frac{\dd}{\dd t}\frac{1}{H}=\frac{\dd\ln H}{\dd N}=\sum_\mu\epsilon_\mu.\label{eq-mul-g-sep2}}

The differential equation \refeq{mul-g-sN1} yields to the same form with \refeq{mul-g-2dN}, with
\eq{u=-\eta_\mu,}
and
\eq{v=\frac{V_{,\mu}}{V}.}
In the case of separable potentials, the integrals in \refeq{mul-g-int} can be worked out analytically, because they only contain total derivatives. For example, $\eta_\mu$ is a total derivative as can be seen from \refeq{mul-g-sep1} and \refeq{mul-g-sep2}. The solution is
\eq{N_\mu=N_\mu\sups{ref}+\frac{8\pi}{\mpl^2}\frac{V_\mu-V_\mu\sups{ref}}{V_{,\mu}}.\label{eq-mul-g-sdN1}}
The superscript ``ref'' is similar to the subscripts in \refeq{mul-g-int}. The same formula is also derived in \cite{GarciaBellido:1995qq,Vernizzi:2006ve,Choi:2007su}.

\ssecs{Spectator fields}
From \ea{the} above, we see that multi-field inflation can be simplified a lot with only two fields, or with uncoupled fields. When we look for the minimal extension of single-field slow-roll inflation, one possible candidate is to introduce only two uncoupled fields (except minimally by gravity). Generically, both fields can participate in inflation, and produce curvature perturbations, but again the simplest possibility is the two fields have separate roles. The \emph{inflaton} field $\phi$ dominates energy density and drives inflation, while it only contributes negligibly to curvature perturbation. The \emph{spectator} field $\sigma$ then remains subdominant in energy density, but is responsible for creating curvature perturbation.

The definition of spectator fields in this thesis follow the criteria below:
\begin{itemize}
\item The spectator field produces \ea{the} observable curvature perturbation, in terms of amount and length scale. Otherwise, we are not motivated to study it.
\item The inflation background dynamics \ea{are hardly changed}\ed{remain unchanged} if the spectator field is modified or removed, at least during the period it produces observable perturbations. This requires the spectator to remain subdominant in energy density. The inflation dynamics should be \ea{hardly affected}\ed{unaffected} by the spectator\ed{, and independent of the spectator potential}.
\end{itemize}

In principle, there are no other limitations, such as when the spectator decays or ends slow roll, or what its (or its decay products') energy density ratio is in the beginning of Hot Big Bang. Most two-field inflation scenarios, such as hybrid inflation \cite{Linde:1993cn} or assisted inflation \cite{Liddle:1998jc,Malik:1998gy,Copeland:1999cs,Mazumdar:2001mm,Jokinen:2004bp,Dimopoulos:2005ac}, do not belong to the inflaton-spectator category. This is because the inflation dynamics cannot stay exactly the same after removing one field. The inflaton and the spectator can even couple as long as the above criteria are satisfied, but we will insist on no coupling in the thesis, purely for minimalism.

The total potential can hence be written as
\eq{V\subs{tot}(\phi,\sigma)\equiv V(\phi)+U(\sigma),\label{eq-mul-spector}}
where $V$ and $U$ are the respective potentials for $\phi$ and $\sigma$. During inflation, the inflaton dominates \ea{the} energy density, and we can neglect $\sigma$ for background evolution. The system then reduces to single-field slow-roll inflation, during which we can directly apply the calculations in \refssec{ssi-Background evolution}. The perturbations, however, would involve model-dependent calculations.

\ssecs{Single inflaton with a perfect fluid}
Let us consider the case where we add a perfect fluid with a constant equation of state \ea{parameter} $w$ into single-field slow-roll inflation. We will only consider the fluid to be matter-like with $w=0$ or radiation-like with $w=1/3$ in the thesis. The perfect fluid can also represent fields, such as those in fast oscillations which behave as matter-like fluids with $w=0$.

Now the energy density of the universe contains two parts
\eq{\rho=V(\phi)+\rho_f,}
where the potential energy density $V(\phi)$ comes from the inflaton $\phi$, and the energy density of the perfect fluid is $\rho_f$. We know that a slow-roll field behaves like a cosmological constant, with
\eq{V(\phi)\propto a^0,}
but the perfect fluid is redshifted according to \cite{Mukhanov:1990me}
\eq{\rho_f\propto a^{-3(1+w)}.}
Due to the distinctive redshift rates, we shall expect the contribution from perfect fluid to be quickly redshifted away, leaving the universe in a single-field slow-roll inflation state. Before the perfect fluid is redshifted away, it however can still contribute to the e-folding of universe expansion\footnote{
We do not address how to establish the initial conditions here, but as an example the perfect fluid can come from the decay of a pre-existing slow-roll field. The preceding stage can be multi-field inflation, or the spectator mechanism which will be discussed in \refsec{Spectator}.}.

Inflation requires the \ea{comoving} Hubble radius to \ea{decrease}\ed{increase} as the universe expands, so the inflaton $\phi$ have to take the major role in energy density. We will work within this constraint, so the Hubble rate of universe expansion satisfies
\eq{H^2=\frac{8\pi}{3\mpl^2}\left(\rho_{fi}e^{3(1+w)(N-N_i)}+V(\phi)\right),\label{eq-mul-f-H}}
where the subscript $i$ indicates the initial condition. Therefore, $\rho_{fi}$ and $N_i$ are regarded as known constants for the scenario.

The equation of motion for inflaton is (after slow-roll approximation)
\eq{3H\dot\phi+V'(\phi)=0,}
or alternatively,
\eq{H^2=\frac{V'(\phi)}{3}\frac{\dd N}{\dd\phi}.\label{eq-mul-f-eomphi}}
By combining \refeq{mul-f-H} and \refeq{mul-f-eomphi}, we can eliminate the Hubble rate $H$, and reduce them to a differential equation of $\phi$ w.r.t $N$ as
\eq{\frac{8\pi}{3\mpl^2}\left[\rho_{fi}e^{3(1+w)(N-N_i)}+V(\phi)\right]=\frac{V'(\phi)}{3}\frac{\dd N}{\dd\phi}.}

The differential equation can be solved \emph{exactly} \ea{with Mathematica,} as
\eq{N_i-N=n(\phi_i,\phi)+\frac{1}{3(1+w)}\ln\frac{1-r_{fi}\alpha}{1-r_{fi}},\label{eq-mul-f-nsol}}
where $n(\phi_1,\phi_2)$ is the number of e-folds of universe expansion if the perfect fluid is not present:
\eq{n(\phi_1,\phi_2)\equiv\int_{\phi_2}^{\phi_1}\frac{8\pi V}{\mpl^2V'}\dd\phi,}
and correspondingly, the second term in \refeq{mul-f-nsol} is the contribution from perfect fluid. Also, $r_f$ is the energy density ratio of the perfect fluid versus total, as
\eq{r_f\equiv\frac{\rho_f}{V(\phi)+\rho_f}<1,}
and
\eq{\alpha\equiv1-\frac{24(1+w)\pi V(\phi_i)}{\mpl^2}\int_{\phi}^{\phi_i}\frac{e^{-3(1+w)n(\phi_i,\phi)}}{V'(\phi)}\dd\phi.\label{eq-mul-f-alpha0}}

To understand how the perfect fluid affects perturbations, let us consider the infinitesimal perturbations in initial condition \ea{on a spatially flat slicing}
\eq{\phi_i\rightarrow\phi_i+\delta\phi_i,\hspace{0.5in}\rho_{fi}\rightarrow\rho_{fi}+\delta\rho_{fi}.}
The parameters $n$, $r_{fi}$ and $\alpha$ will change correspondingly:
\eq{\delta n(\phi_i,\phi)=\frac{8\pi V(\phi_i)}{\mpl^2V'(\phi_i)}\delta\phi_i=2\sqrt\frac{\pi}{\tilde\epsilon_{\phi i}}\,\frac{\delta\phi_i}{\mpl},}
\eq{\delta r_{fi}=r_{fi}(1-r_{fi})\left(\frac{\delta\rho_{fi}}{\rho_{fi}}-4\sqrt{\pi\tilde\epsilon_{\phi i}}\frac{\delta\phi_i}{\mpl}\right),}
and
\eq{\delta\alpha=-\Bigl[4\tilde\epsilon_{\phi i}(1-\alpha)+6(1+w)\alpha\Bigr]\sqrt\frac{\pi}{\tilde\epsilon_{\phi i}}\,\frac{\delta\phi_i}{\mpl}.}
Here the tilded slow-roll parameters are defined as if the inflaton $\phi$ is the only component in the universe, and obtained by replacing the total energy density with $\phi$'s potential energy density in \refeq{ssi-epsilon0}, \refeq{ssi-eta0} and \refeq{ssi-xi}, so that
\eqa{\tilde\epsilon_\phi&\equiv&\frac{\mpl^2V'{}^2}{16\pi V^2}=\frac{\epsilon_\phi}{(1-r_f)^2},\label{eq-mul-f-epsilont}\\
\tilde\eta_\phi&\equiv&\frac{\mpl^2V''}{8\pi V}=\frac{\eta_\phi}{1-r_f},\label{eq-mul-f-etat}\\
\tilde\xi_\phi&\equiv&\frac{\mpl^4V'V'''}{(8\pi V)^2}=\frac{\xi_\phi}{(1-r_f)^2}.\label{eq-mul-f-xit}}

From above, we can find the total change in the number of e-folds of inflation by fixing the reference point at the end of inflation with $N_e=0$, as
\eq{\zeta=\delta N_i=\frac{1}{1-\alpha r_{fi}}\left[2\sqrt\frac{\pi}{\tilde\epsilon_{\phi i}}\,\frac{\delta\phi_i}{\mpl}+\frac{(1-\alpha)r_{fi}}{3(1+w)}\frac{\delta\rho_{fi}}{\rho_{fi}}\right].\label{eq-mul-f-dN}}
Therefore, when the perturbation only exists for $\phi$, the amount of the curvature perturbation produced would be similar to that from single-field slow-roll inflation, except with a small correction from $(1-\alpha r_{fi})$.

When multiple components coexist during inflation, isocurvature perturbations may be produced. In the case of single inflaton with a (matter-like or radiation-like) perfect fluid, the isocurvature perturbations are typically negligible. This is because any energy density perturbation in the perfect fluid is redshifted away quickly during the remaining inflation. Therefore as long as the perfect fluid goes away early enough before the end of inflation, it will not leave any relic in the energy components we see today.

We can also obtain closed form solutions for the background evolution if necessary, by evaluating $\alpha$. For this purpose, we need to calculate the integral in \refeq{mul-f-alpha0}, which converges to the attractor solution quickly as inflation proceeds because of the exponential damping. For the slowly rolling inflaton $\phi$, we can perform a series expansion of the slow-roll parameters around the initial point, i.e.
\eq{\frac{1}{V(\phi)}=\frac{1}{V(\phi_i)}+n\left.\frac{\dd}{\dd n}\frac{1}{V(\phi)}\right|_i+\frac{n^2}{2}\left.\frac{\dd^2}{\dd n^2}\frac{1}{V(\phi)}\right|_i+\ho.\label{eq-mul-f-approx}}
\Refeq{mul-f-alpha0} can then be calculated order by order, in the form
\eq{\alpha=\alpha^{(0)}+\alpha^{(1)}+\alpha^{(2)}+\ho,}
where each $\alpha^{(j)}\sim\cO\biggl(\epsilon_\phi^x\eta_\phi^y\xi_\phi^{\frac{1}{2}(j-x-y)}\biggr)$ corresponds to the slow-roll parameter expansion of the order $j$  (assuming that the first and second slow-roll parameters are small at the same (first) order, while the third slow-roll parameter is small at the second order).

Some calculations then yield $\alpha$ order by order
\eqa{\alpha^{(0)}&=&0,\\
\alpha^{(1)}&=&-\frac{2\tilde\epsilon_{\phi i}}{3(1+w)},\\
\alpha^{(2)}&=&-\frac{4\tilde\epsilon_{\phi i}(3\tilde\epsilon_{\phi i}-\tilde\eta_{\phi i})}{9(1+w)^2},\\
\alpha^{(3)}&=&-\frac{4\tilde\epsilon_{\phi i}}{27(1+w)^3}\Bigl[10\tilde\epsilon_{\phi i}\bigl(3\tilde\epsilon_{\phi i}-2\tilde\eta_{\phi i}\bigr)+2\tilde\eta_{\phi i}^2+\tilde\xi_{\phi i}\Bigr],\\
\dots.\nonumber}
We can also write them in terms of the original slow-roll parameters
\eqa{\alpha^{(0)}&=&0,\\
\alpha^{(1)}&=&-\frac{2\epsilon_{\phi i}}{3(1+w)(1-r_{fi})^2},\\
\alpha^{(2)}&=&-\frac{4\epsilon_{\phi i}}{9(1+w)^2(1-r_{fi})^3}\left(\frac{3\epsilon_{\phi i}}{1-r_{fi}}-\eta_{\phi i}\right),\\
\alpha^{(3)}&=&-\frac{4\epsilon_{\phi i}}{27(1+w)^3(1-r_{fi})^4}\left[\frac{10\epsilon_{\phi i}}{1-r_{fi}}\biggl(\frac{3\epsilon_{\phi i}}{1-r_{fi}}-2\eta_{\phi i}\biggr)+2\eta_{\phi i}^2+\xi_{\phi i}\right],\\
\dots.\nonumber}
Therefore, at the leading order, $\alpha\sim\cO(\epsilon_{\phi i})<0$.

To summarize, in this chapter we have studied multi-field inflation at perturbation level. The existence of a perfect fluid in single-field slow-roll inflation can potentially contribute to cosmological perturbations. Spectator fields arise naturally as a minimal multi-field inflation scenario, which shall be discussed in the next chapter.

%% file: Chapters/Spectator.tex
In this chapter, we first introduce \ea{the} curvaton scenario and recall some of its existing studies. We then proceed to \ea{the} spectator scenario and study its background and perturbation evolutions. In the end, we map out the parameter space of \ea{the} spectator scenario for three example potentials.
\ssecs{Curvaton scenario}
The curvaton scenario was proposed in \cite{Enqvist:2001zp,Moroi:2001ct,Lyth:2001nq,Lyth:2002my}, as an alternative scenario to produce the primordial perturbations. The inflaton $\phi$ drives inflation as in single-field slow-roll inflation, while the curvaton remains subdominant in energy density during inflation. The curvaton is either slowly rolling or frozen during inflation. Its perturbations will be kept, and only transfer to \ea{the} curvature perturbation \ea{after inflation}\ed{when it ends slow roll}. The final energy density ratio of \ea{the} curvaton or its decay products' is not constrained by the scenario. There are two typical curvaton scenarios -- a dominant curvaton and a subdominant curvaton at the time of decay -- which are demonstrated in \refig{cur-evol}. The simplest curvaton potential takes the form\footnote{
Although there are many other choices of the curvaton potential, \refeq{cur-b-U} is advantageous due to its simplicity. The constant mass $m$ allows us to predict when it ends slow roll independently of the field value $\sigma$ or its initial conditions \cite{Lyth:2001nq,Lyth:2002my}. This property also allows the curvaton to be frozen instead of slowly rolling, without changing any of the cosmological predictions. Therefore, we will only investigate the minimal potential \refeq{cur-b-U} for \ea{the} curvaton in the thesis. There is a rich literature in the study of curvaton, such as those with different potentials, kinetic terms, particle physics origins and inflationary models, or the case when both the inflaton and the curvaton contribute to curvature perturbation, or vector curvaton. The curvaton may also drive a second phase of inflation. Some of the relevant literature can be found in \cite{Bartolo:2002vf,Lyth:2001nq,Lyth:2002my,Dimopoulos:2003ii,Postma:2002et,Dimopoulos:2002hm,Dimopoulos:2003az,Lyth:2003ip,Sloth:2002xn,Gordon:2002gv,Enqvist:2003mr,Malik:2002jb,Dimopoulos:2003ss,Dimopoulos:2004uf,Dimopoulos:2002kt,Gupta:2003jc,Chun:2004gx,Lyth:2005fi,Enqvist:2005pg,Dimopoulos:2004yb,Sasaki:2006kq,Valiviita:2006mz,Dimopoulos:2005bp,Dimopoulos:2005bx,Allahverdi:2006dr,Dimopoulos:2006ms,Dimopoulos:2007zb,Assadullahi:2007uw,Dimopoulos:2007ns,BuenoSanchez:2007in,Huang:2008qf,Enqvist:2008gk,Valiviita:2008zb,Dimopoulos:2008rf,Suyama:2008nt,Huang:2008ze,Huang:2008zj,Huang:2008bg,Enqvist:2009zf,Chingangbam:2009xi,Dimopoulos:2008yv,Takahashi:2009cx,Moroi:2008nn,Kawasaki:2008mc,Takahashi:2009dr,Enqvist:2009eq,Li:2008fma,Karciauskas:2008bc,Kawasaki:2008pa,Dimopoulos:2009vu,Dimopoulos:2009am,Feng:2010tf,Chambers:2009ki,Choi:2010re,Enqvist:2009ww,Huang:2010cy,Cai:2010rt,Nakayama:2009ce,Mazumdar:2010sa,Fonseca:2011aa,Chingangbam:2010xn,Kamada:2010yz,Dimopoulos:2011gb,Dimopoulos:2012nj,Mazumdar:2011xe,Dimopoulos:2011ws,Fonseca:2012cj,Choi:2012te,Assadullahi:2012yi,Wang:2013hva,Dimopoulos:2011pe,Assadullahi:2013ey,McDonald:2013aca,Mukaida:2014wma,Byrnes:2014xua}.}
\eq{U(\sigma)\equiv\frac{1}{2}m^2\sigma^2.\label{eq-cur-b-U}}

\figla{cur-evol}{Two possible curvaton scenarios}{The energy density evolutions are shown for the inflaton (red), curvaton (blue) and the total (green), for the two possible curvaton scenarios depending on which field is dominant at the curvaton decay. The specific phases depend on the model, but in general it always starts from end of inflation and lasts until the universe reaches adiabaticity during the radiation dominated epoch. ``Rad'', ``mat'' and ``$\Lambda$'' indicate which perfect fluid the fields should be approximated as -- radiation, matter, or cosmological constant -- in the $m^2\sigma^2$ curvaton model. \ea{The phase of oscillating inflaton behaving as matter is neglected in the figures for simplicity.}}{\figi[0.48]{cur-evol1}{The inflaton or its decay products dominate the universe at the curvaton decay \cite{Enqvist:2001zp,Moroi:2001ct,Lyth:2001nq,Lyth:2002my}.}\hspace{0.04\textwidth}\figi[0.48]{cur-evol2}{The curvaton dominates the universe at the decay \cite{Enqvist:2002rf,Enqvist:2003mr,Allahverdi:2006dr}.}}
The evolution follows the four phases below:
\begin{itemize}
\item During inflation, both fields are slowly rolling, acting as the cosmological constant. The curvaton can even be \emph{frozen}, so its quantum fluctuations are stronger than its slow roll. The existence of \ea{a} curvaton $\sigma$ does not change the inflation dynamics\ed{,} \ea{other than}\ed{but} slightly boost\ea{ing}\ed{s} the Hubble rate by contributing to the total energy density. At the Hubble exit of pivot scales, the perturbation in the field $\sigma$ is preserved and does not affect universe evolution right away.
\item Inflation ends once the inflaton has ended slow roll. It \ed{then }starts to oscillate and \ea{then} decays into relativistic particles \ea{at some point}. The curvaton still remains slowly rolling after inflation because it is much lighter than the inflaton. The universe then consists of two components -- the perfect fluid of radiation coming from $\phi$'s decay products, and the cosmological constant like curvaton $\sigma$. The radiation is expected to dominate the universe for some time, despite the redshift from universe expansion. There can also be a phase of oscillation for $\phi$ right after the end of inflation.
\item As the Hubble rate quickly declines, at some point it will drop below the curvaton mass $m$, so the curvaton starts to oscillate around its vacuum expectation value. The oscillating curvaton behaves like matter, and also starts to be redshifted away, though at a lower rate than radiation. To prevent a second phase of inflation, we require the curvaton to remain subdominant at the transition point where it ends slow roll.
\item When the Hubble rate further drops below the decay rate of the curvaton, the curvaton decay process becomes efficient enough so the energy density of curvaton is assumed to be fully and instantly transferred to its decay products. The decay products are relativistic, so the radiations from inflaton and curvaton are blended together \ea{and are assumed to become indistinguishable}. This forms the adiabatic initial condition for Hot Big Bang. The curvaton's energy density ratio peaks at its time of decay, which consequently leads to the highest conversion rate from the initial curvaton perturbation to curvature perturbation $\zeta$. In the minimal scenario, this is the only source of primordial perturbations, dictating all the inhomogeneities of the universe we see today.
\end{itemize}

After inflation, the curvaton's energy density ratio \ed{consistently }increases, until it decays into radiation. The curvaton may even catch up with the inflaton during the process, after it starts to oscillate. This gives us the two possible scenarios shown in \refig{cur-evol}. If the curvaton fails to catch up with the inflaton in energy density before it decays, then it never will.

The potential \refeq{cur-b-U} has been extensively studied in the past, giving rise to the prediction of power spectrum of curvature perturbation \cite{Lyth:2001nq,Lyth:2002my}
\eq{\P_\zeta=\frac{\tilde r^2}{9\pi^2}\frac{H_*^2}{\sigma_*^2},}
where
\eq{\tilde r\equiv\frac{3\rho_{\sigma d}}{4\rho_d-\rho_{\sigma d}}}
characterizes the conversion rate from the curvaton field perturbation to curvature perturbation. Here the subscripts $*$ and $d$ correspond to the time of pivot scale Hubble exit and the time of curvaton decay respectively, while $\rho_\sigma$ and $\rho$ are the energy densities of the curvaton and the total. Since the curvaton is much lighter than the inflaton, the spectral index only depends on the first slow-roll parameter of the inflaton:
\eq{n_s\approx1-2\epsilon_*\approx1-2\epsilon_{\phi*}.\label{eq-cur-p-ns}}
This constrains the inflaton model by
\eq{\epsilon_{\phi*}\approx0.02.}

The non-Gaussianities of curvature perturbation have also been investigated for curvaton scenario, giving\cite{Lyth:2002my,Lyth:2005fi,Byrnes:2006vq,Sasaki:2006kq}
\eqa{\fNL&=&\frac{5}{4\tilde r}-\frac{5}{3}-\frac{5}{6}\tilde r,\\
\gNL&=&\frac{25}{54}\left(-\frac{9}{\tilde r}+\frac{1}{2}+10\tilde r+3\tilde r^2\right),\\
\tNL&=&\frac{36}{25}\left(\frac{5}{4\tilde r}-\frac{5}{3}-\frac{5}{6}\tilde r\right)^2.}
The curvaton scenario therefore gives two distinct non-Gaussianity predictions for the two possibilities in \refig{cur-evol}:
\begin{itemize}
\item When curvaton always subdominates energy density\ed{ ($r_{\sigma o}<r_{\sigma d}\ll1$)}, we typically acquire large non-Gaussianities ($\gg1$):
\eq{\fNL=\frac{5}{4\tilde r},\hspace{1in}\gNL=-\frac{25}{6\tilde r},\hspace{1in}\tNL=\frac{9}{4\tilde r^2}.\label{eq-cur-p-fNL}}
\item When curvaton is subdominant at the time it ends slow roll\ed{ ($r_{\sigma o}\ll1$)}, but becomes dominant at the time of decay\ed{ ($r_{\sigma d}\approx1$)}, the non-Gaussianities are small ($\sim\cO(1)$):
\eq{\fNL=-\frac{5}{4},\hspace{1in}\gNL=\frac{25}{12},\hspace{1in}\tNL=\frac{9}{4}.\label{eq-cur-p-fNL2}}
\end{itemize}
There is also a possibility in-between where the curvaton contributes significantly but not fully to the total energy density at the time of decay. The non-Gaussianity predictions are also expected to lie in between.

The primordial tensor perturbations only depend on the energy scale of inflation, giving the same power spectrum as in \refeq{ssi-Pt3}. The tensor-to-scalar ratio will be suppressed because scalar perturbations now come from the curvaton
\eq{r=\frac{144\pi\sigma_*^2}{\tilde r^2\mpl^2}.\label{eq-cur-p-rts}}

As a multi-field model, the curvaton scenario can also produce isocurvature perturbations \cite{Lyth:2003ip}. The current observation strongly constrains the existence of isocurvature perturbations. This requires $\tilde r\ll1$ or $1-\tilde r\ll1$, so the universe should be filled with either the inflaton decay products or the curvaton decay products. The only exception \ea{of the requirement} is a fundamental theory that guarantees the same branching ratios for inflaton and curvaton.

The Planck satellite gives accurate measurements on the non-Gaussianities of the CMB temperature perturbation, thus constraining models of inflation and/or the curvature perturbation generation mechanisms. The curvaton models, acting as the source of the curvature perturbation, are also constrained accordingly. Planck sees no local bi-spectrum according to the bound $\fNL=0.8\pm5.0$. This leads to the constraint $\tilde r>0.06$ at over $3\sigma$ significance (see \refeq{cur-p-fNL}), for the quadratic curvaton model that is fully responsible for curvature perturbation. When this is combined with the absence of isocurvature perturbations, the constraint becomes $\tilde r>0.98$ \cite{Ade:2015lrj}. Therefore, the curvaton must also produce the matter contents of our universe.

The large ratio $\tilde r>0.98$ may also cause other issues for the curvaton model. To avoid a second phase of inflation, the curvaton should contribute no more than half to the total energy density at the time it ends slow roll. This means the curvaton must oscillate for more than 4 e-folds before it decays. It raises the question of why its coupling constant \ea{with matter} is so small, and how to suppress possible parametric resonance during the oscillations \cite{Enqvist:2013qba}. 

In order to match the Planck observations, there have been several extensions of curvaton model in recent literature. We will not discuss them in the thesis, but instead, simply list them as below:
\begin{itemize}
\item When the curvaton potential has a self-coupling term, it may produce negative $\fNL$ by itself which cancels out the positive $\fNL$ from the minimal curvaton scenario \cite{Enqvist:2005pg}. The curvaton is thus allowed \ea{not to dominate the energy density of }\ed{to subdominate} the universe at decay. However it produces extra $\gNL$ unavoidably, which can be observed in future measurements. The coupling constant also requires extra tuning for the fine cancellation.
\item If curvaton and inflaton have the same branching ratios, their decay products become indistinguishable and therefore no isocurvature perturbation is produced. The constraint then falls back to $\tilde r>0.06$, and mild non-Gaussianities can be achieved within the observational bound. However the identical branching ratio should come out automatically from the theoretical construction of inflaton and curvaton models, but not as a mere assumption without any justification.
\item The curvaton can also contribute only partially to curvature perturbation, while the rest comes from other fields, such as the inflaton. In such cases, the inflaton perturbations can dilute the characteristics of the curvature perturbations from curvaton, depending on the contribution ratio to curvature perturbation. This can suppress the local $\fNL$ from the curvaton, while the other observables, such as $n_s$ and $\tNL$ are also subject to change \cite{Huang:2008rj,Byrnes:2014xua}.
\end{itemize}

\ssecsx{Spectator scenario}{spe}
Above, we have demonstrated the curvaton scenario, in which a light field seeds the fluctuations and then decays after inflation has come to an end. The minimal curvaton model however \ea{is strongly constrained by}\ed{struggles to find accordance with} the recent Planck data, due to the non-Gaussianity and isocurvature constraints.

On the other hand, if a field exits slow roll or decays much before the end of inflation, it will never influence the thermal history of the universe and will become a \emph{spectator} field. Its decay products will be redshifted away during inflation and the inflaton will be \emph{solely} responsible for creating all the matter. Meanwhile such a spectator field could still be responsible for seeding the CMB anisotropy, provided the relevant scales for the CMB leave the Hubble patch before the spectator field ends slow roll. In this respect, spectator scenarios do not produce any isocurvature perturbations in the universe.

As explained in \refssec{Spectator fields}, we will stick to the separable potential \refeq{mul-spector} where the inflaton and spectator fields do not couple. The spectator \ea{energy density is always subdominant}\ed{always subdominates the total energy density}, giving $V(\phi)\gg U(\sigma)$ at any time. If the spectator field ends slow roll well before the end of inflation but after the Hubble exit of the pivot scales, it gives rise to two consequent phases of inflation as shown in \refig{spe-evol}:

\fig{spe-evol}{Schematic timeline for spectator scenario}{A schematic timeline for spectator scenario is shown above. The solid lines are the background evolutions of the energy densities of inflaton, spectator and the total contribution. The green and blue dashed lines represent the evolution of spectator and total energy densities of the universes with perturbed $\sigma$. The two phases of spectator evolution are separated by the phase boundary when $\sigma$ breaks the slow-roll condition.}

\begin{enumerate}
\item {\bf Phase I}: The inflaton $\phi$ leads inflation. Both $\phi$ and $\sigma$ are slowly rolling. This phase ends as the slow-roll condition for $\sigma$ breaks down. We assume that the relevant perturbations for the CMB leave the Hubble patch in this phase\footnote{
The observed pivot scale actually has a window of several e-folds. Here we consider every mode separately so the window is not shown in \refig{spe-evol}.}.
\item {\bf Phase II}: When $\sigma$ ends slow roll, the inflaton $\phi$ still dominates inflation under slow-roll conditions. Then $\sigma$ either oscillates around the minimum of it's potential, or decays instantly. In either case, $\sigma$ or its decay products are being redshifted away swiftly during this phase, and can be regarded as a perfect fluid with a constant equation of state $w$. For this reason, several e-folds after the beginning of this phase, the dynamics reduce to that of single-field slow-roll inflation of $\phi$.
\end{enumerate}

We will use the subscripts ``$*$'' for the Hubble exit of the mode of our concern, ``$b$'' for the boundary between the two inflationary phases, and ``$e$'' for the end of inflation. In Phase I, slow-roll approximations are satisfied for both fields, giving rise to the equations of motion
\eqa{\frac{\dd\phi}{\dd N}&=&\frac{\mpl^2V'}{8\pi(U+V)},\label{eq-spe-dphi0}\\
\frac{\dd\sigma}{\dd N}&=&\frac{\mpl^2U'}{8\pi(U+V)},\label{eq-spe-dsigma0}}
where we have used the number of the remaining e-folds $N$ as proper time. The potential energy densities $V(\phi)$ and $U(\sigma)$ are defined according to the separable potentials in \refeq{mul-spector}.

\Refeq{spe-dphi0} and \refeq{spe-dsigma0} suggest a simple relation
\eq{\frac{\dd\phi}{V'}=\frac{\dd\sigma}{U'},\label{eq-spe-p1eqd}}
whose integrated form is
\eq{\int_\phi^{\phi_*}\frac{\dd\phi}{V'}=\int_\sigma^{\sigma_*}\frac{\dd\sigma}{U'},\label{eq-spe-p1eq}}
which holds up to the boundary point $b$.

Phase I ends when the slow-roll condition of the spectator breaks down. This can be either first or second slow-roll condition, but here we will only choose the violation of second slow-roll condition. This is because the violation of first slow-roll condition typically leads to negligible curvature perturbation from the spectator field, which will be justified in \refssec{spe-Perturbations}. In order to violate second slow-roll condition before first-order, we can use a plateau potential, such as the one shown in \refig{ssi-f-pot}. For plateau potentials, both slow-roll conditions are well satisfied before the spectator reaches the plateau boundary, but the second slow-roll condition breaks down upon reaching the plateau boundary. The boundary condition of the end of slow roll can then be written as
\eq{\eta_{\sigma b}\equiv\frac{\mpl^2U_b''}{8\pi(U_b+V_b)}=-1.\label{eq-spe-bound}}

After the spectator ends slow roll, it then acts as a perfect fluid with a constant equation of state $w$. The universe enters single-field slow-roll inflation with a perfect fluid, which has been solved in \refssec{Single inflaton with a perfect fluid}. The energy density of spectator (or its decay products) is diluted as the universe expands, according to $\rho_\sigma\propto a^{-3(1+w)}$. If the spectator decays into radiation early on, it will remain subdominant in energy density, and never take any role in the following evolution. However, in the worst case where the spectator \emph{never} decays, it oscillates around its vacuum with $w=0$ while the inflaton decays into radiation immediately after inflation. To prevent the spectator from coming into our sight, it has to end slow roll before the last $\sim20$ e-folds of inflation. We will not consider the worst case, but instead assume that spectator decays reasonably early, so it will not leave any imprint on the energy density as long as it ends slow roll before the end of inflation. This guarantees negligible isocurvature perturbations from the spectator.

\ssecsx{Perturbations}{spe}
The cosmological perturbations arising from the spectator field can be calculated using $\delta N$ formalism. Similar with the calculations in \ea{the} curvaton scenario, we first neglect the perturbations from the inflaton, and only consider the initial spectator perturbation at the Hubble exit, $\delta\sigma_*$. All other perturbations are induced by this initial perturbation.

During \textbf{Phase I}, $\delta\sigma_*$ perturbs \refeq{spe-p1eq} at the boundary by the amount 
\eq{-\frac{\delta\phi_b}{V_b'}=\frac{\delta\sigma_*}{U_*'}-\frac{\delta\sigma_b}{U_b'}.}
The field perturbations at the boundary, $\delta\phi_b$ and $\delta\sigma_b$, should still satisfy the boundary condition \refeq{spe-bound}, giving
\eq{8\pi V_b'\delta\phi_b+(\mpl^2U_b'''+8\pi U_b')\delta\sigma_b=0.}

Combining the above equations gives us the induced field perturbations at the boundary
\eqa{\delta\phi_b&=&-\frac{V_b'}{U_*'}(1-\theta)\delta\sigma_*,\\
\delta\sigma_b&=&\frac{U_b'}{U_*'}\theta\delta\sigma_*,}
where the variable $\theta$ is defined as 
\eq{\theta\equiv\frac{\epsilon_{\phi b}}{\epsilon_{\phi b}+\epsilon_{\sigma b}+\xi_{\sigma b}}\approx\frac{\epsilon_{\phi b}}{\xi_{\sigma b}}\ll1.}
The third slow-roll parameter for $\sigma$ at the boundary, $\xi_{\sigma b}$, is \ea{assumed}\ed{allowed} to be relatively large ($\gtrsim\cO(1)$) for the plateau potentials we consider here. The approximation holds as long as the corresponding slow-roll parameters, $\epsilon_\phi$ and $\epsilon_\sigma$ are both much smaller than unity.

In order to calculate the perturbations in the e-folding, we work in the uniform $\phi$ slicing. \Refeq{spe-p1eq} then tells us how the field perturbation $\delta\sigma$ evolves after the Hubble exit
\eq{\frac{\delta\sigma}{U'}=\frac{\delta\sigma_*}{U_*'}.\label{eq-spe-p1eqp}}
The number of e-folds of Phase I can be written as an integrated form
\eq{N_*-N_b\equiv\int_{N_b}^{N_*}\dd N=\int_{\phi_b}^{\phi_*}\frac{8\pi(U+V)}{\mpl^2V'}\dd\phi.}
The initial perturbation $\delta\sigma_*$ alters it by the amount
\eq{\delta N_*-\delta N_b=\frac{8\pi}{\mpl^2}\left(-\frac{U_b+V_b}{V_b'}\delta\phi_b+\int_{\phi_b}^{\phi_*}\frac{U'}{V'}\delta\sigma\dd\phi\right),}
in which (according to \refeq{spe-p1eqd} and \refeq{spe-p1eqp}) the integral can be simplified to
\eq{\int_{\phi_b}^{\phi_*}\frac{U'}{V'}\delta\sigma\dd\phi=\frac{\delta\sigma_*}{U_*'}\int_{\phi_b}^{\phi_*}\frac{U'{}^2}{V'}\dd\phi=\frac{U_*-U_b}{U_*'}\delta\sigma_*.}
Consequently, we get the final expression for the perturbed number of e-folds in Phase I as
\eq{\delta N_*-\delta N_b=\frac{8\pi\bigl[U_*-U_b+(1-\theta)(U_b+V_b)\bigr]}{\mpl^2U_*'}\delta\sigma_*.\label{eq-spe-p1}}

After the spectator $\sigma$ ends slow roll, the universe enters the second phase of inflation. Single-field slow-roll inflation with a perfect fluid has been solved in \refssec{Single inflaton with a perfect fluid}, which gives the perturbation in the e-folding for Phase II, (according to \refeq{mul-f-dN},) as
\eq{\delta N_b-\delta N_e=\frac{1}{1-\alpha r_{\sigma b}}\left[\frac{8\pi V_b}{\mpl^2V_b'}\delta\phi_b+\frac{(1-\alpha)U_b'}{3(1+w)(U_b+V_b)}\delta\sigma_b\right],\label{eq-spe-p2}}
where
\eq{r_{\sigma b}\equiv\frac{U_b}{U_b+V_b}}
is the energy density ratio of spectator at \ea{the} phase boundary\ea{, and $\alpha$ is defined in \refeq{mul-f-alpha0}}.

Since the energy density of spectator field is redshifted away in Phase II, the universe becomes adiabatic before the end of inflation, without producing any curvature perturbation after inflation:
\eq{\delta N_e=0.\label{eq-spe-p3}}
Combining \refeq{spe-p1}, \refeq{spe-p2} and \refeq{spe-p3} gives the total curvature perturbation $\zeta$ from the initial field perturbation $\delta\sigma_*$
\eq{\zeta=\delta N_*=N_\sigma\delta\sigma_*,}
where
\eq{N_\sigma=\frac{1-\alpha}{1-\alpha r_{\sigma b}}\left\{\frac{8\pi U_*}{\mpl^2 U_*'}+\theta\Bigl[\frac{U_b'{}^2}{3(1+w)(U_b+V_b)U_*'}-\frac{8\pi U_b}{\mpl^2U_*'}\Bigr]\right\}+\frac{\alpha(1-r_{\sigma b})}{1-\alpha r_{\sigma b}}\frac{8\pi(U_*-U_b)}{\mpl^2U_*'}.\label{eq-spe-p-Ns}}

During inflation, most of the slow-roll parameters are much smaller than unity, including $\epsilon_\phi,\epsilon_\sigma,|\eta_\phi|,|\eta_{\sigma *}|,\xi_\phi\ll1$, but with the exceptions of $\eta_{\sigma b}=-1$ and $\xi_\sigma$. Therefore we can perform \ea{a} series expansion w.r.t the slow-roll parameters together with $r_\sigma\ll1$.\footnote{
The slow-roll parameter $\xi_\phi$ does not have to be much smaller than 1, but in many cases it is no larger than $\cO(\epsilon_\phi^2)$. Therefore here we also take it as a small quantity.}
In addition, when the potential $U(\sigma)$ is sharp enough at the phase boundary, i.e.\ $\theta\ll1$ and $r_{\sigma b}\xi_{\sigma b}\gg\epsilon_{\phi b}\epsilon_{\sigma b}$, \refeq{spe-p-Ns} becomes dominated by the very first term. The leading order of \refeq{spe-p-Ns} is then simplified to
\eq{N_\sigma\approx\frac{8\pi U_*}{\mpl^2U_*'}.\label{eq-spe-p-Ns1}}
When the inflaton dominates \ea{the} curvature perturbation, we have
\eq{N_\phi\approx\frac{8\pi V_*}{\mpl^2V_*'}.\label{eq-spe-p-Np1}}
Therefore, by comparing the above two equations, we find that for spectator to dominate \ea{the} curvature perturbation, it simply demands a relatively flatter potential
\eq{\frac{U_*'}{U_*}\ll\frac{V_*'}{V_*}.}

For a simple spectator potential such as the power law potential $U(\sigma)\propto\sigma^p$, it is hardly possible for the spectator to fulfill the three conditions simultaneously -- dominating curvature perturbation, remaining subdominant in energy density, and exiting slow roll before the inflaton does. They can be satisfied altogether, on the other hand, if we use a plateau potential for $U(\sigma)$. For plateau potentials, slow roll is terminated when the spectator reaches the (sharp enough) plateau edge, justifying our previous choice of the plateau potential for \refeq{spe-bound}.

The above calculations would then give the power spectrum of curvature perturbation
\eq{\P_\zeta=\frac{16H_*^2U_*^2}{\mpl^4U_*'{}^2}=\frac{H_*^2r_{\sigma *}^2}{\pi\mpl^2\epsilon_{\sigma *}}.\label{eq-spe-Pz}}
The spectral tilt is given by
\eq{n_s-1\equiv-\frac{\partial\ln\P_\zeta}{\partial N}=-2\epsilon_{\phi*}+2\eta_{\sigma*}-\frac{4\epsilon_{\sigma*}}{r_{\sigma*}},\label{eq-spe-ns0}}
where we have neglected the term $-2\epsilon_{\sigma *}$.

Note from \refeq{spe-Pz}, we can obtain $\epsilon_{\sigma*}$ from the observed power spectrum of the curvature perturbation
\eq{\epsilon_{\sigma*}=\frac{H_*^2r_{\sigma*}^2}{\pi\mpl^2\P_\zeta}.\label{eq-spe-epssx}}
By plugging it back into \refeq{spe-ns0}, we see that as long as the inflation energy scale $H_*\ll10^{-5}\mpl$, the last term in \refeq{spe-ns0} is negligible compared to the observed spectral tilt, $n_s-1\approx-0.033$ \cite{Ade:2015lrj}. So we are left with
\eq{n_s-1=-2\epsilon_{\phi*}+2\eta_{\sigma*}.\label{eq-spe-ns}}
With the help of \refeq{spe-ns0}, the running of spectral tilt can be shown as
\eq{\frac{\dd n_s}{\dd\ln k}=-\frac{\dd n_s}{\dd N}=-\frac{1}{2}(n_s-1)^2-2\Bigl[3\epsilon_*^2-2(\epsilon_{\phi*}\eta_{\phi*}+\epsilon_{\sigma*}\eta_{\sigma*})-\eta_{\sigma*}^2+\xi_{\sigma*}\Bigr].\label{eq-spe-nsr}}

The strength of local bi-spectrum $\fNL$ can be derived from taking the derivative $\partial/\partial\sigma_*$ on \refeq{spe-p-Ns}. The leading terms are\footnote{
\ea{In this thesis some expressions appear different from those in \cite{Wang:2013oea}. However, they are identical because here we use energy density ratios at different times ($r_{\sigma b}$ and $r_{\sigma*}$) to simplify the results.}}
\eqa{\fNL&\equiv&\frac{5}{6}\frac{N_{\sigma\sigma}}{N_\sigma^2}=\frac{10s_\sigma^2\epsilon_{\phi b}}{3r_{\sigma b}}\left[\frac{2\epsilon_{\phi b}-\eta_{\phi b}}{3(1+w)}-\frac{\epsilon_{\phi b}}{\xi_{\sigma b}^2}\biggl(1-\frac{\lambda_{\sigma b}}{\xi_{\sigma b}}\biggr)+\frac{\eta_{\phi b}}{\xi_{\sigma b}}\right]\nonumber\\
&&+\frac{20s_\sigma^2\epsilon_{\sigma b}\epsilon_{\phi b}}{9(1+w)r_{\sigma b}^2\xi_{\sigma b}}\left(2-\frac{4+3w}{\xi_{\sigma b}}-\frac{\lambda_{\sigma b}}{\xi_{\sigma b}^2}\right)+\frac{5\epsilon_{\sigma *}}{3r_{\sigma *}^2}-\frac{5\eta_{\sigma *}}{6r_{\sigma *}}\nonumber\\
&&+\ho,\label{eq-spe-p-Npp}}
where the energy density ratio of spectator between the boundary and the Hubble exit is defined as
\eq{s_\sigma\equiv\frac{U_b}{U_*}<1,}
and the fourth slow-roll parameter is
\eq{\lambda_\sigma\equiv\frac{\mpl^6U'^2U''''}{[8\pi(U+V)]^3}.}

The third order derivative $N\subs{\sigma\sigma\sigma}=\partial^3N_\sigma/\partial\sigma_*^3$ can be calculated in the same way and also $\gNL$. According to \refeq{mul-g-gnl}, we obtain the leading order tri-spectrum of curvature perturbation as
\eqa{\gNL&=&\frac{25}{54}\Biggl\{\frac{2\eta_{\sigma*}}{r_{\sigma*}^2}\biggl(\eta_{\sigma*}-\frac{\epsilon_{\sigma*}}{r_{\sigma*}}\biggr)-\frac{\xi_{\sigma*}}{r_{\sigma*}^2}+\frac{4s_\sigma^3\epsilon_{\phi b}\xi_{\phi b}}{r_{\sigma b}^2}\biggl[\frac{1}{3(1+w)}-\frac{1}{\xi_{\sigma b}}\biggr]\nonumber\\
&&+\frac{8s_\sigma^3\epsilon_{\phi b}\epsilon_{\sigma b}}{3(1+w)r_{\sigma b}^3\xi_{\sigma b}}A\Biggr\},\label{eq-spe-p-Nppp}}
where
\eqa{A&\equiv&\xi_{\phi b}+\eta_{\phi b}\biggl(2\eta_{\phi b}+\frac{3\eta_{\sigma*}}{s}\biggr)-3\epsilon_{\phi b}\biggl(2\eta_{\phi b}+\frac{\eta_{\sigma*}}{s}\biggr)\biggl(2-\frac{4+3w}{\xi_{\sigma b}}-\frac{\lambda_{\sigma b}}{\xi_{\sigma b}^2}\biggr)\nonumber\\
&&+2\epsilon_{\phi b}^2\biggl[6-\frac{2}{\xi_{\sigma b}}-\frac{3(1+w+\lambda_{\sigma b})}{\xi_{\sigma b}^2}+\frac{3(4+3w)\lambda_{\sigma b}-\chi_{\sigma b}}{\xi_{\sigma b}^3}+\frac{3\lambda_{\sigma b}^2}{\xi_{\sigma b}^4}\biggr].}
Here we have also defined another energy density ratio
\eq{s\equiv\frac{U_b+V_b}{U_*+V_*}<1,}
and the fifth slow-roll parameter for $\sigma$
\eq{\chi_\sigma\equiv\frac{\mpl^8U'{}^3U'''''}{[8\pi(U+V)]^4}.}
In the above simplest case, where inflaton perturbation is negligible, curvature perturbation comes solely from the spectator field and $\tNL$ fulfills the simple relation
\eq{\tNL=\frac{36}{25}\fNL^2.}

When the spectator potential is extremely flat on the plateau ($\epsilon_\sigma,\eta_{\sigma*}\rightarrow0$ and $s_\sigma\rightarrow1$) and the plateau edge is extremely sharp ($\xi_{\sigma b}\rightarrow\infty$), the non-Gaussianity predictions are greatly simplified to
\eqa{\fNL&=&\frac{10\epsilon_{\phi b}(2\epsilon_{\phi b}-\eta_{\phi b})}{9(1+w)r_{\sigma b}},\label{eq-spe-p-fnls}\\
\gNL&=&\frac{50\epsilon_{\phi b}\xi_{\phi b}}{81(1+w)r_{\sigma b}^2}-\frac{25\xi_{\sigma*}}{54r_{\sigma*}^2}.\label{eq-spe-p-gnls}}
Therefore the local $\fNL$ \ea{in the}\ed{of} spectator scenario is similar to that of curvaton scenario, because they both are inversely proportional to energy density ratio. However, spectator scenario produces a weaker $\fNL$, because it is also suppressed by the slow-roll parameters of inflaton. The local $\fNL$ in spectator scenario can be large ($\gg O(1)$) or small ($\lesssim O(1)$), depending on which is even smaller, the energy density ratio or the slow-roll parameter suppression.

In the above calculations, we have assumed that classical slow roll dominates over the quantum fluctuations of $\sigma$ for the relevant scales. This requires the classical displacement of $\sigma$ to be larger than the quantum fluctuations per Hubble time, i.e.
\eq{\P_{\delta\sigma*}<\left(\frac{\dd\sigma}{\dd N}\right)^2.} 
Multiplying both sides with $N_\sigma^2$ \ea{and using \refeq{spe-Pz}}, we can convert it to a model independent lower bound on $r_{\sigma*}$
\eq{r_{\sigma*}^2>\P_\zeta.\label{eq-spe-p-rmax}}
According to the Planck observation \cite{Ade:2015lrj}, which gives $\P_\zeta\approx2.5\times10^{-9}$, we obtain a lower bound $r_{\sigma*}>5\times10^{-5}$

On the other hand, when the spectator potential is very flat, such as when the spectator is frozen during inflation, the spectator may fail to end slow roll during inflation. On such occasions, as the Hubble rate decreases after inflation, the spectator will however accelerate exponentially. If the spectator reaches the edge of plateau before taking over the inflaton (or its decay products) in energy density, the model becomes effectively the curvaton scenario. Otherwise, we will encounter a second phase of inflation, which is beyond the scope of this thesis.

In order to achieve an ideal spectator scenario, we will need a plateau that is sufficiently flat to dominate curvature perturbation, but not too flat so the spectator still ends slow roll during inflation. We also require the potential to be smooth enough ($U''$ small enough) on the plateau for the near scale invariant spectrum. There are no \ea{shortage}\ed{dearth} of such fields. Their origin could come from anywhere -- the visible sector or a hidden sector. Such a spectator does not even have to couple to the Standard Model degrees of freedom because it does not leave any trace in the current energy density or isocurvature perturbations. All the onus will be now on the inflaton's coupling to the Standard Model degrees of freedom for creating the right thermal history of the universe.

\ssecs{Spectator models}
\sssecs{Step function spectator}
In certain circumstances, we may encounter a spectator field $\sigma$ with the potential $u(\sigma)$, which slowly rolls at $\sigma>\sigma_0$ and decays instantly at $\sigma=\sigma_0$. Such scenarios can be modelled equivalently with a potential which has a step function multiplier, namely $U(\sigma)$ which is defined as
\eq{U(\sigma)=\cH(\sigma-\sigma_0)u(\sigma)=\left\{\begin{array}{l@{\hspace{0.3in}}l}u(\sigma),&\mathrm{for\ }\sigma>\sigma_0,\\0,&\mathrm{else},\end{array}\right.}
where we assume $u(\sigma)$ is flat and smooth enough to accommodate slow roll for $\sigma>\sigma_0$. \ea{The Heaviside step function is defined as
\eq{\cH(x)\equiv\left\{\begin{array}{l@{\hspace{0.3in}}l}1,&\mathrm{for\ }x>0,\\0,&\mathrm{else},\end{array}\right.}
which}\ed{The step function} can be regarded as a limiting case of the hyperbolic tangent function
\eq{\cH(x)=\lim_{k\rightarrow+\infty}\frac{1+\tanh kx}{2}.}
Such step function potentials can also model composite scenarios, such as using the symmetry breaking mechanism of hybrid inflation as spectators.

We will keep $k$ as a large number here, and only take the $k\rightarrow+\infty$ limit in the final step. The spectator $\sigma$ then leaves slow roll at $\sigma_b>\sigma_0$ when the second slow-roll condition is violated, or $\lim_{k\rightarrow\infty}\sigma_b=\sigma_0^+$. Due to the sharp transition at the plateau edge $\sigma_b$, its whole energy density is instantly transferred to its decay products, which can be regarded as a perfect fluid with a constant equation of state $w$. No energy density is lost in the process, requiring $\tanh k(\sigma_b-\sigma_0)\rightarrow1$, or $k(\sigma_b-\sigma_0)\gg1$ for the sharp transition. This simplifies the derivatives of the step function as
\eq{\frac{\partial^n\tanh k(\sigma-\sigma_0)}{\partial\sigma^n}\approx(-2)^{n+1}k^ne^{-2k(\sigma-\sigma_0)},\hspace{0.3in}\mathrm{for\ }n=1,2,\dots\mathrm{\ under\ }k(\sigma-\sigma_0)\gg1.}

It then yields the second slow-roll parameter at the phase boundary\ea{, for $k(\sigma_b-\sigma_0)\gg1$,}
\eq{\eta_{\sigma b}\equiv\frac{\mpl^2u_b''}{8\pi(u_b+V_b)}=-\frac{k^2\mpl^2u_b}{2\pi(u_b+V_b)}e^{-2k(\sigma_b-\sigma_0)}=-1,}
from which we solve that the spectator ends slow roll at
\eq{\sigma_b-\sigma_0=\frac{1}{2k}\ln\frac{k^2\mpl^2u_b}{2\pi(u_b+V_b)}.}
Based on this, we calculate the rest of the slow-roll parameters
\eqa{\epsilon_{\sigma b}&=&\frac{\mpl^2u_b'{}^2}{16\pi(u_b+V_b)^2},\\
\xi_{\sigma b}&=&2k\mpl\sqrt\frac{\epsilon_{\sigma b}}{\pi},\\
\lambda_{\sigma b}&=&-2k^2\mpl^2\frac{\epsilon_{\sigma b}}{\pi},\\
\chi_{\sigma b}&=&2k^3\mpl^3\left(\frac{\epsilon_{\sigma b}}{\pi}\right)^\frac{3}{2}.}

For a step function potential whose edge is infinitely sharp (i.e. $k\rightarrow\infty$), the local bi-spectrum (\refeq{spe-p-Npp}) and tri-spectrum (\refeq{spe-p-Nppp}) are simplified to
\eqa{\fNL&=&\frac{10s_\sigma^2\epsilon_{\phi b}(2\epsilon_{\phi b}-\eta_{\phi b})}{9(1+w)r_{\sigma b}}+\frac{5\epsilon_{\sigma *}}{3r_{\sigma *}^2}-\frac{5\eta_{\sigma *}}{6r_{\sigma *}},\label{eq-spe-s-fnl}\\
\gNL&=&\frac{25}{54}\left[\frac{2\eta_{\sigma*}}{r_{\sigma*}^2}\biggl(\eta_{\sigma*}-\frac{\epsilon_{\sigma*}}{r_{\sigma*}}\biggr)-\frac{\xi_{\sigma*}}{r_{\sigma*}^2}+\frac{4s_\sigma^3\epsilon_{\phi b}\xi_{\phi b}}{3(1+w)r_{\sigma b}^2}\right].\label{eq-spe-s-gnl}}
If we further assume the potential $u(\sigma)$ is very flat and smooth, (i.e.\ $\epsilon_\sigma\ll1$, $\eta_{\sigma*}\ll1$, and $s_\sigma\approx1$,) they will reduce to \refeq{spe-p-fnls} and \refeq{spe-p-gnls}.

As a naive example, we can consider both fields to have quadratic potentials. The potential $U(\sigma)$ also has a step function for a sharp transition during inflation. In this case, the potentials are written as
\eq{V(\phi)=\hf m_\phi^2\phi^2,\hspace{0.5in}U(\sigma)=\hf m_\sigma^2\sigma^2\,\cH(\sigma-\sigma_0).}

Inflation is driven by $V(\phi)$. In the beginning, when the relevant perturbations leave the Hubble patch, we have $\sigma_*>\sigma_0$, so it stays on the plateau and rolls down slowly. When the spectator field reaches $\sigma_b=\sigma_0$, it decay instantly into radiation which can be modelled by the the sudden change in the potential arising from the step function, which terminates slow roll. The radiation is quickly diluted away by the remaining inflation. Since $\sigma$ rolls very slowly during Phase I, we typically expect $\sigma_*-\sigma_0\ll\sigma_0$. This means the effective potential for $\sigma>\sigma_0$ \ea{can be approximated by a linear potential}\ed{is actually}
\eq{U(\sigma)=\hf m_\sigma^2\sigma_0^2+m_\sigma^2\sigma_0(\sigma-\sigma_0).}

For this model, we have a total of 4 free parameters -- $m_\phi$, $m_\sigma$, $\sigma_0$, and $N_b$ which is the number of e-folds of universe expansion from the phase boundary to the end of inflation (after setting $N_e=0$). The overall energy scale only affects the power spectrum of \ea{the} curvature perturbation by
\eq{\P_\zeta=\frac{8(2N_*+1)m_\phi^2\sigma_0^2}{3\mpl^4}.}
Therefore we can fix $\P_\zeta$ to the observed value and hence reduce the number of free parameters to 3. We want $\sigma$ to dominate curvature perturbation, which requires $\sigma_0$ to be large enough ($\sigma_0\gg\phi_*$). However as long as this condition is satisfied, the value of $\sigma_0$ hardly affects the cosmological predictions of the model. We are left with only two free parameters, which are the mass ratio $m_\sigma/m_\phi$ and $N_b$. For this model, the energy density ratio at the phase boundary is given by
\eq{r_{\sigma b}=\frac{m_\sigma^2\sigma_0^2}{m_\phi^2\phi_b^2}.}
So we will use $r_{\sigma b}$ and $N_b$ for coordinates of the two-dimensional parameter space, replacing $m_\sigma/m_\phi$.

After transforming the parameter space, the free parameters reduce from $m_\phi,m_\sigma,\sigma_0,N_b$ to $r_{\sigma b}$ and $N_b$. The other two degrees of freedom are absorbed in $\P_\zeta$ which is fixed by observation, and $\sigma_0$ which does not change the cosmological predictions. Based on the background solution for single-field slow-roll inflation for $\phi$ (as \refeq{ssi-phiN1}):
\eq{\phi=\frac{\mpl}{2\sqrt\pi}\sqrt{2N+1},}
the transformation in the parameter space has the following relations
\eqa{m_\phi^2&=&\frac{3\mpl^4\P_\zeta}{8(2N_*+1)\sigma_0^2},\label{eq-spe-s-mphi}\\
m_\sigma^2&=&\frac{3(2N_b+1)\mpl^6\P_\zeta r_{\sigma b}}{32(2N_*+1)\pi\sigma_0^4}.}
The local bi-spectrum in \refeq{spe-s-fnl} is now simplified to
\eq{\fNL=\frac{5}{6(2N_b+1)^2r_{\sigma b}}.}

The two dimensional parameter space $(r_{\sigma b},N_b)$ is constrained by the following conditions
\begin{enumerate}
\item The Hubble exit of pivot scale, the phase boundary, and the end of inflation are all well separated. So typically we choose $3\le N_b\le N_*-3$.
\item The spectator field should remain subdominant in energy density, so $r_{\sigma b}\ll1$.
\item The inflaton should provide suppressed curvature perturbation compared to that from spectator, which means \ea{$r_{\sigma b}^2\gg s_\sigma^2\epsilon_{\sigma*}/(s^2\epsilon_{\phi*})$. For the specific model, it corresponds to} $\sigma_0\gg\phi_*$ \ea{and does not constrain the free parameters}.
\item The first slow-roll parameter is smaller than unity when $\sigma$ stays on the flat potential, i.e. $\epsilon_{\sigma *}<1$.
\item The quantum fluctuations of $\sigma$ should not dominate over its classical slow roll. This means \refeq{spe-p-rmax} is valid. \ea{Here, it means}
\eq{r_{\sigma b}>\sqrt{\frac{2N_*+1}{2N_b+1}P_\zeta}.}
\end{enumerate}

\fig{spe-Step}{The local bi-spectrum of the spectator field with a step potential}{The local bi-spectrum $\fNL$ is shown for a step potential spectator field with a quadratic inflaton, as discussed in \refsssec{Step function spectator}. The yellow shaded region is excluded from parameter space by the five criteria discussed in \refsssec{Step function spectator}. Darker regions indicate a higher $\fNL$. The red dashed contours are for $\fNL=1,5,10,20$ from top right to bottom left.}
Under the above conditions, we can calculate the spectral index $n_s$, its running $\dd n_s/\dd\ln k$, the local bi-spectrum $\fNL$, and tri-spectrum $\gNL$. A specific example of $\fNL$ is shown in \refig{spe-Step}, for the parameters $N_*=50$, $\sigma_0=10\mpl$ and $w=1/3$. The regions violating any of the above five conditions are excluded and shown by the yellow shaded region in \refig{spe-Step}. We can read out from \refeq{spe-s-mphi} the mass for $\phi$ here is $m_\phi\approx3.0\times10^{-7}\mpl$, so the curvature perturbation from $\phi$ is indeed negligible. In addition, for these parameters we have $n_s=0.98$, $\dd n_s/\dd\ln k=-3\times10^{-4}$ and $\gNL\ll1$, all of which hardly depend on the choice of $N_b$ or $r_{\sigma b}$. They all fall within the observational bounds \cite{Ade:2015lrj,Ade:2015ava}.

From \refig{spe-Step}, we see that the parameter space is limited. In particular, $r_{\sigma b}$ is constrained on both sides because we need $\sigma$ to be subdominant and its quantum fluctuations not to overcome slow-roll motion. Moreover, $\sigma$ hardly contributes to $n_s$, $\fNL$ or $\gNL$ because its slow-roll parameters are tiny. With the inflaton $\phi$ being the only contribution to spectral index, we get $n_s\approx0.98$, a small running, a small local bi-spectrum $\fNL$, and a smal tri-spectrum $\gNL$. For these parameters, we can see the local bi-spectrum strength $\fNL$ has a maximum value around 20, which is capped because the classical slow roll has to dominate over quantum fluctuations. It also agrees well with the Planck observations, which limit $\fNL<16$ at $\approx3\sigma$ \cite{Ade:2015ava}. In this case, the major contribution to $\fNL$ comes from the conversion from non-adiabatic perturbations to curvature perturbation, which becomes non-Gaussian after the spectator ends slow roll and decays into perfect fluid, even though this non-Gaussian conversion only lasts for one e-fold or so.

\sssecs{Inflection point spectator}
Flat directions naturally arise in string theory and supersymmetric theories \cite{Mazumdar:2010sa}. These flat directions can also be candidates for the spectator potential. In most cases, such flat directions can be written locally as an effective scalar potential in the form~\cite{Allahverdi:2006we,Allahverdi:2006iq}
\eq{U(\Delta\sigma)=U_0\left(1+\gamma_1\frac{\Delta\sigma}{\mpl}+\frac{\gamma_3}{6}\frac{\Delta\sigma^3}{\mpl^3}\right)+\ho,\label{eq-spe-i-U}}
where
\eq{\Delta\sigma\equiv\sigma-\sigma_0}
is the displacement of the scalar spectator $\sigma$ from its inflection/saddle point $\sigma_0$.

Therefore at $\Delta\sigma=0$, i.e.\ $\sigma=\sigma_0$, we will get an inflection/saddle point where $U=U_0$ and $U''=0$. For inflection and saddle points, we will have $\gamma_1>0$ and $\gamma_1=0$ respectively, and we always have $\gamma_3>0$. In general, the higher order terms in the effective potential \refeq{spe-i-U}, e.g.\ $\Delta\sigma^4$, also provide a small contribution to the potential or its derivatives. Here we assume their contributions vanish for the sake of simplicity.

In this respect the motion of $\sigma$ can be solved as follows. We first obtain the field displacement at the phase boundary $\Delta\sigma_b$, from the breakdown of the second slow-roll condition $\eta_{\sigma b}=-1$
\eq{\Delta\sigma_b=-\frac{8\pi\mpl(U_b+V_b)}{\gamma_3 U_0}.\label{eq-spe-i-eta}}
With this we can introduce a very helpful parameter $\gamma_0$, which tells us how ``flat'' the potential is at the inflection point
\eq{\gamma_0\equiv\sqrt{\frac{\gamma_1}{\mpl}\frac{2\mpl^3}{\gamma_3\Delta\sigma_b^2}}=\sqrt\frac{\gamma_1\gamma_3}{2}\,\frac{U_0}{4\pi(U_b+V_b)}.}
Therefore the ratio of $U'(\sigma)$ between the inflection point and the phase boundary is $\gamma_0/(1+\gamma_0)$. For the inflection point potential we are interested in here, typically $\gamma_0\ll1$.

As long as we specify the inflaton potential and $N_b$, we are able to solve the equation of motion for $\Delta\sigma$. The slow-roll approximation in Phase I gives the l.h.s of \refeq{spe-p1eq} as
\eqa{\int_{\Delta\sigma_b}^{\Delta\sigma}\frac{\dd\Delta\sigma}{U'}&=&\frac{\mpl^2}{U_0}\sqrt\frac{2}{\gamma_1\gamma_3}\left.\arctan\sqrt\frac{\gamma_3}{2\gamma_1}\,\frac{\Delta\sigma}{\mpl}\right|_{\Delta\sigma_b}^{\Delta\sigma}\nonumber\\
&=&\frac{\mpl^2}{4\pi\gamma_0(U_b+V_b)}\left(\arctan\frac{1}{\gamma_0}+\arctan\frac{x}{\gamma_0}\right),\label{eq-spe-i-sil}}
where the temporary variable
\eq{x\equiv\frac{\Delta\sigma}{|\Delta\sigma_b|},}
is the relative displacement from the inflection point.

Since $\sigma$ always subdominates energy density, we can neglect its contribution to the Hubble rate when solving the background evolution for $\phi(N)$. By equating \refeq{spe-i-sil} (as the l.h.s of \refeq{spe-p1eq}) with the r.h.s of \refeq{spe-p1eq}, we derive the evolution of $\sigma(N)$ in terms of $x(N)$, as
\eq{\arctan\frac{x}{\gamma_0}=-\arctan\frac{1}{\gamma_0}+\frac{4\pi\gamma_0(U_b+V_b)}{\mpl^2}\int_{\phi_b}^{\phi}\frac{\dd\phi}{V'(\phi)}.\label{eq-spe-i-x}}

We can also derive the slow-roll parameters that are needed to calculate the cosmological observables
\eqa{\epsilon_\sigma&=&\frac{64\pi^3(U_b+V_b)^4}{\gamma_3^2U_0^2(U+V)^2}(\gamma_0^2+x^2)^2,\\
\eta_\sigma&=&\frac{U_b+V_b}{U+V}x,\\
\xi_\sigma&=&\frac{(U_b+V_b)^2}{2(U+V)^2}(\gamma_0^2+x^2),}
and all the higher slow-roll parameters vanish by our initial assumption. When the plateau formed by the inflection point is flat but has a sharp transition at the boundary ($\gamma_3$ is large and $\gamma_0<1$), and the plateau is narrow ($|\Delta\sigma_b|\ll\mpl$), the observables then can be simplified. For example, the spectral index and its running become
\footnote{\ea{In the limit $s\rightarrow1$, they reduce to the results in \cite{Wang:2013oea}.}}
\eqa{n_s-1&=&-2\epsilon_{\phi*}+2sx_*,\label{eq-spe-i-ns1}\\
\frac{\dd n_s}{\dd\ln k}&\approx&-\frac{1}{2}(n_s-1)^2+2\epsilon_{\phi*}(2\eta_{\phi*}-3\epsilon_{\phi*})+s^2(\gamma_0^2-x_*^2).\label{eq-spe-i-dns1}}

As a simple example, we consider the quadratic inflaton potential
\eq{V(\phi)=\frac{1}{2}m^2\phi^2,}
and we assume it dominates over the spectator potential $U(\sigma)$. After setting the end of inflation at $N_e=0$ and taking the pivot scale as $N_*=50$, we have a total of $5$ free parameters $m,U_0,\gamma_1,\gamma_3$ and $N_b$. By using the same trick as in \refsssec{Step function spectator}, we can fix the overall energy scale to match $\P_\zeta$, and switch to the parameter space $r_{\sigma b},\gamma_0,\gamma_3,N_b$. Then the cosmological observables and the solution $x_*(N_*)$ are independent of $\gamma_3$ when $\gamma_3$ is large, as can be seen from \refeq{spe-i-x}, \refeq{spe-i-ns1}, and \refeq{spe-i-dns1}. Therefore, here we take a large $\gamma_3$ and further reduce the parameter space to $r_{\sigma b},\gamma_0,N_b$.

The background evolution of $\phi$ can be worked out as a function of $N$
\eq{\phi(N)=\mpl\sqrt\frac{2N+1}{4\pi}.\label{eq-spe-i-phi}}
This reduces \refeq{spe-i-x} to
\eq{\arctan\frac{x_*}{\gamma_0}=-\arctan\frac{1}{\gamma_0}+\frac{\gamma_0(2N_b+1)}{4}\ln\frac{2N_*+1}{2N_b+1}.\label{eq-spe-i-x2}}

We can make some predictions from the expressions already. In \refeq{spe-i-x2} when $N_*=50$ and $\gamma_0$ are fixed, the relative displacement $x_*$ at the Hubble exit is maximized when the last term in \refeq{spe-i-x2} reaches maximum at $N_b\approx18$. Since the field $\sigma$ contributes to the spectral index by $2sx_*$ as in \refeq{spe-i-ns1}, we should expect the spectral index to also reach maximum at $N_b\approx18$. The last term of \refeq{spe-p-Npp} contributes $-5x_*/6r_{\sigma b}$ to the local $\fNL$, which should reach \ea{the} minimum at $N_b\approx18$.

\figa{spe-I}{Parameter space for the inflection point spectator with a quadratic inflaton (Part I).}{Cosmological observables and $\gamma_1$ are drawn for the inflection point spectator field with a quadratic inflaton potential. The $x$ axis is the logarithmic of $r_{\sigma b}$, the energy density ratio of the spectator field w.r.t the total at the phase boundary. The $y$ axis is $N_b$, the number of e-folds of inflation from the phase boundary ``$b$'' to the end of inflation. The shaded yellow regions are excluded by the five constraints discussed in \refsssec{Step function spectator}. The shaded green regions are observationally favoured by the spectral index within $0.9667\pm0.0080$ whereas the solid green lines indicate the central value, based on the $2\sigma$ constraint by Planck \cite{Ade:2015lrj}. Two additional subfigures are placed in \refig{spe-I2} due to page limit.}{\figi[0.45]{spe-IGamma1}
{The model parameter $\gamma_1$.}\hspace{0.07\textwidth}\figi[0.45]{spe-Idns}{The running of the spectral index $\dd n_s/\dd\ln k$.}}
\figa{spe-I2}{Parameter space for the inflection point spectator with a quadratic inflaton (Part II).}{Parameter space for the inflection point spectator with a quadratic inflaton, followed from \refig{spe-I}.}{\figi[0.45]{spe-Ifnl}{The local bi-spectrum $\fNL$.}\hspace{0.07\textwidth}\figi[0.45]{spe-Ignl}{The local tri-spectrum $\gNL$.}}
Our analytical predictions can be verified by considering the typical case $N_*=50$, $\gamma_3=10^{10}$, $\gamma_0=0.15$ and $w=1/3$. The only remaining free parameters are $r_{\sigma b}$ and $N_b$. We can hence plot the CMB observables on the $(r_{\sigma b},N_b)$ parameter space, as in \refig{spe-I} and \refig{spe-I2}, where the parameter $\gamma_1$ is also shown. The parameter space has the same exclusion conditions as discussed in \refsssec{Step function spectator}, which are shaded in yellow. \ea{Here the third constraint for the specific model can be shown as}
\eq{r_{\sigma b}^4\gg\frac{64\pi^3(\gamma_0^2+x_*^2)}{\gamma_3^2\epsilon_{\phi*}}.}
\ed{Here t}\ea{T}he energy scale for the spectator field, $U_0$, varies from $10^{-26}\mpl^4$ to $10^{-17}\mpl^4$.

In \refig{spe-I} and \refig{spe-I2}, we find agreement with our analytical predictions. Around $N_b\approx18$, the spectral index peaks and drops out of the $2\sigma$ confidence level, whereas the local bi-spectrum reaches its minimum. The running of the spectral index is typically small. The local $\fNL$ and $\gNL$ can both attain small values within the current observational bound.

\ea{It is worth noting that the fine-tuning of inflection/saddle point potentials is alleviated on spectator fields. This is simply because the inflaton dominated energy density allows for a less flat inflection/saddle point potential.}

\sssecs{Hyperbolic tangent spectator}
In this section, we consider a smoother plateau for the spectator potential, which is the hyperbolic tangent function, in the form
\eq{U(\sigma) = \frac{U_0}{2}\left(1+\tanh\frac{\sigma}{\sigma_0}\right)\,.}
For simplicity, we favour a small enough parameter for $\sigma_0$, so the slow-roll phase for the spectator takes place at $\sigma>2\sigma_0$, but $\sigma_0$ should not be so small that it becomes a step-like potential. The effective potential is
\eq{U(\sigma) = U_0\left(1-2e^{-\frac{2\sigma}{\sigma_0}}\right)\,.}
We can then write down the slow-roll parameters
\eqa{\epsilon_\sigma&=&\frac{\mpl^2U_0^2}{4\pi\sigma_0^2V^2}e^{-\frac{4\sigma}{\sigma_0}},\label{eq-spe-ht-epsilon}\\
\eta_\sigma&=&-\frac{\mpl^2U_0}{2\pi\sigma_0^2V}e^{-\frac{2\sigma}{\sigma_0}},\label{eq-spe-ht-eta}\\
\xi_\sigma&=&\frac{\mpl^2U_0^2}{4\pi^2\sigma_0^4V^2}e^{-\frac{4\sigma}{\sigma_0}},\label{eq-spe-ht-xi}}
where we have neglected the spectator energy density because it is subdominant.

The spectator ends slow roll at $\eta_{\sigma b}=-1$, i.e.
\eq{e^\frac{2\sigma_b}{\sigma_0}=\frac{\mpl^2U_0}{2\pi\sigma_0^2V}.\label{eq-spe-ht-sb}}
Here we assume the inflaton energy density $V(\phi)$ to be almost constant during inflation. This can be guaranteed by the saddle point potential $V(\phi)$. The background evolution of the spectator field can then be solved as
\eq{e^\frac{2\sigma}{\sigma_0}=\frac{\mpl^2U_0}{2\pi\sigma_0^2V}(N-N_b+1).}
The solution then gives the simple expressions for the slow-roll parameters, \refeq{spe-ht-epsilon}, \refeq{spe-ht-eta} and \refeq{spe-ht-xi}, as
\eqa{\epsilon_\sigma&=&\frac{\pi\sigma_0^2}{\mpl^2}\frac{1}{(N-N_b+1)^2},\label{eq-spe-ht-epsilon2}\\
\eta_\sigma&=&-\frac{1}{N-N_b+1},\label{eq-spe-ht-eta2}\\
\xi_\sigma&=&\frac{1}{(N-N_b+1)^2}.\label{eq-spe-ht-xi2}}
The condition $\sigma>2\sigma_0$ then demonstrates the relation (together with \refeq{spe-ht-sb})
\eq{\epsilon_\sigma\ll r_\sigma\eta_{\sigma}^2<|\eta_\sigma|,\label{eq-spe-ht-epsmall}}
which guarantees the slow roll is terminated by the second order condition.

When the spectator $\sigma$ reaches the edge of the hyperbolic tangent plateau at $\sigma_b$, we assume it decays instantly and completely into relativistic species, thus leaving no residual isocurvature perturbations. All the matter is then created by the decay of the inflaton field, such as the saddle point inflaton similar with that in \refsssec{Inflection point inflation}, which can be written as \cite{Allahverdi:2006cx,Allahverdi:2007wt,Wang:2013hva}
\eq{V(\phi)=\frac{1}{2}m^2|\phi|^2-\frac{Ah}{6\sqrt{3}}\phi^3+\frac{h^2}{12}|\phi|^4\,.}

For $A=4m$, we can find the saddle point at
\eq{\phi_0=\sqrt{3}\frac{m}{h}.}
The effective potential around the saddle point then becomes
\eq{V(\phi)=V(\phi_0)+\frac{V'''(\phi_0)}{6}(\phi-\phi_0)^3+\ho,}
where
\eqa{V(\phi_0)&=&\frac{m^4}{4h^2},\\
V'''(\phi_0)&=&\frac{2}{\sqrt3}hm.}

The spectator field can always be dropped when considering the inflaton dynamics. Also neglecting the higher order terms and taking the inflaton energy density to be near constant ($V(\phi)\approx V(\phi_0)$) during inflation, the slow-roll parameters for the inflaton can be written as
\eqa{\epsilon_\phi&=&\frac{9\mpl^2(\phi-\phi_0)^4}{\pi\phi_0^6},\\
\eta_\phi&=&\frac{3\mpl^2(\phi-\phi_0)}{\pi\phi_0^3}.}
At the end of inflation, the second slow-roll condition is violated with $\eta_{\phi e}=-1$, giving
\eq{\frac{\phi_0-\phi_e}{\phi_0}=\frac{\pi\phi_0^2}{3\mpl^2}\ll1,}
where the inequality holds because of our sub-Planckian assumption $\phi_0\ll\mpl$. The small relative field displacement then justifies our approximation of $V(\phi)\approx V(\phi_0)$. Then we can solve the background evolution of the inflaton
\eq{\phi=\phi_0-\frac{2\pi\phi_0^3}{3\mpl^2(2+N-N_e)}.}
For practical purposes, the inflaton perturbations could be assumed to be subdominant in producing curvature perturbation, as compared to that of the spectator's. 

Assuming the saddle point potential of the inflaton is very flat (\ea{$\epsilon_{\phi*}\ll\eta_{\sigma*}$}\ed{$V'$ is small}) and smooth (\ea{$\eta_{\phi*}\ll\eta_{\sigma*}$}\ed{$V''$ is small}) near the Hubble exit of the pivot scales, we can neglect the inflaton's slow-roll parameters in the cosmological observables. The slow-roll parameter $\epsilon_\sigma$ is also negligible according to \refeq{spe-ht-epsmall}. At the pivot scale $N=N_*$, the spectral index $n_s$, the local bi-spectrum $\fNL$, and the local tri-spectrum $\gNL$ are then only determined by the slow-roll parameters of the spectator $\sigma$, giving the leading order terms
\eqa{n_s-1&=&2\eta_{\sigma*}=-\frac{2}{N_*-N_b+1},\\
\fNL&=&-\frac{5\eta_{\sigma*}}{6r_{\sigma*}}=\frac{5}{6(N_*-N_b+1)r_{\sigma*}},\label{eq-s-fNL}\\
\gNL&=&\frac{25(2\eta_{\sigma*}^2-\xi_{\sigma*})}{54r_{\sigma*}^2}=\frac{25}{54(N_*-N_b+1)^2r_{\sigma*}^2},}
where, due to the flatness of the hyperbolic tangent potential and the saddle point potential, the energy density ratio is simply
\eq{r_{\sigma*}\approx r_{\sigma b}\approx\frac{U_0}{V(\phi_0)}.}

The power spectrum of curvature perturbation then becomes
\eq{\P_\zeta=\frac{2(N_*-N_b+1)^2r_{\sigma*}^2V_0}{3\pi\mpl^2\sigma_0^2}.}
Therefore to obtain the observed power spectrum for the curvature perturbation, which requires $\sigma_0$ to take the value
\eq{\frac{\sigma_0^2}{H_*^2}=\frac{r_{\sigma*}^2}{4\pi^2\P_\zeta}(N_*-N_b+1)^2.}

\figa{spe-ht}{Parameter space for the hyperbolic tangent spectator with a saddle point inflaton (Part I).}{The parameter space for the hyperbolic tangent spectator model with a saddle point inflaton. The yellow shaded regions are excluded due to the multiple constraints in \refsssec{Step function spectator}. The green bands comply with the Planck observations at $3\sigma$ confidence level for all the observables, including the spectral index and its running, and the local non-Gaussianities. The red contour lines are for the values of the respective parameters. Here we have taken the pivot scale e-folding $N_*=50$. Two additional subfigures are placed in \refig{spe-ht2} due to page limit.}{\figi[0.48]{spe-htsigma0}{The relative scale $\sigma_0/H_*$.}\hspace{0.04\textwidth}\figi[0.48]{spe-htdns}{The running of the spectral index.}}
\figa{spe-ht2}{Parameter space for the hyperbolic tangent spectator with a saddle point inflaton (Part II).}{Parameter space for the hyperbolic tangent spectator with a saddle point inflaton, followed from \refig{spe-ht}.}{\figi[0.48]{spe-htfnl}{The local bi-spectrum $\fNL$.}\hspace{0.04\textwidth}\figi[0.48]{spe-htgnl}{The local tri-spectrum $\gNL$.}}

When the inflaton model is given, the parameters $N_*$ and $V(\phi_0)$ are fixed. The spectral index $n_s$, the local bi-spectrum $\fNL$, the local tri-spectrum $\gNL$, and the relative value $\sigma_0/H_*$ then only depend on $r_{\sigma*}$ and $N_b$. We can then plot the two-dimensional phase space $(r_{\sigma*},N_b)$ in \refig{spe-ht} and \refig{spe-ht2}, where $N_e=0$ is taken.

One can see that the model predicts the spectral index, its negligible running, and the local bi-spectrum as shown in the green shade which depicts the $3\sigma$ range of the current Planck data \cite{Ade:2015lrj,Ade:2015ava}. We have also shown the values of $\gNL$ in \refig{spe-ht} and \refig{spe-ht2}.  Since $\sigma$ decays into radiation, there is no residual isocurvature perturbations, which matches with observation.

In this chapter, we have summarized the previous studies of the minimal curvaton scenario, which has been severely constrained by the Planck observations. We then moved on to spectator scenario, and obtained good agreement with the CMB. For the spectator models with inflection point potentials, hyperbolic tangent potentials, and those with a sudden phase transition, we have mapped out their parameter space in accordance with the CMB.

\ea{Although the second half of this chapter has been devoted to producing the CMB fluctuations with the spectator scenario, it is merely a possibility for certain model potentials under certain initial conditions. For this reason, spectator scenario does not suffer from fine tuning in initial condition. Initial conditions that fail to allow the spectator to end slow roll during inflation may instead lead to curvaton scenario or a second phase of inflation, which can also seed the desired cosmological observables in their own ways. Undesired spectator scenario may also appear in many-field inflation, indicating another future direction of spectator studies.}

%% file: Chapters/Asymmetry.tex
As discussed in \refssec{CMB power asymmetry}, the power asymmetry of the CMB temperature fluctuations should come from the inflationary epoch or other alternative scenarios. There have been plenty of relevant analyses and proposals to explain the power asymmetry \cite{Flender:2013jja,Dai:2013kfa,Lyth:2013vha,Wang:2013lda,Kohri:2013kqa,Chen:2013eaa,Liu:2013kea,McDonald:2013aca,Chang:2013vla,McDonald:2013qca,Namjoo:2013fka,Zhao:2013jya,Liddle:2013czu,Cai:2013gma,Mazumdar:2013yta,Kanno:2013ohv,Jazayeri:2014nya,McDonald:2014lea,Liu:2013iha,Assadullahi:2014pya}. We first discuss the generic conditions for producing a large CMB power asymmetry from inflation, showing the deficiency of CMB power asymmetry from (nearly) scale invariant perturbations. We then demonstrate how a tachyonic fast roll phase may enhance power asymmetry. Combining with other observational constraints, we construct a minimal model for the observed CMB power asymmetry based on spectator scenario.

\ssecs{Primordial origins of the CMB power asymmetry}
\sssecs{Consistency relation for single-source perturbations}
\emph{Single-source} here indicates that the primordial perturbations originated from only a single field. It can be single-field slow-roll inflation, where only one field is present, but hybrid inflation, curvaton or spectator scenarios also count, as long as only one field contributes to primordial perturbations. This is in contrast with the primordial perturbations from multiple sources, which will be discussed in \refsssec{Consistency relation for multi-source perturbations}.

Using the separate universe approach \cite{Sasaki:1995aw,Wands:2000dp}, we can regard the opposite sides of our Hubble patch as independent separate universes for the CMB power spectra. Assuming the curvature perturbation is determined by only one field $\sigma$, (which may participate in inflation,) the power spectrum of the curvature perturbation can be written as (at the leading order)
\footnote{
Starting from here, all primordial variables would indicate their values at Hubble exit unless otherwise specified.}
\eq{\P_\zeta=\P_{\delta N}=N_\sigma^2\P_{\delta\sigma}.\label{eq-pa-po-Pz}}
As an example, the $\sigma$ field can be one of the inflatons, a curvaton, or a spectator.

As explained in \refssec{ssi-First order perturbations}, field perturbations, such as $\delta\sigma_\vs k$, can exist on scales which are even much larger than the size of our current Hubble patch. Such \emph{very large scale perturbations} can bring about a non-vanishing (vector) gradient\footnote{
$\nabla\equiv(\partial/\partial x^1,\partial/\partial x^2,\partial/\partial x^3)$ is the vector representation of the gradient operator for three spatial dimensions.}
$\nabla\sigma$ at the Hubble exit of the pivot scales, which can be regarded as a constant in our observable universe. Consider two opposite local patches on the CMB map, separated by $2\vs r\subs{ls}$ directed from one patch to the other, where $r\subs{ls}=1/a_*H_*$ is our comoving distance to the Last Scattering Surface, defined in \refeq{ssi-cosav2}. The gradient $\nabla\sigma$ then yields a background field asymmetry between the two opposite local patches at the Hubble exit, by the amount
\eq{\Delta\sigma=2\vs r\subs{ls}\cdot\nabla\sigma.}
Along the maximal direction, the field asymmetry is most significant by the amount
\eq{\Delta\sigma=2r\subs{ls}|\nabla\sigma|.}

The background field asymmetry $\Delta\sigma$ can then result in the asymmetry in the curvature perturbation. According to \refeq{pa-po-Pz}, on the opposite sides the power spectra of curvature perturbation should differ by
\eq{\frac{\Delta\P_\zeta}{\P_\zeta}=\frac{\partial\P_\zeta}{\P_\zeta\partial\sigma}\Delta\sigma=\frac{2N_{\sigma\sigma}}{N_\sigma}\Delta\sigma,\label{eq-pa-po-DPz1}}
where we have neglected the possible change in Hubble rate due to $\Delta\sigma$, because during inflation $H$ should remain almost constant. Since we are only interested in the absolute amount of asymmetry, we can redefine $\Delta\P_\zeta$ and $\Delta\sigma$ as their absolute values. This transforms \refeq{pa-po-DPz1} into
\eq{\frac{\Delta\P_\zeta}{\P_\zeta}=\left|\frac{2N_{\sigma\sigma}}{N_\sigma}\right|\Delta\sigma.\label{eq-pa-po-DPz2}}

Since $\sigma$ is the only field that creates cosmological perturbations, we can write $N_\sigma$ and $N_{\sigma\sigma}$ in terms of cosmological observables. This will create a link between the asymmetries and other cosmological observables, such as
\eq{\frac{\Delta\P_\zeta}{\P_\zeta}=\frac{12}{5}\frac{\Delta\sigma}{\sqrt{\P_{\delta\sigma}}}|\fNL|\sqrt{\P_\zeta},}
where $\Delta\sigma/\sqrt{\P_{\delta\sigma}}$ can be regarded as the \emph{relative strength} of very large scale perturbations, compared with those at the pivot scales. The CMB power asymmetry factor $A$ can then be related to
\eq{A=\frac{\Delta\P_{\Delta T}}{4\P_{\Delta T}}=\frac{\Delta\P_\zeta}{4\P_\zeta}=\frac{3}{5}\frac{\Delta\sigma}{\sqrt{\P_{\delta\sigma}}}|\fNL|\sqrt{\P_\zeta}.\label{eq-pa-po-A1}}

From the latest Planck observations \cite{Ade:2015lrj,Ade:2015kab}, we know $\P_\zeta=(2.142\pm0.0040)\times10^{-9}$ and $A=0.07\pm0.02$. For simplicity, we can take their central values. The local $\fNL=0.8\pm5.0$ constraint at $1\sigma$ corresponds to $|\fNL|<10.8$ at over $95\%$ CL. Putting together all these constraints, \refeq{pa-po-A1} then leads to the lower bound for very large scale perturbations
\eq{\frac{\Delta\sigma}{\sqrt{\P_{\delta\sigma}}}=\frac{5A}{3|\fNL|\sqrt{\P_\zeta}}>233,\label{eq-pa-po-Dsigma1}}
at over $95\%$ CL.

For compatibility with forthcoming sections, we also use expectation values as the measure for the amount of asymmetry. The expectation versions of \refeq{pa-po-A1} and \refeq{pa-po-Dsigma1} are
\eq{\langle A^2\rangle=\frac{9}{25}\frac{\left\langle|\Delta\sigma|^2\right\rangle}{\P_{\delta\sigma}}\fNL^2\P_\zeta,\label{eq-pa-po-A11}}
and
\eq{\frac{\left\langle|\Delta\sigma|^2\right\rangle}{\P_{\delta\sigma}}=\frac{25\langle A^2\rangle}{9\fNL^2\P_\zeta}>5.4\times10^4,\label{eq-pa-po-Dsigma2}}
where for simplicity, we have taken the observed central value $A\approx0.07$ as the standard deviation of $A$.

\sssecs{Consistency relation for multi-source perturbations}
In the more generic scenario of multi-field slow-roll inflation, where perturbations of all fields may contribute to curvature perturbation, \refeq{pa-po-Pz} becomes
\eq{\P_\zeta=\P_{\delta N}=\P_{\delta\phi}\sum_\mu N_\mu^2,\label{eq-pa-po-Pz2}}
where $\mu=0,1,\dots,n-1$ for the canonical slow-roll scalar fields $\phi^\mu$ and $n$ is the number of $\phi$ fields. The very large scale perturbations then bring about the gradient $\nabla\phi^\mu$, whose maximum directions can differ for each field. By comparing the opposite local patches on the CMB map along an arbitrary direction, which are separated by $2\vs r\subs{ls}$, we can find the asymmetries in the background field evolutions by the amount
\eq{\Delta\phi^\mu=2\vs r\subs{ls}\cdot\nabla\phi^\mu.\label{eq-pa-po-Dphim}}
The asymmetry in the curvature perturbation can then be calculated as
\eq{\frac{\Delta\P_\zeta}{\P_\zeta}=\frac{4N_\mu N_{\mu\nu}}{\sum_\lambda N_\lambda^2}\vs r\subs{ls}\cdot\nabla\phi^\nu=\frac{4r\subs{ls}}{\sum_\lambda N_\lambda^2}\uv r\subs{ls}\cdot\sum_{\mu,\nu}N_\mu N_{\mu\nu}\nabla\phi^\nu.\label{eq-pa-po-DPP}}
The maximal asymmetry of curvature perturbation is achieved when the unit vector $\uv r\subs{ls}$ aligns with the vector it is multiplied with in \refeq{pa-po-DPP}. In the following calculations, we are only interested in the direction with the \emph{strongest} asymmetry signal, which has
\eq{\frac{\Delta\P_\zeta}{\P_\zeta}=\frac{4r\subs{ls}}{\sum_\lambda N_\lambda^2}\left|\sum_{\mu,\nu}N_\mu N_{\mu\nu}\nabla\phi^\nu\right|.\label{eq-pa-po-DPz3}}
where
\eq{\left|\sum_{\mu,\nu}N_\mu N_{\mu\nu}\nabla\phi^\nu\right|\equiv\sqrt{\sum_{i=1}^3\left(\sum_{\mu,\nu}N_\mu N_{\mu\nu}\frac{\partial\phi^\nu}{\partial x^i}\right)^2}\;.}

In the simplest case where perturbative evolutions are identical for all the fields $\phi^\mu$ on very large scales, the expectation of the gradients of all the fields should be independently Gaussian and isotropic, so they are all equal to
\eq{\left\langle\frac{\partial\phi^\mu}{\partial x^i}\frac{\partial\phi^\nu}{\partial x^i}\right\rangle=\delta^{\mu\nu}\left\langle\left(\frac{\partial\phi^\mu}{\partial x^i}\right)^2\right\rangle,}
and
\eq{\left\langle|\nabla\phi^\mu|^2\right\rangle\equiv\left\langle\sum_{i=1}^3\left(\frac{\partial\phi^\mu}{\partial x^i}\right)^2\right\rangle=\left\langle|\nabla\phi|^2\right\rangle,\hspace{0.5in}\mathrm{for\ any\ }\mu.}
These properties allow us to calculate with ease the expectation value of the absolute in \refeq{pa-po-DPz3}
\eqa{\left\langle\left|\sum_{\mu,\nu}N_\mu N_{\mu\nu}\nabla\phi^\nu\right|^2\right\rangle&=&\left\langle\sum_{i=1}^3\left(\sum_{\mu,\nu}N_\mu N_{\mu\nu}\frac{\partial\phi^\nu}{\partial x^i}\right)^2\right\rangle\\
&=&\sum_{i=1}^3\sum_{\mu,\nu,\lambda,\eta}N_\mu N_{\mu\nu}N_\lambda N_{\lambda\eta}\left\langle\frac{\partial\phi^\nu}{\partial x^i}\frac{\partial\phi^\eta}{\partial x^i}\right\rangle\\
&=&\sum_{\mu,\nu,\lambda}N_\mu N_{\mu\nu}N_{\nu\lambda}N_\lambda\left\langle\sum_{i=1}^3\left(\frac{\partial\phi^\nu}{\partial x^i}\right)^2\right\rangle\\
&=&\tNL\left\langle|\nabla\phi|^2\right\rangle\left(\sum_\mu N_\mu^2\right)^3.}
Then we know the expectation of power asymmetry in the CMB perturbation spectrum, according to \refeq{pa-po-DPz3}, as
\eqa{\langle A^2\rangle&=&\frac{\langle\Delta\P_\zeta^2\rangle}{16\P_\zeta^2}\\
&=&\tNL r\subs{ls}^2\left\langle|\nabla\phi|^2\right\rangle\sum_\mu N_\mu^2\\
&=&\frac{1}{4}\tNL\P_\zeta\frac{\left\langle|\Delta\phi|^2\right\rangle}{\P_{\delta\phi}},\label{eq-pa-po-A2}}
where
\eq{\left\langle|\Delta\phi|^2\right\rangle\equiv4r\subs{ls}^2\left\langle|\nabla\phi|^2\right\rangle}
is the expectation value of the field asymmetry along \emph{its} maximal direction (see \refeq{pa-po-Dphim}). Then \refeq{pa-po-A2} yields the lower bound of the strength of very large scale perturbations
\eq{\frac{\left\langle|\Delta\phi|^2\right\rangle}{\P_{\delta\phi}}=\frac{4\langle A^2\rangle}{\tNL\P_\zeta}>3200,\label{eq-pa-po-Dphi2}}
at over $95\%$ CL.

Comparing it with \refeq{pa-po-Dsigma2}, we find that in multi-source scenario, the applicable non-Gaussianity parameter is instead $\tNL$. This also states that the single-source scenario \refeq{pa-po-Dsigma2} is a special case of the multi-source version \refeq{pa-po-Dphi2}, because of the consistency relation \refeq{ssi-cons2}.

\ssecs{Lack of asymmetry from scale invariant perturbations}
It still remains a question if \refeq{pa-po-Dsigma2} or \refeq{pa-po-Dphi2} can be satisfied in order to explain the CMB power asymmetry, especially for slow-roll scalar fields during inflation. In this section, we will calculate the expectation of field gradient from very large scale perturbations. The very large scale perturbations produce a non-vanishing gradient along any arbitrary $z$ direction, whose expectation can be calculated as
\eqa{\left\langle\left|\frac{\partial\delta\phi(\vs x)}{\partial z}\right|^2\right\rangle&=&\left\langle\int_{k,k'\le k_*}\frac{\dd^3\vs k\dd^3\vs k'}{(2\pi)^3}k_zk_z'\delta\phi_\vs k\delta\phi_{\vs k'}^\dag e^{i(\vs k-\vs k')\cdot\vs x}\right\rangle\\
&=&\int_{k\le k_*}k_z^2P_{\delta\phi}(k)\dd^3\vs k\\
&=&\frac{4\pi}{3}\int_0^{k_*}k^4P_{\delta\phi}(k)\dd k,}
where the upper bound of the integral
\eq{k_*\equiv a_*H_*,}
is the comoving wavenumber of the pivot scale, only scales larger than which can make a significant contribution to the overall gradient in our current Hubble patch. The lower bound of the integral is treated as zero as an approximation, because a more careful choice of the lower bound (that is $\ll k_*$) only changes the integral negligibly in most cases. The vanishing lower bound does not indicate any past completeness for inflation.

We can then derive the expectation of field gradient
\eq{\left\langle|\nabla\phi|^2\right\rangle=4\pi\int_0^{k_*}k^4P_{\delta\phi}(k)\dd k,}
and that of field asymmetry
\eq{\left\langle|\Delta\phi|^2\right\rangle=16\pi r\subs{ls}^2\int_0^{k_*}k^4P_{\delta\phi}(k)\dd k.\label{eq-pa-la-Dphi}}

In the simplest case, if all the fields $\phi^\mu$ are slowly rolling in the last several e-folds before the Hubble exit of the CMB scales, we can obtain the (almost) scale invariant field perturbations\footnote{
We assume that perturbations in the infrared limit ($k\rightarrow0$) do not affect local physics, so any possible deviation from scale invariance in the infrared limit would be negligible.}
\eq{k^3P_{\delta\phi}(k)=k_0^3P_{\delta\phi}(k_0)=2\pi^2\P_{\delta\phi},\hspace{0.5in}\mathrm{for\ }k\lesssim k_*.\label{eq-pa-la-si}}
This greatly simplifies \refeq{pa-la-Dphi}, yielding the relative strength of field asymmetry for scale invariant perturbations:
\eq{\frac{\left\langle|\Delta\phi|^2\right\rangle}{\P_{\delta\phi}}=16\pi^3,\label{eq-pa-la-Dphi2}}
which is insufficient, according to \refeq{pa-po-Dphi2}, to produce the observed CMB power asymmetry. Also, \refeq{pa-po-A2} becomes
\eq{\langle A^2\rangle=4\pi^3\tNL\P_\zeta<9\times10^{-4},\label{eq-pa-la-A2}}
at $>95\%$ CL.

The statement \ea{still holds}\ed{remains to hold} when the spectral index of field perturbations are non-vanishing but small. The power spectrum of field perturbations can be parameterized w.r.t the reference scale $k_0$, as
\eq{k^3P_{\delta\phi}(k)=k_0^3P_{\delta\phi}(k_0)\left(\frac{k}{k_0}\right)^{n_{\delta\phi}-1}\;,\label{eq-pa-ndsigmadef}}
where $n_{\delta\phi}\approx1$ is the spectral index of field perturbations, \ed{similarly }defined \ea{similarly to}\ed{with} $n_s$ in \refeq{cmb-Pz} for curvature perturbation $\zeta$. \ed{\Refeq{pa-la-si} and }\refeq{pa-la-Dphi2} then become\ea{s} (assuming $n_{\delta\phi}\ne-1$)
\eq{\frac{\left\langle|\Delta\phi|^2\right\rangle}{\P_{\delta\phi}}=\frac{32\pi^3}{n_{\delta\phi}+1}\frac{k_0^2}{a_*^2H_*^2}\left(\frac{k_*}{k_0}\right)^{n_{\delta\sigma}+1}.\label{eq-pa-la-Dphi3}}
\Refeq{pa-la-A2} in the weak scale dependence case then becomes
\eq{\langle A^2\rangle=\frac{8\pi^3}{n_{\delta\phi}+1}\frac{k_0^2}{a_*^2H_*^2}\left(\frac{k_*}{k_0}\right)^{n_{\delta\sigma}+1}\tNL\P_\zeta.\label{eq-pa-la-A3}}
As a good approximation for practical purposes, here we choose the reference scale $k_0=k_*$, at the pivot scale. \Refeq{pa-la-A3} is then simplified to
\eq{\langle A^2\rangle=\frac{8\pi^3}{n_{\delta\phi}+1}\tNL\P_\zeta.\label{eq-pa-la-A4}}

As we can see from \refeq{pa-la-A4}, the weak scale dependence only introduces the extra factor $2/(n_{\delta\phi}+1)$. This cannot change the order of magnitude for a nearly scale invariant spectrum with $n_{\delta\phi}\approx1$. In the single-source case where \refeq{pa-po-A11} should be applied, the inconsistency with the observed central value $A=0.07$ is even stronger. We therefore conclude that for canonical slow-roll scalar fields with nearly scale invariant perturbations, any mechanism cannot produce the CMB power asymmetry with central value $A=0.07$, while satisfying the local non-Gaussianity constraints from $\fNL$ and $\tNL$.

\ssecs{Enhancing very large scale perturbations from a tachyonic fast-roll phase}
In the previous section, we have shown that canonical slow-roll scalars cannot produce the observed CMB power asymmetry because of (near) scale invariant spectrum. This suggests \ea{to} us straightforwardly to violate scale invariance using non-slow-roll fields. It is well known that perturbations can get enhanced during a tachyonic fast-roll phase \cite{Leach:2001zf,Linde:2001ae,Mazumdar:2013yta}. Therefore we wish to investigate how a tachyonic fast-roll phase before the Hubble exit of \ea{the} pivot scale may enhance CMB power asymmetry.

In the simplest case, let us consider a scalar field $\sigma$ that acquires a tachyonic mass for a brief period. The total effective action for the tachyonic phase is given by
\eq{S=\int\sqrt{-g}\,\dd^4x\left(-\frac{1}{2}\partial_\mu\sigma\partial^\mu\sigma+\frac{1}{2}m^2\sigma^2+{\cal L}\subs{else}\right),\label{eq-pa-tf-S}}
where $m$ is the effective tachyonic mass of $\sigma$ during the tachyonic phase, which is regarded as a constant for simplicity. All other components of the universe are then contained in the ${\cal L}\subs{else}$ term. For simplicity, we also assume that the Hubble rate $H$ remains almost constant, which may come from the ${\cal L}\subs{else}$ term. Then the Hubble rate $H$ determines a unique energy scale, w.r.t which all the dimensional variables can be expressed. As an example, the tachyonic mass $m$ can be written as
\eq{m^2\equiv(e^{2N_m}-1)2H^2,\label{eq-pa-tf-Nm}}
where $N_m$ is defined as such.

Before or after the brief tachyonic fast-roll phase, we simply assume the scalar $\sigma$ becomes light and enters slow-roll phase. For perturbative calculations in slow-roll phase, we can simply take $m=0$ in \refeq{pa-tf-S}, or $N_m=0$ in \refeq{pa-tf-Nm}, because the effective mass is much smaller than the Hubble rate. Here we do not investigate how the brief tachyonic fast-roll phase may be motivated in any specific particle physics theory, but instead only discuss the phenomenological consequences. For simplicity, we will also assume that the fast-roll phase (or the tachyonic mass) is switched on and off instantly.

During tachyonic fast-roll phase, we will not solve the perturbative evolutions exactly, such as in \refeq{ssi-sol}. Instead and as an approximation, we can find the attractor solutions for slow-roll and fast-roll phases and concatenate them. For this purpose, we define the redshift damping rate for perturbation mode $k$ as
\eq{\alpha_k(N)\equiv\frac{1}{2}\frac{\partial\ln P_{\delta\sigma}(k,N)}{\partial N}.}
Note here we have used $N\equiv\ln\frac{a}{a_0}$, the past e-folds of the universe expansion w.r.t a certain reference scale $a_0$, as proper time\footnote{
Note that $N$ here has a different sign with the previous convention in this thesis.}
. Also, $P_{\delta\sigma}(k,N)$ is the realtime power spectrum, i.e.\ a function of the e-folding $N$, instead of $P_{\delta\sigma}(k)$ which is its late time value after freezing. We can then write the power spectrum of \ea{the} field perturbation at any e-folding $N$ w.r.t its value at a reference e-folding $N_0$ as
\eq{P_{\delta\sigma}(k,N)=e^{2\int_{N_0}^N\alpha_k(N)\dd N}P_{\delta\sigma}(k,N_0).\label{eq-pa-tf-alpha}}

As an example, the slow-roll solution \refeq{ssi-PdPhi} of field perturbation yields the redshift rate
\eq{\alpha_k(N)=-\frac{2e^{2(N_k-N)}}{1+2e^{2(N_k-N)}},}
where $N_k$ is the dimensionless relative wave number, defined as
\eq{k^2\equiv2a_0^2H^2e^{2N_k}.}
Obviously then super-Hubble modes correspond to $k^2\ll a^2H^2$ or $N_k<N$, and for sub-Hubble modes we have $k^2\gg a^2H^2$ or $N_k>N$.

With a brief tachyonic phase, the exact solution no longer holds. Any mode $k$ or $N_k$ may experience three possible phases -- the sub-Hubble, slow-roll, and tachyonic phases. As an approximation, we can assume when any mode enters any phase, it immediately behaves as the attractor solution of this phase. For simplicity, we also assume no energy is lost during phase transitions. The evolutions of power spectra then follow \refeq{pa-tf-alpha}, where $\alpha_k(N)$ can be a piecewise function for different phases.

Using the tachyonic action \refeq{pa-tf-S}, we can write down the equation of motion for the field perturbation $\delta\sigma$ as
\eq{\ddot{\delta\sigma}_{\vs k}+3H\dot{\delta\sigma}_{\vs k}+\left(\frac{k^2}{a^2}-m^2\right)\delta\sigma_{\vs k}=0.}
The redshift rate $\alpha_k(N)$ of the three phases can then be calculated as follows. (The slow-roll phase is just the special case of $N_m=0$.)
\begin{itemize}
\item For sub-Hubble modes, the momentum is much larger than the Hubble rate and tachyonic mass combined, with $k^2\gg a^2(2H^2+m^2)$. Therefore we can define $\tau$ and $\psi$ according to \refeq{ssi-tau} and \refeq{ssi-psi}. The equation of motion then transforms into
\eq{\psi_\vs k''+\left[k^2-a^2(2H^2+m^2)\right]\psi_\vs k=0,}
or
\eq{\psi_\vs k''+\left[1-e^{2(N+N_m-N_k)}\right]k^2\psi_\vs k=0.}
Therefore, sub-Hubble modes correspond to $N_k>N+N_m$, which has the harmonic oscillator solution $\psi_\vs k\propto\tau^0$ or $\delta\sigma\propto a^{-1}$. The damping rate is simply
\eq{\alpha_k(N)=-1,\hspace{1in}\mathrm{for\ }N_k>N+N_m.}
\item For tachyonic modes, the relatively small momentum \ea{($k\ll a^2(2H^2+m^2)$)} allows us to define
\eq{\psi_\vs k\equiv a^{\frac{3}{2}}\delta\sigma_\vs k.}
The equation of motion then becomes
\eq{\ddot\psi_\vs k-\left(m^2+\frac{9}{4}H^2-\frac{k^2}{a^2}\right)\psi_\vs k=0,}
or
\eq{\ddot\psi_\vs k-\left[\frac{1}{4}+2\left(e^{2N_m}-e^{2(N_k-N)}\right)\right]H^2\psi_\vs k=0.}
The attractor solution corresponds to the redshift rate
\eq{\alpha_k(N)=-\frac{3}{2}+\sqrt{\frac{1}{4}+2\left[e^{2N_m}-e^{2(N_k-N)}\right]},\hspace{0.5in}\mathrm{for\ }N_k<N+N_m.}
\end{itemize}

Combining the two phases, we arrive at the full expression for redshift rate (while approximately applying the solutions near phase boundary)
\eq{\alpha_k(N)=\left\{\begin{array}{l@{\hspace{0.5in}\mathrm{for\ }}l}
-1,&N_k\ge N+N_m,\\
-\frac{3}{2}+\sqrt{\frac{1}{4}+2\left[e^{2N_m}-e^{2(N_k-N)}\right]},&N_k<N+N_m.
\end{array}\right.\label{eq-pa-tf-asol1}}
It can be verified that the piecewise solution is continuous at the boundary $N_k=N+N_m$ with $\alpha_k=-1$. During slow roll, we can take the $m\rightarrow0$ or $N_m\rightarrow0$ limit, reducing \refeq{pa-tf-asol1} to
\eq{\alpha_k(N)|\subs{sr}=\left\{\begin{array}{l@{\hspace{0.5in}\mathrm{for\ }}l}
-1,&N_k\ge N,\\
-\frac{3}{2}+\sqrt{\frac{9}{4}-2e^{2(N_k-N)}},&N_k<N.
\end{array}\right.\label{eq-pa-tf-asol2}}

\fig{pa-Timeline}{Timeline for the brief tachyonic fast-roll scenario}{The hierarchy of scales, and the timeline of the tachyonic fast-roll scenario. The $x$ axis is the number of e-folds of inflation as time measure, whose zero point is chosen at the end of the tachyonic phase. The $y$ axis is the effective mass of the field of our concern. The tachyonic fast-roll phase lasts during $N_i<N<0$, with the mass $m^2=(e^{2N_m}-1)2H^2$ where $N_m\le N_*$.}
\fig{pa-alpha}{Enhancement rates for selected tachyonic masses}{A demonstration of $\alpha_k(N)$ and the enhancement rate $\Delta\alpha_k(N)$ for some typical values of $N_m$. From bottom to top, black, green, yellow and red correspond to $N_m=0,1,1.5,2$ respectively. Dashed curves represent $\alpha_k(N)$, and solid ones represent $\Delta\alpha_k(N)$.}
Outside the tachyonic phase, the field $\sigma$ is always light, yielding \refeq{pa-tf-asol2} as the solution for $\alpha_k(N)$. In this case, we know already that the power spectrum of the field perturbation $\delta\sigma_\vs k$ is almost scale invariant. However when a brief tachyonic fast-roll phase is present, as shown in \refig{pa-Timeline}, the tachyonic phase will provide a different redshift rate for field perturbations. The difference $\Delta\alpha_k(N)\equiv\alpha_k(N)-\alpha_k(N)|\subs{sr}$ is therefore defined as the relative \emph{enhancement rate}, due to the tachyonic mass $m$. It can be expressed as
\eq{\Delta\alpha_k(N)=\left\{\begin{array}{l@{\hspace{0.2in}\mathrm{for\ }}l}
0,&N_k\ge N+N_m,\\
\sqrt{\frac{1}{4}+2\left[e^{2N_m}-e^{2(N_k-N)}\right]}-\frac{1}{2},&N+N_m>N_k\ge N,\\
\sqrt{\frac{1}{4}+2\left[e^{2N_m}-e^{2(N_k-N)}\right]}-\sqrt{\frac{9}{4}-2e^{2(N_k-N)}},&N_k<N.
\end{array}\right.\label{eq-pa-tf-dasol}}
This is plotted with some typical values of $N_m$ in \refig{pa-alpha}, from which we see that the enhancement can be quite significant ($\Delta\alpha_k\sim e^{N_m}$). Also, from \refeq{pa-tf-dasol} we can see that the scales with $N_k\ge N+N_m$ are not affected by the tachyonic phase. Remembering that the tachyonic phase lasts from $N=N_i<0$ to $N=0$, the scales $N_k\ge N_m$ will be totally unaffected, which is where we want the pivot scale to lie ($N_*\ge N_m$).

The relative enhancement from the fast-roll phase changes \ea{the} CMB power asymmetry by
\eq{\frac{\left\langle|\Delta\sigma|^2\right\rangle}{\P_{\delta\sigma}}=\int_{-\infty}^{N_m}32\pi^3e^{(n_{\delta\sigma}+1)(N_k-N_*)+2\int_{N_i}^0\Delta\alpha_k(N)\dd N}\dd N_k,\label{eq-pa-tf-Ds2}}
where $n_{\delta\sigma}\approx1$ is the spectral index of $\delta\sigma$ in the absence of tachyonic enhancement, as defined in \refeq{pa-ndsigmadef}. Here we have neglected the integral region $N_m<N_k<N_*$, because it is not enhanced by the tachyonic fast-roll scenario, and has been shown in \refssec{Lack of asymmetry from scale invariant perturbations} to generate only a small CMB power asymmetry. The inner integral of $\Delta\alpha_k$ is performed for the tachyonic fast-roll phase $N_i<N<0$ only. \Refeq{pa-tf-Ds2} can be recast into
\eq{\frac{\left\langle|\Delta\sigma|^2\right\rangle}{\P_{\delta\sigma}}=32\pi^3e^{(n_{\delta\sigma}+1)(N_m-N_*)}\int_{-\infty}^{N_m}e^{2\beta_k}\dd N_k,\label{eq-pa-tf-Ds22}}
where we have defined
\eq{\beta_k\equiv\frac{1}{2}(n_{\delta\sigma}+1)(N_k-N_m)+\int_{N_i}^0\Delta\alpha_k(N)\dd N.\label{eq-pa-tf-b}}
Since the mode dependence in \refeq{pa-tf-Ds22} has been absorbed into $\beta_k$, the scale $k\subs{max}$ which maximizes $\beta_k$ will contribute most to the CMB asymmetry. The overall exponential coefficient in \refeq{pa-tf-Ds22} simply means that a longer slow roll after the tachyonic phase will stretch the initial perturbation modes, and lead to a weaker CMB power asymmetry.

\fig{pa-beta}{The tachyonic fast-roll enhancement $\beta_k$}{The enhancement $\beta_k$ is shown in the black solid curve, where its components $N_k-N_m$ and the integral in \refeq{pa-tf-b} are shown in the green and the red dashed curves respectively. The blue shaded region is the number of e-folds of the universe expansion during the fast-roll phase. We have taken the parameter values $N_i=-2, N_m=1.2$.}
We would like $\beta_k$ to peak at some scale $N_{k\subs{max}}$, or otherwise it is difficult to produce sufficient CMB power asymmetry. The peak mode $N_{k\subs{max}}$ can be solved from $\partial\beta_k/\partial N_k=0$. Noticing $\Delta\alpha_k(N)$ is only a function of $N-N_k$, this yields
\eq{\Delta\alpha_{k\subs{max}}(0)=\Delta\alpha_{k\subs{max}}(N_i)+\frac{1}{2}(1+n_{\delta\sigma}).\label{eq-pa-tf-beq}}
Since $n_{\delta\sigma}\approx1$, there would be no peak if $\Delta\alpha_k(0)$ is always less than 1. The above condition requires the tachyonic mass to be large enough. According to \refeq{pa-tf-dasol}, we obtain
\eq{m^2\ge2H^2.}

The contribution to the CMB power asymmetry would then mostly come from around the peak scale $N_{k\subs{max}}$. If we know the full width at half maximum (FWHM) of the peak, namely $\Delta N$, we can have a good estimation for the integral, hence writing \refeq{pa-tf-Ds22} as
\eq{\frac{\left\langle|\Delta\sigma|^2\right\rangle}{\P_{\delta\sigma}}\approx32\pi^3\Delta Ne^{(n_{\delta\sigma}+1)(N_m-N_*)+2\beta_{k\subs{max}}}.\label{eq-pa-tf-Ds23}}

A typical example of $\beta_k$ is shown in \refig{pa-beta}, for $n_{\delta\sigma}=1$, $N_i=-2$, and $N_m=1.2$, in which $\beta_k$ peaks at about $N_k\approx-1$, with $\beta_{k\subs{max}}\approx4.7$. The half maximum lies at $\beta_{k\subs{max}}-\frac{1}{2}\ln2$ with $\Delta N\approx0.8$. We do not want the pivot scale spectrum to be modified by the fast-roll phase, so we require $N_*>N_m$. As a consequence, we take $N_*=N_m+2.5$. Plugging these numbers into \refeq{pa-tf-Ds23} will give
\eq{\frac{\left\langle|\Delta\sigma|^2\right\rangle}{\P_{\delta\sigma}}\approx6.4\times10^4.\label{eq-pa-tf-Dsn}}
This result satisfies both necessary conditions \refeq{pa-po-Dsigma2} and \refeq{pa-po-Dphi2}, yielding a much stronger CMB power asymmetry than the slow-roll scenario.

We also have to make sure that the perturbations remain small throughout the dynamics. This typically requires the curvature perturbation from the $\sigma$ field to have a power spectrum $k^3P_{\zeta_{\delta\sigma}}(k)<1$. Since there can be other sources of curvature perturbation, we define a ratio for $\sigma$ at the pivot scale
\eq{R^2(k)\equiv\frac{P_{\zeta_{\delta\sigma}}(k)}{P_{\zeta}(k)}\le1.\label{eq-pa-tf-R}}
The constraint $k^3P_{\zeta_{\delta\sigma}}(k)<1$ then becomes
\eqa{k^3P_{\zeta_{\delta\sigma}}(k)&=&\left.k^3P_{\zeta_{\delta\sigma}}(k)\right|\subs{sr}e^{2\int_{N_i}^0\Delta\alpha_k(N)\dd N}\nonumber\\
&=&k_*^3P_{\zeta}(k_*)R^2(k)e^{2\int_{N_i}^0\Delta\alpha_k(N)\dd N}\nonumber\\
&=&2\pi^2\P_{\zeta}R^2(k)e^{2\int_{N_i}^0\Delta\alpha_k(N)\dd N}<1,\label{eq-pa-tf-pze}}
where we have used the scale invariance $k^3P_{\zeta_{\delta\sigma}}(k)|\subs{sr}=k_*^3P_{\zeta_{\delta\sigma}}(k_*)$ for any mode $k$. This constrains the total amount of asymmetry enhancement, i.e.\ the height of the red dashed curve in \refig{pa-beta},  by
\eq{\int_{N_i}^0\Delta\alpha_k(N)\dd N<-\frac{1}{2}\ln2\pi^2\P_\zeta-\ln R,\hspace{0.3in}\mathrm{for\ any\ }N_k<0.\label{eq-pa-tf-pac}}

Since $R\le1$ and $\P_\zeta=2.142\times10^{-9}$, in the example of \refig{pa-beta} the red curve is lower than $-\frac{1}{2}\ln\P_\zeta\approx8.5$, and therefore the condition \refeq{pa-tf-pac} is well satisfied.

\ssecs{Other observational constraints}
\sssecs{Quasars}
\fig{pa-CMBq}{Power asymmetries of the CMB and quasar perturbations}{A schematic figure on the power asymmetries of the CMB and quasars spectrum. The outer sphere is the LSS and the inner one contains all the observed quasars. Therefore, quasar observation can only constrain the asymmetry in the distance scales smaller than $r_q$, whilst the asymmetry in the distance scale $r\subs{ls}$ can be much larger. In this sense, we need running in the asymmetry factor $A$ in order to satisfy the quasar constraint.}
Quasar observation \cite{Hirata:2009ar} constrains the power asymmetry of the universe in the quasar scale $N_q>N_*$. If we define $r_q$ as our physical distance to the farthest quasar, we can write $N_q$, the length scale of our distance to the quasars as (see also \refig{pa-CMBq})
\eq{r_q\equiv\frac{e^{-N_q}}{H}.}

The quasar observation finds no asymmetry, requiring $A<0.02$ in the quasar scale $N_q$; see \cite{Hirata:2009ar}. From \refeq{pa-po-A11} and \refeq{pa-po-A2}, we find that it may be accommodated with the CMB scale power asymmetry $A\sim0.07$, if the non-Gaussianity parameter, $\fNL$ or $\tNL$, has a running\footnote{
The discussion of the running of the local non-Gaussianities, and how it may affect the scale dependence of the CMB power asymmetry, is beyond the scope of the thesis. Some of the relevant discussions can be found in \cite{Byrnes:2015asa,Cai:2015xba,Adhikari:2015yya,Lyth:2014mga}.}
. The amount of non-Gaussianity and its running depend very much on the inflationary model, but many existing models can provide such a running. For example, in spectator scenario, a large running can be achieved if the effective mass of the spectator field runs between the Hubble exits of the pivot and the quasar scales.

\sssecs{Quadrupole and octupole}
The source of CMB power asymmetry should not generate any excessive quadrupole or octupole in the CMB. Following the conventions in \cite{Erickcek:2008jp,Erickcek:2008sm}, we replicate their derived constraints here, from \refeqraw{4}, and \refeqraw{5} of \cite{Erickcek:2008sm}
\eqa{(kx_d)^2|\Phi_{\vec k}(\tau_d)\sin\overline{\omega}|&\lesssim&5.8{\cal Q},\hspace{0.3in}\mathrm{for\ quadrupole,}\\
(kx_d)^3|\Phi_{\vec k}(\tau_d)\cos\overline{\omega}|&\lesssim&32{\cal O},\hspace{0.34in}\mathrm{for\ octupole,}}
where ${\cal Q}=1.8\times10^{-5}$ and ${\cal O}=2.7\times10^{-5}$. We rephrase them with our convention, with $k x_d|\Phi_{k}(\tau_d)|=\frac{1}{3}|\Delta\zeta|=\sqrt{\P_\zeta}\frac{|\Delta\sigma|}{3\sqrt{\P_{\delta\sigma_*}}}$ where $k x_d=\sqrt2e^{N_{k\subs{max}}-N_*}$. After neglecting the $\sin$ and $\cos$ functions, these two inequalities become
\eqa{N_{k\subs{max}}-N_*&\lesssim&\ln\frac{17.4{\cal Q}}{\sqrt{2\P_\zeta}}\frac{\sqrt{\P_{\delta\sigma_*}}}{|\Delta\sigma|},\hspace{0.355in}\mathrm{for\ quadrupole},\label{eq-pa-cons-quad}\\
N_{k\subs{max}}-N_*&\lesssim&\frac{1}{2}\ln\frac{48{\cal O}}{\sqrt{\P_\zeta}}\frac{\sqrt{\P_{\delta\sigma_*}}}{|\Delta\sigma|},\hspace{0.3in}\mathrm{for\ octupole}.}

Therefore the quadrupole and octupole constraints put a lower bound on $N_*$, the e-folding of the second slow-roll phase (see \refig{pa-Timeline}). In the example shown in \refig{pa-beta}, we have $N_{k\subs{max}}\approx-1$. By plugging in the values of ${\cal O,\ Q},\ \P_\zeta,\ N_*$ and $|\Delta\sigma|/\sqrt{\P_{\delta\sigma}}$ from the example \refig{pa-beta} and \refeq{pa-tf-Dsn}, we can see that both the quadrupole and the octupole constraints are satisfied.

\ssecs{CMB power asymmetry from spectator scenario}
The tachyonic enhancement discussed in \refssec{Enhancing very large scale perturbations from a tachyonic fast-roll phase} in principle applies to any scenario, by providing a large enough field asymmetry on the opposite sides of our Hubble patch, while satisfying all the observational constraints such as quasars, quadrupole and octupole. In this section, we will however only discuss how to produce the observed CMB power asymmetry with spectator mechanism, where the spectator field receives the enhanced field asymmetry from a brief tachyonic fast-roll phase. There are several reasons for the consideration.
\begin{itemize}
\item When the tachyonic enhancement applies on an inflaton field, (or one or more of the inflaton fields,) the newly introduced tachyonic fast-roll phase may change the original behaviour of the inflaton, and hence inflation itself. The tachyonic fast-roll phase may totally destroy inflation, change the dynamics or predictions of inflation dramatically, etc. Therefore, the tachyonic fast-roll phase is preferably applied on a field which does not participate in inflation, e.g.\ a curvaton or spectator field.
\item When the tachyonic enhancement applies on an inflaton field, even if the calculations have been adjusted to take into account \ed{of} the tachyonic phase, it still remains a problem how to motivate the tachyonic potential (and the transitions) within a well-established particle physics framework. This argument also applies to curvaton scenario because it has to produce a significant amount of the current universe components, according to non-Gaussianity and isocurvature perturbation constraints. (See \refssec{Curvaton scenario}.)
\end{itemize}

Having all these said, in this thesis we would only seek an explanation of the observed CMB power asymmetry from spectator scenario, on top of single-field slow-roll inflation. The role of driving inflation and producing matter is then separated from producing the perturbations. The single-field slow-roll inflation can be chosen with a particle physics foundation, while all the perturbations come from the spectator field and contain a mild non-Gaussianity $\fNL$. In principle, there can be more than one spectator field or inflaton, but as a minimum we only consider one of them each.

\fig{pa-s1}{CMB power asymmetry from spectator mechanism}{The parameter space for the inflaton and spectator model in \refssec{CMB power asymmetry from spectator scenario}. The blue line shows the maximum CMB power asymmetry that can be reached by any given $\fNL$. The vertical red bands indicate the latest Planck bounds for local bi-spectrum $\fNL$, for $1\sigma$, $2\sigma$ and $3\sigma$ regions \cite{Ade:2013ydc}. The horizontal blue bands indicate the Planck observational bounds for CMB power asymmetry, also for $1\sigma$, $2\sigma$ and $3\sigma$ regions \cite{Ade:2013nlj}. The neighbouring area of $\fNL\approx 6$ and $A\approx0.06$ is within $1\sigma$ C.L.\ for both observables.}
We then combine the expressions of the CMB power asymmetry \refeq{pa-po-A11} and the field asymmetry \refeq{pa-tf-Ds23}, with the inequalities \refeq{pa-tf-pac} and \refeq{pa-cons-quad}. This gives the upper bound of CMB power asymmetry
\footnote{
The quadrupole constraint is stronger than the octupole constraint in most cases. Therefore we drop the octupole constraint in the discussion.}
\eq{\langle A^2\rangle<\frac{9}{25}\times17.4{\cal Q}\,\fNL^2\sqrt{2\Delta N}.\label{eq-pa-s-1}}
After substituting in the values for $Q\lesssim1.8\times10^{-5}$, $\Delta N\approx1$ and $\fNL<10.8$, we find the upper bound $A\lesssim0.14$, which allows the observed central value $A\approx0.07$. The parameter space for $A$ and $\fNL$ is shown in \refig{pa-s1}.

In another slightly more complicated scenario, \ea{we may suppose}\ed{think} that the spectator partially contributes to \ea{the} curvature perturbation but is still the only field that receives the tachyonic enhancement. The other source(s) of \ea{the} curvature perturbation can be the inflaton(s) or some other spectator/curvaton field(s). The upper bound then becomes
\eq{\langle A^2\rangle<8.7{\cal Q}\,\tNL R^{-1}\sqrt\frac{\Delta N}{2},}
where $R$ is defined in \refeq{pa-tf-R}. We substitute the values of $Q\lesssim1.8\times10^{-5}$, $\Delta N\approx1$ and $\tNL<2800$, so the above equation yields \eq{A<0.56R^{-\frac{1}{2}}.}
Therefore $A=0.07$ is also allowed, and this model can generate a much stronger CMB power asymmetry.

To summarize, this chapter has provided an explanation for the observed CMB power asymmetry with the spectator scenario. The (nearly) scale invariant and (almost) Gaussian primordial fluctuation cannot produce the observed CMB power asymmetry. We have proposed a viable solution through the violation of scale invariance on scales much larger than the pivot scales, which can be realized with a tachyonic fast-roll phase. In order to avoid disruptions in inflationary dynamics or in the production of visible matter, \ea{the} spectator scenario \ea{could be}\ed{has become} an ideal candidate.

%% file: Chapters/Conclusions.tex
This thesis started from the observational aspects of the CMB, and investigated early universe theories step by step from single-field slow-roll inflation, to multi-field inflation, and then to \ea{the} spectator scenario. During the investigation, we have compared cosmological predictions of relevant models with observations. \ea{The s}\ed{S}pectator scenario has been found \ea{to be} in agreement with cosmological observations, being able to produce large or small local non-Gaussianity without any isocurvature perturbations. This thesis has also proposed an explanation for the CMB power asymmetry with the spectator scenario.

To be specific, in \refsec{CMB}, we have studied CMB temperature fluctuation. Its angular power spectrum \ea{is consistent with}\ed{has indicated} a nearly scale invariant primordial perturbation. Its angular bi-spectrum and angular tri-spectrum are compatible with zero, suggesting a (mostly) Gaussian primordial perturbation. CMB observations \ea{have not yet detected}\ed{are still yet to discover} the \ea{primordial} $B$ mode and isocurvature perturbations, but the CMB power asymmetry has drawn community interest.

The CMB favors a featureless, ordinary beginning of the universe. In \refsec{Single}, single-field slow-roll inflation has been \ea{shown}\ed{proven} to agree well with observations. We have demonstrated the slow-roll mechanism for single-field inflation. Single-field slow-roll inflation can attain a nearly scale invariant and almost Gaussian primordial perturbation. It does not produce any isocurvature perturbation either. We have listed the observational constraints in \reftab{ssi-obs}. Models of single-field slow-roll inflation have been tested against \reftab{ssi-obs}, such as power-law and inflection point potentials.

In \refsec{Multi}, we have derived the generic observational predictions of multi-field inflation. Multi-field inflation can provide richer features than single-field slow-roll inflation, for example a significant non-Gaussianity with nearly scale invariant spectrum. The generic predictions of multi-field inflation can be greatly simplified for two-field inflation or scenarios with separable potentials. When an additional perfect fluid coexists during single-field inflation, we have found it may induce significant curvature perturbation.

We started \refsec{Spectator} by summarizing the cosmological predictions of \ea{the} minimal curvaton scenario. We then proposed the spectator scenario, and derived its evolution at background and perturbation levels. The spectator scenario can produce large or small non-Gaussianities, whilst not generating any isocurvature perturbations. As typical examples of \ea{the} spectator scenario, we have examined step function potential\ea{s}, inflection point potential\ea{s}, and hyperbolic tangent potential\ea{s}, all of which have demonstrated satisfactory agreements with observations.

The spectator field is able to dominate cosmological perturbations without \ea{significantly} affecting inflationary dynamics or matter production. This advantage makes it an ideal candidate for an explanation of the CMB power asymmetry. With a brief tachyonic fast-roll phase well before the Hubble exit of the pivot scales, \ea{the} spectator scenario has been shown \ea{to be} capable \ea{of}\ed{in} bringing about the observed CMB power asymmetry in \refsec{Asymmetry}. This realization also agrees with other cosmic observations. On the other hand, with a nearly scale invariant and Gaussian primordial perturbation, generation of the observed CMB power asymmetry has been shown \ea{to be} difficult.

In the future, we may expect the discovery of \ea{primordial} $B$ mode\ea{s} in the CMB \ea{due to gravitational waves from inflation}, which will surely bring about a great change the field of cosmology. Potential discoveries in particle physics, by the Large Hadron Collider or future particle accelerators, may change the way we understand High Energy Physics. Dark matter observations are also trying to accumulate evidence for detection, hopefully to expand the Standard Model in \ea{the} near future. Broadly speaking, all the above knowledge we gain are vital towards a unified and deeper understanding of particle physics and cosmology. Specifically, they will also further distinguish early universe models including the spectator scenario.

%% file: Chapters/Funcs.tex
\ssecs{Spherical harmonic functions}
Spherical harmonic functions form an orthogonal and complete base of spherical functions, i.e.
\eq{\int \dd^2\uv nY_{lm}^*(\uv n)Y_{l'm'}(\uv n)=\delta_{ll'}\delta_{mm'}.}
So any spherical function can be expanded in terms of spherical harmonics
\eq{f(\uv n)=\sum_{lm}f_{lm}Y_{lm}(\uv n),}
where
\eq{f_{lm}\equiv\int\dd^2\uv nY_{lm}^*(\uv n)f(\uv n).}
We can also decompose the product of two spherical harmonics
\eq{\Ga{1}{2}{3}\equiv\int\dd^2\uv nY_{l_1m_1}(\uv n)Y_{l_2m_2}(\uv n)Y_{l_3m_3}(\uv n),\label{eq-asf-Gaunt}}
So $\Ga{1}{2}{3}$ is called the \emph{Gaunt integral}.

The spherical harmonics have other properties such as
\eq{Y_{lm}^*(\uv n)=(-1)^mY_{l-m}(\uv n),}
\eq{Y_{lm}(-\uv n)=(-1)^lY_{lm}(\uv n),}
\eq{\sum_mY_{lm}^*(\uv n)Y_{lm}(\uv n')=\frac{2l+1}{4\pi}P_l(\uv n\cdot\uv n').}

\ssecs{3j symbols}
The $3j$ symbols are used to characterize the coupling between different angular momenta. The reader can find the precise definition and properties in \cite{Bartolo:2004if,Olver:2010:NHMF,NIST:DLMF}.
\begin{enumerate}
\item Triangle conditions

The $3j$ symbol $\Wj{1}{2}{3}$ is nonzero if and only if all of the following conditions are satisfied:
\begin{itemize}
\item $2l_1,2l_2,2l_3\in\mathbb{N}_0$.
\item $|l_1-l_2|\le l_3\le l_1+l_2$.
\item $m_i=-l_i,-l_i+1,\dots,l_i$, for $i=1,2,3$.
\item $m_1+m_2+m_3=0$.
\end{itemize}
\item Symmetries
\eqa{\Wj{1}{2}{3}&=&\Wj{2}{3}{1},\\
\Wj{1}{2}{3}&=&(-1)^{l_1+l_2+l_3}\Wj{1}{3}{2},\\
\Wj{1}{2}{3}&=&(-1)^{l_1+l_2+l_3}\Wigner{l_1}{l_2}{l_3}{-m_1}{-m_2}{-m_3}.}
\item Orthogonalities
\eqa{\sum_{m_1m_2}(2l_3+1)\Wj{1}{2}{3}\Wigner{l_1}{l_2}{\tilde l_3}{m_1}{m_2}{\widetilde{m}_3}&=&\delta_{l_3\tilde l_3}\delta_{m_3\widetilde m_3},\\
\sum_{l_3m_3}(2l_3+1)\Wj{1}{2}{3}\Wigner{l_1}{l_2}{l_3}{\widetilde m_1}{\widetilde m_2}{m_3}&=&\delta_{m_1\widetilde m_1}\delta_{m_2\widetilde m_2}\,.\hspace{1cm}.}
\item Other relations
\eqa{\Ga{1}{2}{3}&=&\sqrt\frac{(2l_1+1)(2l_2+1)(2l_3+1)}{4\pi}\nonumber\\
&&\times\Wjz{1}{2}{3}\Wj{1}{2}{3},}
\eqa{&&\Wjz{1}{2}{3}^2\nonumber\\
&=&\frac{1}{2}\int_{-1}^1P_{l_1}(x)P_{l_2}(x)P_{l_3}(x)\,\dd x\nonumber\\
&=&\left\{\begin{array}{l@{\hspace{0.3in}}l}
0,&L\mathrm{\ odd},\\
\displaystyle\frac{(L-2l_1)!(L-2l_2)!(L-2l_3)!(L/2)!^2}{(L+1)!(L/2-l_1)!^2(L/2-l_2)!^2(L/2-l_3)!^2},&L\mathrm{\ even},\end{array}
\right.\hspace{0.3in}}
where $L\equiv l_1+l_2+l_3$ in the above expression.
\end{enumerate}

\ssecs{6j symbols}
The $6j$ symbols can be defined from $3j$ symbols as
\eqa{\left\{\begin{array}{ccc}L_1&L_2&L_3\\l_1&l_2&l_3\end{array}\right\}&\equiv&\sum_{M_im_j}(-1)^{l_1+l_2+l_3+m_1+m_2+m_3}\Wigner{L_1}{L_2}{L_3}{M_1}{M_2}{M_3}\nonumber\\
&&\times\Wigner{L_1}{l_2}{l_3}{M_1}{m_2}{-m_3}\Wigner{l_1}{L_2}{l_3}{-m_1}{M_2}{m_3}\nonumber\\
&&\times\Wigner{l_1}{l_2}{L_3}{m_1}{-m_2}{M_3},}
where $i,j=1,2,3$. Its precise definition and properties can also be found in \cite{Bartolo:2004if,Olver:2010:NHMF,NIST:DLMF}.